\documentclass{statsoc}

\usepackage{amsfonts,amstext,amssymb}
\usepackage{multirow}
\usepackage{subfigure}
\usepackage[utf8]{inputenc} 
\usepackage[T1]{fontenc}    
\usepackage{hyperref}  
\hypersetup{
	colorlinks=true,       
	linkcolor={blue!66!black},        
	citecolor={blue!66!black},        
	filecolor=magenta,     
	urlcolor=magenta
}
\usepackage{url}            
\usepackage{booktabs}       
\usepackage{nicefrac}       
\usepackage{microtype}      
\usepackage{lipsum}
\usepackage{graphicx}
\usepackage{wrapfig}
\graphicspath{ {./images/} }
\usepackage{algorithm,algorithmic}
\usepackage{footmisc}
\usepackage[a4paper]{geometry}
\usepackage{amsmath}
\usepackage{natbib}
\usepackage{dsfont}
\usepackage[dvipsnames]{xcolor}

\newcommand{\cB}{\mathcal{B}}
\newcommand{\cC}{\mathcal{C}}

\newcommand{\cN}{\mathcal{N}}
\newcommand{\cO}{\mathcal{O}}

\newcommand{\cX}{\mathcal{X}}

\newcommand{\EE}{\mathbb{E}}

\newcommand{\NN}{\mathbb{N}}
\newcommand{\PP}{\mathbb{P}}

\newcommand{\RR}{\mathbb{R}}

\newcommand{\norm}[1]{\|#1\|}
\def\T{{ \mathrm{\scriptscriptstyle T} }}

\newcommand{\rhod}{{\rho}\mathcal{D}({p},{q})}

\ifx\proof\undefined
\newenvironment{proof}{\par\noindent{\bf Proof\ }}{\hfill\BlackBox\\[2mm]}
\fi

\ifx\theorem\undefined
\newtheorem{theorem}{Theorem}
\fi
\ifx\example\undefined
\newtheorem{example}{Example}
\fi
\ifx\property\undefined
\newtheorem{property}{Property}
\fi
\ifx\lemma\undefined
\newtheorem{lemma}{Lemma}
\fi
\ifx\proposition\undefined
\newtheorem{proposition}{Proposition}
\fi
\ifx\remark\undefined
\newtheorem{remark}{Remark}
\fi
\ifx\corollary\undefined
\newtheorem{corollary}{Corollary}
\fi
\ifx\definition\undefined
\newtheorem{definition}{Definition}
\fi
\ifx\conjecture\undefined
\newtheorem{conjecture}{Conjecture}
\fi
\ifx\fact\undefined
\newtheorem{fact}{Fact}
\fi
\ifx\claim\undefined
\newtheorem{claim}{Claim}
\fi
\ifx\assumption\undefined
\newtheorem{assumption}{Assumption}
\fi
\ifx\condition\undefined
\newtheorem{condition}{Condition}
\fi

\usepackage{enumitem}
\usepackage{tikz}

\def\mycolumn{1}

\def\mytitle{Online Kernel CUSUM for Change-Point Detection}
\title{\mytitle}

\title[Online Kernel CUSUM]{\mytitle}
\author[Wei and Xie]{Song Wei}
\address{Georgia Institute of Technology,  
Atlanta, Georgia, 30332,
U.S.A. }
\author[Wei and Xie]{Yao Xie}
\address{Georgia Institute of Technology,  
Atlanta, Georgia, 30332,
U.S.A. }
\email{yao.xie@isye.gatech.edu}

\begin{document}
\begin{abstract}
We present a computationally efficient online kernel Cumulative Sum (CUSUM) method for change-point detection that utilizes the maximum over a set of kernel statistics to account for the unknown change-point location. Our approach exhibits increased sensitivity to small changes compared to existing kernel-based change-point detection methods, including the Scan-B statistic, corresponding to a non-parametric Shewhart chart-type procedure. We provide accurate analytic approximations for two key performance metrics: the Average Run Length (ARL) and Expected Detection Delay (EDD), which enable us to establish an optimal window length to be on the order of the logarithm of ARL to ensure minimal power loss relative to an oracle procedure with infinite memory. Moreover, we introduce a recursive calculation procedure for detection statistics to ensure constant computational and memory complexity, which is essential for online implementation. Through extensive experiments on both simulated and real data, we demonstrate the competitive performance of our method and validate our theoretical results.
\end{abstract}

\section{Introduction}\label{sec:intro}

Online change-point detection is a fundamental and classic problem in statistics and related fields. The goal is to detect a change in the underlying data distribution as quickly as possible after the change has occurred while maintaining a false alarm constraint. Traditional approaches are based on parametric models that detect changes in distribution parameters, such as the Shewhart chart \citep{shewhart1925application}, Cumulative Sum (CUSUM) \citep{page1954continuous}, and Shiryaev-Roberts (SR) \citep{shiryaev1963optimum}. For comprehensive and systematic reviews of parametric change-point detection, readers are referred to \citet{basseville1993detection,siegmund2013sequential,tartakovsky2014sequential,xie2021sequential}.

In modern applications, especially those involving high-dimensional data settings, specifying the exact probability distribution can be challenging, necessitating non-parametric and distribution-free methods. Non-parametric kernel-based approaches do not require distributional assumptions, allow flexible kernel choices, and have gained significant attention due to their flexibility and wide applicability. Notably, the kernel Maximum Mean Discrepancy (MMD) for two-sample tests \citep{gretton2006kernel,smola2007hilbert,gretton2012kernel,muandet2017kernel} has been extensively studied.

Literature on kernel-based change-point detection has focused on the offline setting, where the objective is detecting and locating change points retrospectively. Offline change-point detection procedures typically involve two-sample tests for samples with multiple candidate change-point locations. Notable examples of such procedures include the maximum Kernel Fisher Discriminant Ratio-based procedure \citep{harchaoui2008kernel}, the kernel MMD-based Scan $B$-procedure \citep{li2019scan}, and a recent kernel-based scan procedure designed to enhance detection power in high-dimensional settings \citep{song2022new}. 

Online change-point detection seeks to detect changes as soon as possible using sequential data. Existing online kernel-based change-point detection procedures are mainly the Shewhart chart-type, such as \citet{li2019scan}. The Shewhart chart-type procedure differs from the CUSUM-type procedure, which relies on CUSUM recursion and is generally more sensitive to detecting small changes \citep{xie2021sequential}. The CUSUM-type kernel procedure was rarely considered in the literature, possibly due to its higher computational and memory cost than Shewhart chart-type procedures. Recently, \citet{flynn2019change} presented a Kernel CUSUM procedure by replacing the likelihood ratio in the CUSUM recursion with the {\it linear-time} MMD statistic and derived crude bounds for both ARL and EDD. However, the linear-time MMD statistics used therein to achieve the recursive computation of the detection statistic have a lower statistical detection power compared with the full MMD statistic considered in this paper. It remains challenging to apply full MMD statistics in the CUSUM procedure due to increased computational and memory costs as the time horizon expands, and to analyze its theoretical properties, which we aim to tackle in this paper.  
Here, we present a summary of existing kernel-based online change-point detection methods in Table~\ref{table:kernelcpd}, and a more comprehensive literature survey can be found in Section~\ref{subsec:related_work}.

\begin{table}
\caption{\label{table:kernelcpd} Summary of online kernel change-point detection procedures.}
\centering
\begin{tabular}{c|l} 
\hline
 Detection procedure type  & Existing literature  \\ 
\hline
 \multirow{3}{*}{Shewhart chart}  &  \citet{huang2014high,chang2019kernel}; \\
   &  \citet{li2019scan,bouchikhi2019kernel}; \\
    &  \citet{cobb2022sequential} \\
\cline{2-2}
 CUSUM & \citet{flynn2019change} \\
\hline
\end{tabular}
\end{table}

In this paper, we present a kernel-based CUSUM procedure for online change-point detection, referred to as the {\it Online Kernel CUSUM}, which differs from the Shewhart chart-type procedures (e.g., the Scan $B$-procedure \citep{li2019scan}) in the search for the unknown change-point location, and leads to improved performance in detecting small changes. At each time step, the procedure forms a set of self-normalizing kernel test statistics, one for each candidate change-point location, takes the maximum over the detection statistics, and detects a change when the maximum exceeds a pre-specified threshold value. We find numerically accurate approximations for two standard performance metrics: the Average Run Length (ARL), which is useful for calibrating the procedure to control false alarms, and the Expected Detection Delay (EDD). The EDD analysis is essential for determining the optimal window length, but such an analysis is not present in closely related prior works (e.g., the Scan $B$-procedure). Based on the ARL and EDD analysis, we establish the optimal window length, defined as the minimal samples needed to be kept in the memory for our procedure to achieve a similar performance as the oracle procedure with infinite memory, to be on the order of the logarithm of ARL, which interestingly matches the classic result for the window-limited Generalized Likelihood Ratio procedure \citep{lai1995sequential}. Algorithmically, we present a recursive implementation of our procedure to achieve a constant computational and memory cost at each time step, which is essential for online change-point detection. We demonstrate the performance gain of our proposed approach using simulation and real-data experiments.

\subsection{Literature}\label{subsec:related_work}

\vspace{.1in}

\noindent \textit{Kernel methods in change-point detection.}
Early work on kernel change-point detection is \citet{harchaoui2007retrospective} for offline multiple change-point estimation methods, and the method was extended by \citet{arlot2019kernel} to handle an unknown number of change points via a model selection penalty. 
\citet{garreau2018consistent} performed a non-asymptotic analysis to show that the multiple change-point estimations by model selection \citep{arlot2019kernel} can estimate the correct number of change-points with high probability and locate them at the optimal rate.
The problem of calibrating detection procedures also exists for the maximum Kernel Fisher Discriminant Ratio-based approach \citep{harchaoui2008kernel}, where they applied the bootstrap re-sampling method to compute the detection threshold.
In addition, \citet{bouchikhi2019kernel,ferrari2023online} considered a kernel density ratio \citep{huang2006correcting,nguyen2010estimating} based Shewhart chart-type method for online change-point detection.

\vspace{.1in}

\noindent \textit{Kernel MMD and its application in change-point detection.}
\citet{gretton2006kernel,smola2007hilbert,gretton2012kernel} consider a measure of discrepancy between distributions via the Reproducing Kernel Hilbert Space (RKHS) embedding of distributions that leads to the kernel MMD statistic, which can be linked back to the classic U-statistic \citep{serfling2009approximation}. Since then, kernel MMD two-sample tests \citep{li2019optimality,balasubramanian2021optimality} have become popular and have been applied in various settings, ranging from anomaly detection \citep{zou2014nonparametric},
model criticism and goodness-of-fit test \citep{lloyd2015statistical}, robust hypothesis testing \citep{sun2021data}, as well as change-point detection.

There are various offline kernel MMD-based change-point detection: apart from the aforementioned work, such as  \citet{truong2019greedy}, which developed a greedy single change-point estimation procedure, \citet{jones2020kernel}  developed a multiple change-point estimation method leveraging Nystr\"{o}m kernel approximation \citep{williams2001using} to achieve linear complexity.

Online kernel MMD-based change-point detection procedures typically test for changes between a window of reference data and a window of sequential data. 
Recently, there has been much effort focusing on kernel MMD-based Shewhart chart-type procedures, which consider fixed-window length scan statistics, such as the Scan $B$-procedure \citep{li2019scan}. Other contributions include \citet{huang2014high}, which considers the RKHS-based control chart for online change-point detection, using full MMD statistics computed using sequential data (it differs from the Scan $B$-procedure, which is a block-based MMD statistic to enable quick detection), and utilized bootstrap to calibrate the detection, 
\citet{chang2019kernel} that leveraged neural network-based data-driven kernel selection \citep{gretton2012optimal} to boost detection power, \citet{cobb2022sequential} that considered a memory-less geometric distribution of the stopping time to control the false alarm rate and calibrated the detection via bootstrap. However, an issue with the fixed-length scanning window of Shewhart chart-type procedures is that they do not account for unknown change-point location, which may decrease the detection power; \citet{cobb2022sequential} indeed tackled this problem by considering a heuristic time-varying threshold without theoretical analysis.

\section{Methodology}\label{sec:formulation}

\subsection{Preliminaries}\label{sec:preliminary}

Consider independent and identically distributed (i.i.d.) reference samples from the domain $\cX$ (usually taken to be $\mathbb{R}^d$) following an unknown pre-change distribution with density $p$: 
$$X_1,\dots,X_M \overset{i.i.d.}{\sim} {p}.$$ 
At time $t$, we observe i.i.d. data $Y_1,\dots,Y_t$ sequentially. The online change-point detection problem tests the null hypothesis
$$H_0: Y_1,\dots,Y_t \overset{i.i.d.}{\sim} {p},$$ 
against the alternative hypothesis
$$H_1: \exists \  \kappa< t, \  Y_1,\dots,Y_\kappa \overset{i.i.d.}{\sim} {p}, \  Y_{\kappa+1},\dots,Y_t \overset{i.i.d.}{\sim} {q},$$
where ${q}$, distinct from ${p}$, is the density of an unknown post-change distribution and $\kappa>0$ is the unknown change-point location. We aim to detect the change as soon as possible after it has occurred while controlling the false alarms.

Our notations are standard. Under the null hypothesis, we use $\PP_{\infty}$ and $\mathbb{E}_{\infty}$ to denote the probability and expectation when there is no change. Under the alternative hypothesis, when there is a change at time $\kappa$, $\kappa = 0, 1, \ldots$, we use $\PP_{\kappa}$ and $\mathbb{E}_{\kappa}$ to denote the probability and expectation in this case. In particular, $\kappa = 0$ denotes an immediate change. 
We denote $\mathbf{1}_d = (1,\dots,1)^\T \in \RR^d$, $\mathbf{0}_d = (0,\dots,0)^\T \in \RR^d$ and $I_d \in \RR^{d \times d}$ as the identity matrix, where superscript $^\T$ stands for the vector or matrix transpose. For integers $0 < m \leq n$, let $[m:n] = \{m,\dots,n\}$, and $[n] = \{1,\dots,n\}$. We denote $a \wedge b = \min\{a,b\}$, $a \vee b = \max\{a,b\}$ and $(a)^+ = \max\{a,0\}$.
For asymptotic notations, we adopt standard definitions:
$f(m) = o(g(m))$ means for all $c > 0$ there exists $k > 0$ such that $0 \leq f(m) < cg(m)$ for all $m \geq k$;
$f(m) =  \cO(g(m))$ means there exist positive constants $c$ and $k$, such that $0 \leq f(m) \leq cg(m)$ for all $m \geq k$;
$f(m) \sim g(m)$ means there exist positive constants $c$ and $k$, such that $0 \leq cg(m) \leq f(m)$ for all $m \geq k$ and at the same time $f(m) =  \cO(g(m))$.

\subsection{Classic Parametric Procedures}\label{sec:para_methods}

Shewhart chart \citep{shewhart1925application} uses a sliding window to compute detection statistics, which can be either parametric, e.g., based on likelihood ratio, or non-parametric. For the parametric approach, at time step $t$, the Shewhart chart uses the log-likelihood ratio of the most recent sequential observation $Y_t$ as the detection statistic, i.e., $\log ({q(Y_t)}/{p(Y_t)})$. 
However, due to the sliding window approach, the Shewhart chart disregards the past sequential observations, and there can be an ``information loss'' that can lower the detection power (see, e.g., \citet{xie2021sequential}).

The likelihood-ratio based on the CUSUM procedure \citep{page1954continuous} assumes that densities $p$ and $q$ are both known. Given the sequential data $Y_1,\dots,Y_t$, the CUSUM procedure computes the cumulative log-likelihood ratio while taking the maximum with respect to the unknown change-point location, i.e.,
\begin{equation}\label{eq:cusum_original_form}
\begin{split}
    S_t    &= \max_{1 \leq \kappa \leq t+1} \sum_{s=\kappa}^{t} \log \frac{q(Y_s)}{p(Y_s)}\\
    &= \sum_{s=1}^t \log \frac{q(Y_s)}{p(Y_s)} - \min_{0 \leq \kappa \leq t} \sum_{s=1}^\kappa \log \frac{q(Y_s)}{p(Y_s)}.
\end{split}
\end{equation}
The CUSUM procedure utilizes complete history information and is particularly popular for online change-point detection due to its recursive implementation:
\begin{equation}\label{eq:cusum}
    S_t = \left(S_{t-1} + \log \frac{q(Y_t)}{p(Y_t)}\right)^+ , \quad S_0 = 0.
\end{equation}
The CUSUM procedure is defined as the first time $t$ that the statistic $S_t$ exceeds a pre-set threshold.
Notably, \citet{lorden1971procedures, moustakides1986optimal} showed that the likelihood ratio-based CUSUM enjoys the optimality property.

When the post-change distribution has unknown parameters $\theta$, the Generalized Likelihood Ratio (GLR) procedure \citep{lorden1971procedures,siegmund1995using} is used: $$\max_{1\leq \kappa \leq t-1} \sup_{\theta \in \Theta}\sum_{s= \kappa }^{t} \log \frac{q(Y_s;\theta)}{p(Y_s)},$$
which involves maximum likelihood estimation of both change-point location $\kappa$ and parameter $\theta$. Although the inner supremum accounts for the unknown post-change distribution parameter,  the GLR statistic loses the recursive implementation. To reduce the computational cost, a {\it window-limited}~ GLR (W-GLR) is adopted by restricting the search for the potential change-point to the most recent $w$ data points, i.e., the outer maximization is over $\kappa \in [t-w:t-1]$. The window length parameter $w$ is crucial to the success of the W-GLR procedure, and \citet{lai1995sequential} proved that, as $\gamma \rightarrow \infty$, the optimal window length under the Gaussian mean shift assumption is $w \sim \log \gamma.$
According to \citet{lai2001sequential}, by choosing such a window length, ``there is little loss of efficiency in reducing the computational complexity by the window-limited modification''.

\subsection{Kernel MMD-Based Scan $B$-Procedure}

Kernel-based statistics are popular for non-parametric two-sample tests of high-dimensional data (e.g., \cite{gretton2012kernel}) due to their flexibility. The Shewhart chart type of kernel-based online change-point detection procedure, Scan $B$, has been considered in \citep{li2019scan}. In the online Scan $B$-procedure, $N$ {\it pre-change blocks} with equal block size $B$, denoted by $\color{black}\mathbf{X}_B^{(n)}, n\color{black} \in [N]$, are built by randomly sampling $N B$ samples from the reference data $X_1,\dots,X_M$ without replacement, assuming the reference sample size $M$ is large enough such that $M > N B$. At time $t$, the {\it post-change block} consists of the most recent $B$ sequential data, i.e., $\mathbf{Y}_B(t) = (Y_{t-B+1},\dots, Y_t)$. The detection statistic is obtained by computing the unbiased MMD statistic between the post-change block and each pre-change block and taking their average:
\begin{equation}\label{eq:scan-b}
    {Z}_B(t) = \frac{\hat{\mathcal{D}}_B(t)}{\sqrt{\operatorname{Var}_{\infty}(\hat{\mathcal{D}}_B(t))}},
\end{equation}
where $\operatorname{Var}_{\infty} (\cdot)$ denotes the variance under $H_0$ and
\begin{equation}\label{eq:scan-b_unmoralized}
    \hat{\mathcal{D}}_B(t) = \frac{1}{N}\color{black} \sum_{n=1}^N \color{black} {\hat{\mathcal{D}}(\color{black}\mathbf{X}_B^{(n)}\color{black},\mathbf{Y}_B(t))}.
\end{equation}
For $\mathbf X = (X_1,\dots,X_B)$ and $\mathbf Y = (Y_1,\dots,Y_B)$, here we use the unbiased estimator of MMD \citep{gretton2012kernel}:
\begin{equation}\label{eq:MMD_estimator}
\begin{split}
    \hat{\mathcal{D}}(\mathbf X, \mathbf Y)&= \frac{1}{B(B-1)} \sum_{i=1}^{B} \sum_{j \neq i}^{B} k(X_i,X_j) + \frac{1}{B(B-1)} \sum_{i=1}^{B} \sum_{j \neq i}^{B} k(Y_i,Y_j) \\
    &\quad \ -  \frac{2}{B(B-1)} \sum_{i=1}^{B} \sum_{j \neq i}^{B} k(X_i,Y_j) \\ 
    &= \frac{1}{B(B-1)}\sum_{i=1}^{B} \sum_{j \neq i}^{B} h\left(X_{i}, X_{j}, Y_{i}, Y_{j}\right),
\end{split}
\end{equation}
where
\if\mycolumn1
 \begin{equation}\label{eq:kernel_h}
    h(x_{1}, x_{2}, y_{1}, y_{2})=k\left(x_{1}, x_{2}\right)+k\left(y_{1}, y_{2}\right)-k\left(x_{1}, y_{2}\right)-k\left(x_{2}, y_{1}\right),
 \end{equation}
\else
 \begin{equation}\label{eq:kernel_h}
 \begin{split}
    h&(x_{i}, x_{j}, y_{i}, y_{j})\\
    &=k\left(x_{i}, x_{j}\right)+k\left(y_{i}, y_{j}\right)-k\left(x_{i}, y_{j}\right)-k\left(x_{j}, y_{i}\right),
 \end{split}
 \end{equation}
\fi
and $k(\cdot,\cdot)$ is the user-specified kernel function:
\begin{equation*}
    k(\cdot,\cdot): \cX \times \cX \rightarrow \RR.
\end{equation*}
Commonly used kernel functions include 
Gaussian radial basis function (RBF) $k(x,y) = \exp \{-{\norm{x-y}^2}/{{{r}}^2}\},$ where $\norm{\cdot}$ is the vector $\ell_2$ norm and ${{r}} > 0$ is the bandwidth parameter.

Scan $B$-procedure uses one post-change block $\mathbf{Y}_B(t)$ in computing MMD with respect to all the reference blocks and only requires $\cO(N B^2)$ computations. Since the block size $B$ is typically held constant and $N$ will depend on the reference sample size $M$, the Scan $B$-procedure achieves linear complexity, addressing the quadratic computational complexity issue in MMD statistics.

The normalizing constant, i.e., the standard deviation of $\hat{\mathcal{D}}_B(t)$ under $H_0$, has an analytic expression and thus can be estimated using the reference samples prior to the online implementation of the detection procedure. 
\begin{lemma}[Lemma 3.1 \citep{li2019scan}]\label{lma:var_H0}
Under $H_0$, for block size $B \geq 2$ and the number of pre-change blocks $N > 0$, we have
\begin{equation*}
    \begin{split}
        &\operatorname{Var}_{\infty} (\hat{\mathcal{D}}_B(t)) \\
  &  \quad = \frac{2\left(\mathbb{E}\left[h^{2}\left(X, X^{\prime}, Y, Y^{\prime}\right)\right] + {(N-1)}\operatorname{Cov}\left[h(X,X',Y,Y'), h\left(X^{\prime \prime}, X^{\prime \prime \prime}, Y, Y^{\prime}\right)\right]\right)}{NB(B-1)},
    \end{split}
\end{equation*}
where $X, X^{\prime}, X^{\prime \prime}, X^{\prime \prime \prime}, Y, Y^{\prime} $ are i.i.d. random variables following ${p}$.
\end{lemma}

The online Scan $B$-procedure with block size $B$ is defined by the first time $t$ that $Z_B(t)$ exceeds a pre-set detection threshold. 
\textcolor{black}{In practical implementation, one can either wait for \(B\) observations before starting the procedure or begin calculating the detection statistic from \(t = 1\) and only ``forget'' data once \(t > w\); this can allow a quicker ``start'' initially. We adopt the latter approach in our experiments.} For brevity, we refer to the online Scan $B$-procedure as the Scan $B$-statistic or procedure, since this work does not consider the offline setting.

\section{Proposed Detection Procedure} \label{sec:detect_stat}

Inspired by the original form of recursive online CUSUM \eqref{eq:cusum_original_form}, we propose to consider a parallel set of Scan $B$-statistics ${Z}_B(t)$ in \eqref{eq:scan-b} with block size $B$ taking values from $[2:w]$ and take their maximum. 
For a pre-selected threshold $b$, the {\it online kernel CUSUM} procedure is defined via the following stopping time:
\begin{equation}\label{eq:stopping_time}
    T_{w} = \inf \left\{t\geq 2 : \max_{B \in \big[2: \color{black} \min{\{w,t\}} \color{black}\big]} {Z}_B(t) \geq b\right\}.
\end{equation}
where $ w $ is the {\it window length} parameter.

\begin{figure}[H]
\centerline{
\includegraphics[width = .95\textwidth]{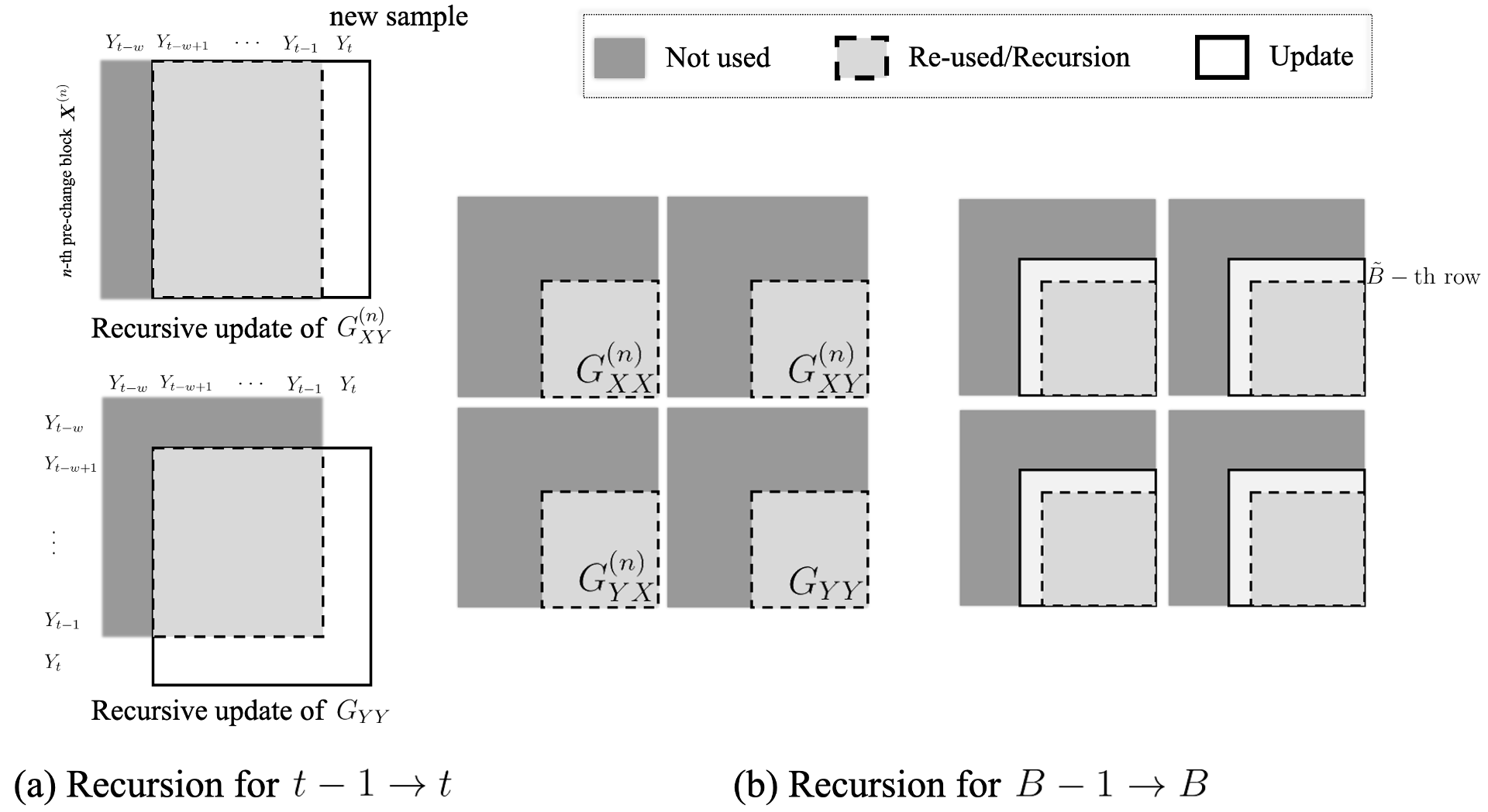}
}
\caption{Graphical illustration of the recursive calculation: $\color{black}G_{XX}^{(n)}\color{black}, \color{black}G_{XY}^{(n)}\color{black}$ and $G_{YY}$ denote the Gram metrics and $\tilde B = w - B + 1$; see definitions in Algorithm~\ref{algo:kernel_CUSUM}. The light gray part of the Gram matrix can be readily obtained from previous calculations, and only the white part needs to be updated (or recalculated).}
\label{fig:online_update_illus}
\end{figure}

The detection statistic in \eqref{eq:stopping_time} can be computed recursively, which is crucial for online implementation: (a) Specifically, the computation of the MMD statistic involves evaluating the Gram matrix, which can be updated recursively as we receive sequential data. Such a recursive update with increasing time $t$ was first considered in \citet{li2019scan}. (b) The main difference between our CUSUM-type procedure and the Shewhart chart-type Scan $B$-procedure is the window-limited maximization with respect to the block size $B$. Importantly, this additional window-limited maximization can be implemented in $\cO(N w^2)$ computations, ensuring that the overall computational and memory complexity of our procedure remains the same as that of the Scan $B$-procedure with block size $w$. Moreover, since the Gram matrix can be updated recursively, the overall computational and memory complexity remains linear in sample size, and it does not grow with time, ensuring computational efficiency in practice. One can see the complete details of our detection procedure in Algorithm~\ref{algo:kernel_CUSUM} and the graphical illustration of the recursive calculations in Figure~\ref{fig:online_update_illus}. \textcolor{black}{Appendix~\ref{appendix:moment} provides detailed derivations for the updates in Algorithm~\ref{algo:kernel_CUSUM}}.

Compared to the Scan $B$-procedure, our proposed procedure offers improved performance by enhancing detection power under the same false alarm constraint, achieved by taking the maximum over detection statistics with different block sizes $B$. Since the change-point $\kappa$ is unknown, the maximum is likely to be achieved for the block size $B$ such that $[t-B+1, t]$ contains all post-change samples. In contrast, the Scan $B$-procedure employs a fixed post-change block size and may include some pre-change samples in the post-change sample block, which can diminish its detection power. Maximizing the utilization of a small post-change sample size is crucial to detecting the change as soon as possible after it occurs. Moreover, our proposed detection statistic has a constant expected increment value, facilitating the derivation of the analytic EDD approximation. However, the Scan $B$ statistic includes varying post-change samples as it scans through the change-point over time, making it challenging to derive an analytic EDD approximation.

\begin{figure}[h!]
\centerline{
\includegraphics[width = 0.5\textwidth]{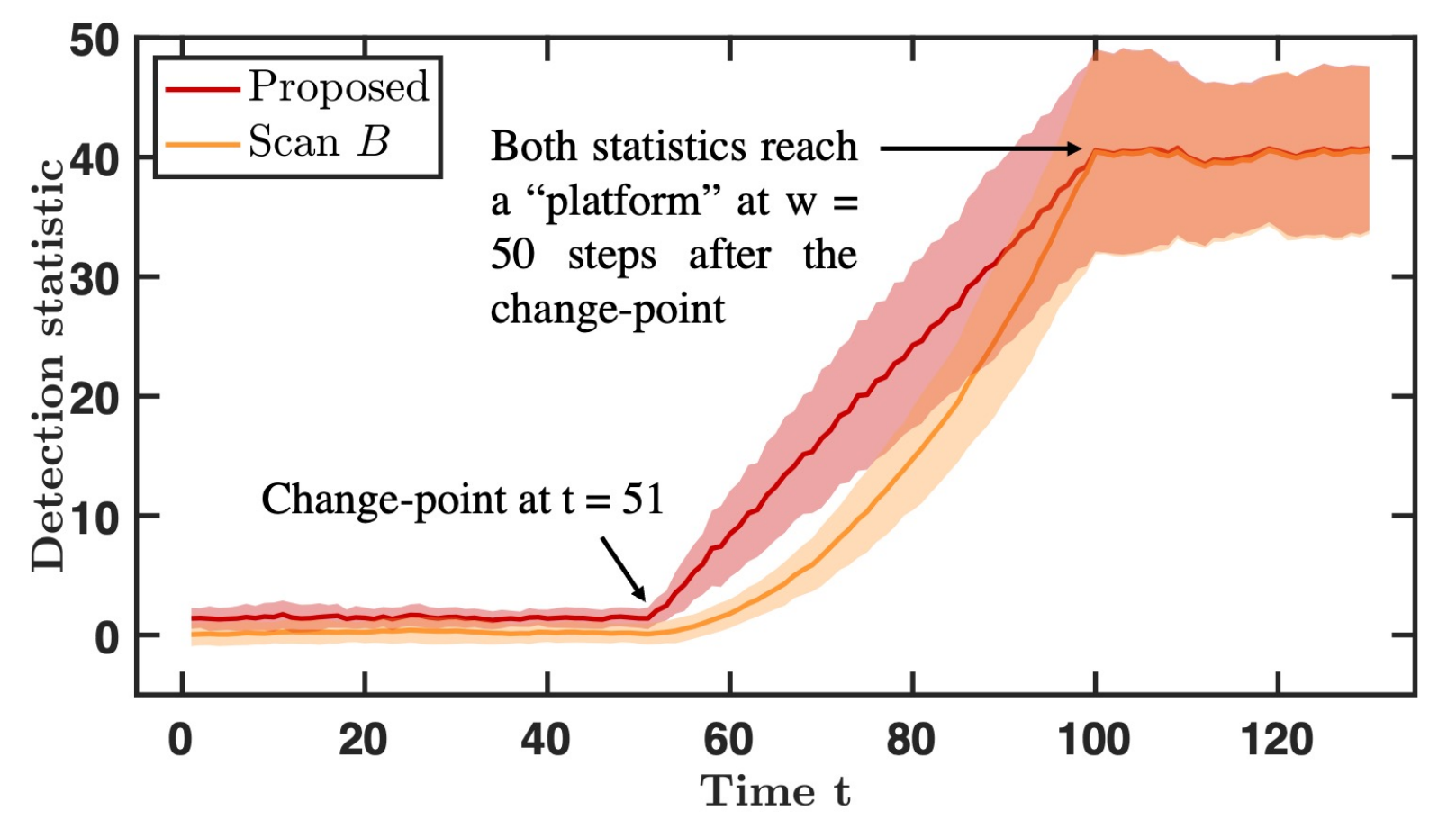}
}
\caption{Trajectories of the mean (solid line) and standard deviation (shaded region) of our proposed detection statistic and the Scan-$B$ statistic over 100 independent trials. A distributional change occurs at $t = 51$, from ${p} = {\cN}(\mathbf{0}_{20}, I_{20})$ to a Gaussian mixture $0.3\, {\cN}(\mathbf{0}_{20}, I_{20}) + 0.7\, {\cN}(\mathbf{0}_{20}, 4I_{20})$. Our proposed method exhibits a noticeably sharper increase at the change-point compared to the Scan-$B$ statistic, demonstrating the effectiveness of constructing the detection statistic via the maximum over a range of set sizes $B$.}
\label{fig:detect_stat_illus_more}
\end{figure}

Figure~\ref{fig:detect_stat_illus_more} shows numerical evidence for the advantages of our proposed online kernel CUSUM over the Scan $B$-procedure, in which the mean trajectories over 100 independent trials are plotted; we highlight the mean $\pm$ standard deviation region. Detailed experimental settings can be found in Section~\ref{sec:numerical}. Our detection statistic has a larger slope after the change-point than the Scan $B$-procedure, indicating a quicker detection. Additionally, both statistics stop increasing after $w=50$ steps from the change-point and plateau around similar values, which validates the necessity of choosing a large $w$ according to \eqref{eq:Bmax_choice} in our following EDD analysis to ensure the procedure stops with a moderate EDD.

\begin{algorithm}[!htp]

 \caption{Online Kernel CUSUM}\label{algo:kernel_CUSUM}
 \begin{algorithmic}[1]
 \renewcommand{\algorithmicrequire}{\textbf{Input:} }

  \STATE Initialization: estimate $\rho$ \eqref{eq:cpq} using historical data, take post-change block $\mathbf{Y} = \mathbf{Y}_w(t)$, and calculate the Gram matrices $$\color{black}G_{XX}^{(n)}\color{black} = k(\color{black}\mathbf{X}^{(n)}\color{black},\color{black}\mathbf{X}^{(n)}\color{black}), \quad \color{black}G_{XY}^{(n)}\color{black} = k(\color{black}\mathbf{X}^{(n)}\color{black},\mathbf{Y}), \quad \color{black}n\color{black}=1,\dots,N, \quad G_{YY} = k(\mathbf{Y},\mathbf{Y})$$
 \\ 
  \WHILE {True}
  \STATE receive sample $Y_t$
  \STATE update post-change block: 
      $$\mathbf{Y} (1:w-1) \leftarrow \mathbf{Y} (2:w), \quad 
      \mathbf{Y} (w) \leftarrow Y_t$$
  \STATE update Gram matrix $G_{YY}$: 
  \begin{align*}
      &G_{YY} (1:w-1, 1:w-1)  \leftarrow G_{YY} (2:w, 2:w), \\
      &G_{YY} (:, w)  \leftarrow k (\mathbf{Y}, Y_t), \quad G_{YY} (w, :)  \leftarrow G_{YY}^\T (:, w)
  \end{align*} 
  \FOR {$i = 1,\dots,N$}
  \STATE update Gram matrix $\color{black}G_{XY}^{(n)}\color{black}$:
  \begin{align*}
      \color{black}G_{XY}^{(n)}\color{black} (:, 1:w-1)  \leftarrow \color{black}G_{XY}^{(n)}\color{black} (:, 2:w), \quad
      \color{black}G_{XY}^{(n)}\color{black} (:, w)  \leftarrow k (\color{black}\mathbf{X}^{(n)}\color{black}, Y_t)
  \end{align*}
  \ENDFOR
  \STATE calculate 
  \begin{align*}
      & z = \color{black} \sum_{n=1}^N \color{black} \color{black}G_{XX}^{(n)}\color{black} (w-1, w) + G_{YY} (w-1, w) - \color{black}G_{XY}^{(n)}\color{black} (w-1, w) - \color{black}\left(G_{XY}^{(n)}\right)^\T\color{black} (w-1, w), \\
      & Z_t =  \frac{\sqrt{2}\rho}{N}z 
  \end{align*}
  \FOR {$B = 3, \dots, w$}
  \STATE calculate $\tilde B = w - B + 1$ 
  \STATE update 
 \begin{align*}
      &z \leftarrow z + \sum_{i = \tilde B + 1}^{w} \color{black} \sum_{n=1}^N \color{black}   \color{black}G_{XX}^{(n)}\color{black} (\tilde B, i) + G_{YY} (\tilde B, i) -  \color{black}G_{XY}^{(n)}\color{black} (\tilde B, i) - \color{black}\left(G_{XY}^{(n)}\right)^\T\color{black} (\tilde B, i) \\
      &Z_t \leftarrow \max\left\{Z_t, \ \frac{2\rho}{N\sqrt{B(B-1)}}z\right\}
  \end{align*}
  \ENDFOR
  \IF {$Z_t > b$}
  \STATE raise an alarm, set $T_{w} = t$ and break
  \ENDIF
  \ENDWHILE
 \end{algorithmic} 
 \end{algorithm}

\subsection{Recursive Computation of Detection Statistic}\label{appendix:procedure_detail}

For a matrix $G \in \RR^{m \times n}$, $G(i_1:i_2,j_1:j_2)$ represents a slice of matrix $G$ containing its $i_1$-th to $i_2$-th rows and $j_1$-th to $j_2$-th columns; in particular, when we want to include all rows (or columns) in this slice, we use ``$:$'' instead of ``$1:m$'' (or ``$1:n$'').
Similarly, for a vector $g \in \RR^{m}$, $g(i:j)$ represents a slice of $g$, containing its $i$-th to $j$-th elements. 
In Algorithm~\ref{algo:kernel_CUSUM}, the recursive computation of detection statistics comes from two parts: a recursive update of Gram matrices (lines $5 - 8$) and a recursive formulation to find the window-limited maximum (lines $9 - 13$), which are direct benefits of CUSUM and closely resemble that of the original CUSUM \eqref{eq:cusum}. 

\subsection{Complexity Analysis}

Even though the online update of Gram matrices only takes $ \cO(Nw)$ computational complexity, the whole procedure still takes $ \cO(N w^2)$ computations to calculate the detection statistic in lines $9 - 13$, which is the same as the Scan $B$-procedure.
To be precise, for fixed $t$, our detection procedure scans block size $B$ to calculate the maximum of the Scan $B$-statistics, where the recursive update in line $12$ only requires additional $\cO(NB)$ computations and does not require additional memory to store those intermediate calculations; such a recursive computation results in $\cO(Nw^2)$ total computations.
We need to clarify that Algorithm~\ref{algo:kernel_CUSUM} requires $\cO(Nw^2)$ memory to store the Gram matrix. However, in theory, we can only use $\cO(Nw)$ memory to store raw observations and recompute the whole Gram matrix with $\cO(Nw^2)$ computations at each time step.
Table~\ref{table:complexity} provides a comparison among the Shewhart chart, CUSUM, W-GLR, Scan $B$, our proposed online kernel CUSUM, and its oracle variant \eqref{eq:stopping_time_noWL}. 

\begin{table}
\caption{\label{table:complexity}Comparison of online change-point detection methods. Here, $w$ is the window length parameter, and $N$ is the number of pre-change blocks.}
\centering
\begin{small}
\resizebox{.95\textwidth}{!}{%
    \begin{tabular}{lccc}
\hline
 ~ & Assumptions & Computational Complexity & Memory  \\ \hline
 Shewhart chart  & ${p}$ and ${q}$ are both known & $ \cO(1)$ & $-$  \\  
 CUSUM  & ${p}$ and ${q}$ are both known & $ \cO(1)$ & $ \cO(1)$  \\ 
 W-GLR  & ${p}$ is known and ${q}$ has known form & $ \cO(w^2)$ & $ \cO(w)$  \\ 
 Scan $B$  & ${p}$ and ${q}$ are both unknown &  $ \cO(Nw^2)$ & $ \cO(Nw)$  \\ 
 Proposed  & ${p}$ and ${q}$ are both unknown  & $ \cO(Nw^2)$ & $ \cO(Nw)$  \\ 
 Oracle  & ${p}$ and ${q}$ are both unknown & $ \cO(Nt^2)$ & $ \cO(Nt)$  \\ \hline
\end{tabular}
}
\end{small}
\end{table}

\section{Theoretical Properties}\label{sec:theory}

In this section, we derive analytic approximations for two standard performance metrics in online change-point detection \citep{xie2013sequential}: the Average Run Length (ARL) (i.e., $\mathbb{E}_{\infty}[T_{w}]$, the expected stopping time when there is no change), and the Expected Detection Delay (EDD) (i.e., $\mathbb{E}_{0}[T_{w}]$, the expected stopping time when the change occurs immediately at $\kappa = 0$).
Our ARL approximation is shown to be accurate numerically, enabling us to calibrate the detection procedure without relying on time-consuming Monte Carlo simulations.
These results also allow us to establish the asymptotically optimal window length $w^\star$ in the same sense as that in \citet{lai1995sequential}.

\subsection{Assumptions on kernels}

We begin with the assumptions on the kernel. Consider the probability measure $P$ (corresponding to the pre-change distribution) on a measurable space $(\cX,\cB)$. For a symmetric, positive semi-definite, and square-integrable (with respect to measure $P$) kernel function $k(\cdot,\cdot)$, Mercer's theorem gives the following decomposition:
\begin{equation*}
    k(x,x^\prime) = \sum_{j = 1}^\infty \lambda_j \varphi_j(x) \varphi_j(x^\prime),
\end{equation*}
where the limit is in $L^2({p})$, $\lambda_1 \geq \lambda_2 \geq \dots \geq 0$ are eigenvalues of
the integral operator induced by kernel $k$, and $\{\varphi_j(\cdot): j \geq 1\}$ are the corresponding orthonormal eigenfunctions such that
\begin{equation*}
    \int_{\cX} \varphi_j(x) \varphi_{j^\prime} (x) d P(x) = \left\{\begin{array}{ll}
1 & \text{ if } j = j^\prime, \\
0 & \text{ otherwise}.
\end{array}\right. 
\end{equation*}
To avoid considering the meaningless zero kernel, we assume the largest eigenvalue is positive, i.e., $\lambda_1 > 0$. In addition, we further impose a technical assumption that all eigenfunctions are uniformly bounded, i.e.,
$$\sup_{j \geq 1} \norm{\varphi_j}_{\infty} < \infty,$$
where $\norm{\varphi_j}_{\infty} = \sup_{x \in \cX} |\varphi_j(x)|$.
This assumption ensures that the kernel $k(\cdot,\cdot)$ is uniformly bounded on domain $\cX \times \cX$, i.e., there exists a constant $K > 0$ such that
\begin{equation}
    0 \leq k(x,y)  \leq K, \quad \forall x, y \in \cX. \label{assumption:bounded_kernel}\tag*{(A1)}
\end{equation}
Our approach does not impose any restrictions on the rank of the kernel function, meaning that the kernel does not necessarily need to be characteristic. For instance, on the domain $\cX = \RR^d$, the polynomial kernel function $k_\ell(x,y) = (x^\T y + c)^\ell$ is finite-rank and cannot differentiate between probability distributions ${p}$ and ${q}$ if they have the same first $\ell$-th order moments. To avoid such undesirable cases, we will impose an assumption on the post-change distribution to ensure that the change can be detected with our chosen kernel at the population level.

Next, we consider the assumption on the post-change distribution ${q}$. To ensure the change from ${p}$ to ${q}$ is detectable with a chosen kernel $k$, we require the population MMD to be positive. 
The (squared) population MMD, denoted by $\mathcal{D}({p},{q})$, can be expressed as:
\begin{align}
    \mathcal{D}({p},{q}) &= \int_{\cX} \int_{\cX} k(x,x^\prime) ({p} - {q}) (x) ({p} - {q}) (x^\prime) d x d x^\prime \nonumber \\
    & = \sum_{j = 1}^\infty \lambda_j \left(\int_{\cX} \varphi_j(x) (p(x) - q(x)) dx \right)^2. \label{eq:eigen_decomp}
\end{align}
The {\it detectability assumption} on ${q}$ is that, for pre-change distribution ${p}$ and kernel $k$ satisfying all aforementioned assumptions, there exists $j$ such that $\lambda_j > 0$ and
\begin{equation}\label{assumption:detection}\tag*{(A2)}
    a_j = \int_{\cX} \varphi_j(x) (p(x) - q(x)) dx \not= 0.
\end{equation}
Here, $a_j$ represents the ``projection'' of the departure $p - q$ along the ``direction'' $\varphi_j$, and Assumption~\ref{assumption:detection} ensures that
$$\mathcal{D}({p},{q}) = \sum_{j = 1}^\infty \lambda_j a_j^2 > 0.$$
We want to remark that, if we assume that kernel $k$ is universal, i.e., $\{\varphi_j(\cdot): j \geq 1\}$ forms an orthonormal basis of $L^2({p})$, the kernel will be full-rank and characteristic, i.e., $\mathcal{D}({p},{q}) > 0$ when ${p} \not= {q}$. However, such an assumption is stronger than we need in the following analysis. Additionally, we want to mention that a characteristic kernel does not necessarily imply its universality unless the kernel is translation-invariant or radial; a more detailed discussion can be found in \citet{sriperumbudur2011universality}.

\subsection{Average Run Length (ARL)}\label{sec:arl_approx}

Approximating ARL $\mathbb{E}_{\infty}[T_{w}]$ in closed-form requires studying the behaviors of extremes of random fields, and we use the techniques developed in \citet{siegmund2010tail,yakir2013extremes}: (i) change-of-measure via exponential tilting, (ii) applying likelihood ratio identity to change the probability of interest into expectation, and (iii) invoking localization theorem. 
We obtain the ARL approximation as follows:
\begin{lemma}[ARL approximation]\label{lma:arl}
As $b \rightarrow \infty$, an approximation to ARL for our proposed  online kernel CUSUM procedure $T_{w}$ \eqref{eq:stopping_time} is given by:
\if\mycolumn1
\begin{align}
    \mathbb{E}_{\infty}[T_{w}]=\frac{\sqrt{2 \pi}}{b} \Bigg\{  \sum_{B = 2}^w e^{\psi_B(\theta_B) - \theta_B b} \frac{\left(2 B-1\right)}{ B\left(B-1\right)} \label{eq:arl_approx}   \nu\left(\theta_B \sqrt{\frac{2\left(2 B-1\right)}{B\left(B-1\right)}}\right)\Bigg\}^{-1} [1+o(1)], 
\end{align}
\else
\begin{align}
    \mathbb{E}_{\infty}[T_{w}]=\frac{\sqrt{2 \pi}}{b} \Bigg\{ & \sum_{B = 2}^w e^{\psi_B(\theta_B) - \theta_B b} \frac{\left(2 B-1\right)}{ B\left(B-1\right)} \label{eq:arl_approx} \\
    &  \nu\left(\theta_B \sqrt{\frac{2\left(2 B-1\right)}{B\left(B-1\right)}}\right)\Bigg\}^{-1} [1+o(1)], \nonumber
\end{align}
\fi
where $\psi_B(\cdot)$ is defined as:
\begin{equation*}
    \psi_B(\theta) = \sum_{n=1}^\infty \frac{\mathbb{E}_{\infty} [Z_B^n(t)]}{n!} \theta^n,
\end{equation*}
and $\theta_B$ is obtained via solving $\dot{\psi_B}(\theta_B) = b$. Moreover, the function $\nu(\cdot)$ is given by (c.f. \cite{siegmund2007statistics}, p. 112)
\begin{equation}
    \nu(\mu) \color{black} = \frac{2}{\mu^2} \exp \left\{-2 \sum_{n=1}^{\infty} \frac 1 n \Phi\left(- \frac{\mu \sqrt{n}}{2}\right)\right\} \color{black} \approx \frac{(2 / \mu)(\Phi(\mu / 2)-0.5)}{(\mu / 2) \Phi(\mu / 2)+\phi(\mu / 2)},
    \label{nu_def}
\end{equation}
where $\phi(\cdot)$ and $\Phi(\cdot)$ are the probability density function and the cumulative distribution function of the standard normal distribution, respectively.
\end{lemma}
Proof of Lemma \ref{lma:arl} can be found in Appendix~\ref{appendix:thm1}. \textcolor{black}{Here, the special function \( \nu(\cdot) \) arises primarily because our analysis of the ARL is based on a change-of-measure argument (see, e.g., \cite{yakir2013extremes}); it was originally introduced in the analysis of the overshoot of the detection statistic at the stopping time via nonlinear renewal theory (see, e.g., \cite{siegmund1985sequential}).}

In practice, incorporating the information of the first two order moments will suffice for ARL approximation, i.e.,
\begin{equation}\label{eq:theta_B_Gaussian_assumption}
    \psi_B(\theta_B) \approx \theta_B^2/2, \quad \theta_B \approx b,
\end{equation}
since under the null hypothesis, the detection statistic has an expectation of zero, this gives us
\begin{equation}\label{eq:ARL_Gaussian_assumption}
    \mathbb{E}_{\infty}[T_{w}]=\sqrt{2 \pi} \frac{b  e^{b^2/2}}{ w } [1+o(1)].
\end{equation}
A detailed derivation of the approximation in \eqref{eq:ARL_Gaussian_assumption} can be found in Appendix~\ref{appendix:thm1}.

To obtain a more precise ARL approximation, one approach would be to incorporate higher order moments while solving for $\theta_B$ using the equation $\dot{\psi_B}(\theta_B) = b$. We use {\it skewness correction} to consider the third order moment $\mathbb{E}_{\infty} [Z_B^3(t)]$ in the following way:
\begin{equation}\label{eq:theta_B_skewness_correction}
\begin{split}
    \psi_B(\theta_B) &\approx \theta_B^2/2 + \mathbb{E}_{\infty} [Z_B^3(t)] \theta_B^3/6, \\
    \theta_B &\approx \frac{-1 + \sqrt{1+2b\mathbb{E}_{\infty} [Z_B^3(t)]}}{\mathbb{E}_{\infty} [Z_B^3(t)]}.
\end{split}
\end{equation}
Similar to the second order moment, the third order moment $\mathbb{E}_{\infty} [Z_B^3(t)]$ can be pre-estimated using the reference data; see Lemma~\ref{lma:third_moment_H0} in Appendix~\ref{appendix:moment}.

\begin{figure}[!htp]
\centerline{
\includegraphics[width = 0.5\textwidth]{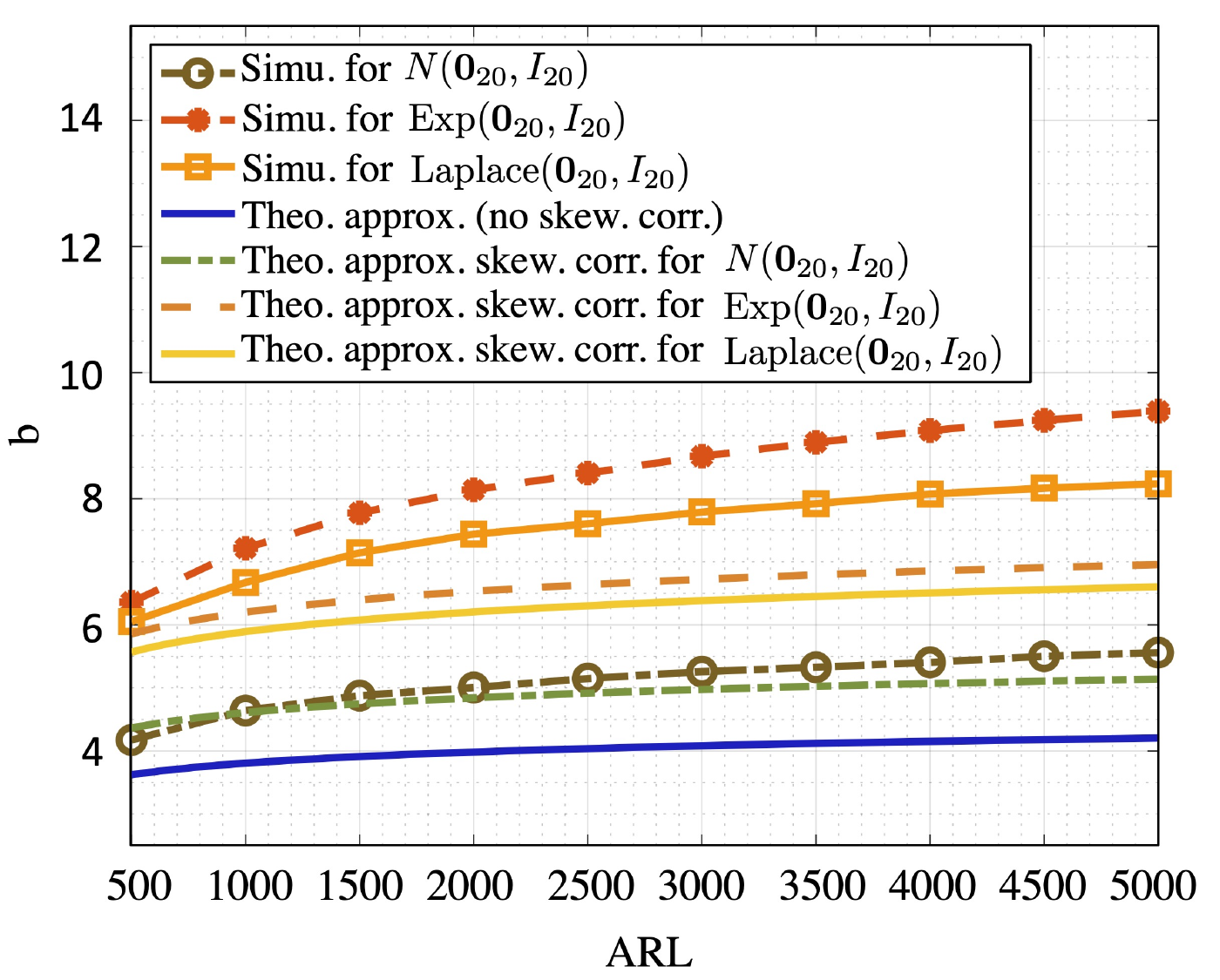} }
\caption{Comparison of the detection threshold $b$ obtained by the Monte Carlo simulation (Simu.) and the theoretical approximation (with and without skewness correction) for a given ARL.
}
\label{fig:arl_approx}
\end{figure}

To demonstrate the accuracy of our ARL approximation, we compare it with Monte Carlo simulation results for three types of pre-change distributions (i.e., Gaussian, Exponential, and Laplace distributions) in Figure~\ref{fig:arl_approx}. \textcolor{black}{The result shows that the theoretical prediction for the threshold is reasonably accurate in some cases (e.g., Gaussian) after the skewness correction. For other cases, such as when the pre-change distribution is exponential or Laplace, the gap between the theoretical prediction and Monte Carlo simulation is larger, particularly for large ARLs. This discrepancy may arise because these distributions deviate more significantly from Gaussian assumptions. In such cases, the theoretical results serve as a reasonable starting point for fine-tuning the threshold in practice, to avoid the expensive complete grid search.} 

\subsection{Expected Detection Delay (EDD)}\label{sec:edd_approx}

\textcolor{black}{Since the procedure compares the tail block of the data sequence against available pre-change sequences, we can show that the worst-case EDD, the commonly used performance metric, as defined in \cite{lorden1971procedures}. The proof by adopting an argument for a different CUSUM procedure (Lemma 4 in \cite{xie2023window}):
\begin{lemma}[Worst-case expected detection delay]\label{WEDD}
For any change-point $\kappa \geq 0$, we have that
\[
\mbox{ess} \mbox{sup}~\mathbb E_\kappa [T_w-\kappa|T_w>\kappa,\mathcal F_\kappa] \leq \mathbb E_0[T_w].
\]
where $F_\kappa$ is the filtration of observations up to time $\kappa$.
\end{lemma}
Proof can be found in Appendix \ref{appendix:WEDD}. }

\textcolor{black}{Due to Lemma \ref{WEDD}, we will focus on establishing $\mathbb E_0[T_w]$, the EDD when the change happens at the beginning.} 
Define 
\begin{equation}\label{eq:cpq}
 \color{black} {\rho} = \frac 1 2\left(\frac{\mathbb{E}\left[h^{2}\left(X, X^{\prime}, Y, Y^{\prime}\right)\right]}{N} + \frac{N-1}{N}\operatorname{Cov}\left[h(X,X',Y,Y'), h\left(X^{\prime \prime}, X^{\prime \prime \prime}, Y, Y^{\prime}\right)\right]\right) ^{-1/2}. \color{black}
\end{equation}
Then we have:
\begin{lemma}[EDD approximation]\label{lma:EDD}
Under Assumptions~\ref{assumption:bounded_kernel} and~\ref{assumption:detection}, for window length 
\begin{equation}\label{eq:Bmax_choice}
     w  \geq \frac{7 b}{{\rho}\mathcal{D}({p},{q})},
\end{equation} 
and $N > 0$, as $b \rightarrow \infty$, the first-order approximation to EDD for   online kernel CUSUM \eqref{eq:stopping_time} is given by:
\begin{equation}\label{eq:first_edd_approx}
    \mathbb{E}_{0} [T_{w}] = \frac{b}{{\rho}\mathcal{D}({p},{q})  } \ (1+o(1)).
\end{equation}
\end{lemma}
The proof of Lemma~\ref{lma:EDD} can be found in Appendix~\ref{appendix:thm2}. 

\begin{remark}[Effect of pre-change block number $N$]\label{rmk:Neffect} 
The parameter $N$ has a two-fold impact in our analysis: (1) there is one remainder term in our proof, which affects the accuracy of the approximation, which 
can be neglected only when $N$ is sufficiently large. (2) it affects the EDD in \eqref{eq:first_edd_approx} through the normalizing constant ${\rho}$ in \eqref{eq:cpq}. To showcase the impact on EDD, let us consider the example in which the pre-change distribution is ${\cN}(\mu \mathbf{1}_{d}, \sigma^2 I_{d})$ and we use Gaussian RBF kernel with bandwidth ${{r}}$ selected by the median heuristic \citep{scholkopf2002learning}. The Taylor expansion (see detailed derivations in Appendix~\ref{appendix:moment}) gives us:
$$0 < \mathbb{E}\left[h^{2}\left(X, X^{\prime}, Y, Y^{\prime}\right)\right] - \operatorname{Cov}\left[h(X,X',Y,Y'), h\left(X^{\prime \prime}, X^{\prime \prime \prime}, Y, Y^{\prime}\right)\right] = \cO(1/d).$$
This implies that ${\rho}$ monotonically increases with increasing $N$, i.e.,
\[{\rho} \rightarrow ({\operatorname{Cov}\left[h(X,X',Y,Y'), h\left(X^{\prime \prime}, X^{\prime \prime \prime}, Y, Y^{\prime}\right)\right]}/2)^{-1/2}  \ \text{\rm as} \ N \rightarrow \infty,\]
meaning that larger $N$ leads to smaller EDD according to \eqref{eq:first_edd_approx}. This finding agrees with the intuition: as $N$ grows larger, the variance of $\hat{\mathcal{D}}_B(t)$ reduces,
which leads to a quicker detection (this is also consistent with Lemma~\ref{lma:var_H0} in Appendix~\ref{appendix:moment}).
Moreover, this shows ${\rho}$ cannot increase unboundedly with increasing $N$. 
\end{remark}

\begin{remark}[Lower bound on EDD] We can establish a lower bound for the EDD. Under the same assumption of  Lemma~\ref{lma:EDD}), in the proof, Equation \eqref{eq:pTb<t1} in appendix, we have an upper bound $\PP_{0} (T_{w} < t_1) \leq 2t_1^2 O(e^{-b^2}) = O(e^{-b^2})$, for $t_1 = b/(4 \rhod)$ and when $N$ held finite. Then, 
\begin{equation}
    \EE_{0} [T_{w}]
    \geq \EE_{0} [T_{w} | T_{w} \geq t_1] \PP_{0} (T_{w} \geq t_1) \geq  t_1(1- \PP_{0} (T_{w} < t_1)) 
      \geq \frac{b}{4 \rhod}(1 - o(e^{-b^2})).
\end{equation}
Therefore, we have a lower bound on EDD that matches the order of \eqref{eq:first_edd_approx}, which implies that the EDD approximation is asymptotically tight. 
\end{remark}

\subsection{Optimal Window Length}\label{subsec:optimal_Bmax}

We now present the optimal window length parameter $w$ due to ARL and EDD analysis by comparing our procedure with an {\it oracle procedure} using infinite past data up to time $t$. The stopping time of the oracle procedure with detection threshold $b>0$ is defined as follows:
\begin{equation}\label{eq:stopping_time_noWL}
T_{\rm o} = \inf \left\{t : \max_{B \in [2:t]} {Z}_B(t) > b \right\}.
\end{equation} There is a trade-off between the computational and memory efficiency as well as the detection power: On one hand, larger $w$ intuitively leads to smaller EDD, but the benefit diminishes as $w$ further increases. On the other hand, the oracle procedure is computationally expensive as it involves $\cO(t^2)$ computational complexity and $\cO(t)$ memory up to time $t$, which is unsuitable for online settings. Thus, the fundamental question is: when holding ARL constant, what is the optimal window length $w^\star$ without sacrificing much performance?

Define a set of detection procedures with a constant ARL constraint $\gamma > 0$:
\begin{equation}\label{eq:constant_arl_procedure}
    \cC_{\gamma} = \{T : \mathbb{E}_{\infty} [T] \geq \gamma\}.
\end{equation}
Following the optimality definition in \citet{lai1995sequential}, we are interested in
\begin{itemize}
    \item [(i)]  Choosing thresholds $b$'s for each procedure respectively to ensure $T_{w}, T_{\rm o} \in \cC_{\gamma}$, and
    \item [(ii)] for tolerance $\varepsilon > 0$, choosing an optimal window length $w^\star$ that incurs a $\varepsilon$-performance loss, i.e.,
    \begin{equation}
            w^\star = \ \min w,   
 \quad {\rm s.t. \ } 0 \leq \mathbb{E}_0[T_{w}] - \mathbb{E}_0[T_{\rm o}] \leq \varepsilon. \label{eq:opt_w_def}
    \end{equation}
\end{itemize}
We have the following result:
\begin{theorem}[Optimal window length]\label{thm:opT_{w}max}
Under Assumptions~\ref{assumption:bounded_kernel} and~\ref{assumption:detection}, for any $N > 0$, as $\gamma \rightarrow \infty$, when choosing $b \sim {\log \gamma}$, we have $T_{w}, \  T_{\rm o} \in \cC_{\gamma}$, and furthermore the optimal window length
\begin{equation}\label{eq:w_star}
    w^\star \sim \frac{6b}{{\rho}\mathcal{D}({p},{q}) } + \frac{512K^2 \log ({3}/{\varepsilon})}{b^2 \left(\frac{N}{4} \wedge \frac{b}{{\rho}\mathcal{D}({p},{q})} \right)}.
\end{equation}
\end{theorem}

Proof can be found in Appendix~\ref{appendix:opT_{w}max}.

The main message from Theorem~\ref{thm:opT_{w}max} is that the optimal window length is $w^\star \sim {\log \gamma}$, as the first term in \eqref{eq:w_star} dominates the second term when $\gamma$ (and $b$) becomes large. While we only characterize the order of the optimal window length due to the complexity of our ARL approximation \eqref{eq:arl_approx}, the result $w^\star \sim {\log \gamma}$ is informative to make a connection with classic parametric results. Specifically, our result is in a similar form as the classic result for window-limited GLR \citep{lai1995sequential}. Furthermore, under the optimal choice $w^\star$, the EDD of our proposed procedure will be:
\begin{equation*}
\mathbb{E}_{0} [T_{w^\star}] \sim \frac{{\log \gamma}}{\mathcal{D}({p},{q})},
\end{equation*}
which closely resembles the EDD of the classic CUSUM procedure: $\mathcal O(\log \gamma/{\rm KL}(q||p))$, where the denominator is the Kullback-Leibler divergence between two distributions $q$ and $p$.

\begin{remark}[Choosing $b$ to control ARL]
In Theorem~\ref{thm:opT_{w}max}, we choose the same threshold \( b \) for both \( T_{w} \) and \( T_{\rm o} \) and keep it fixed across different values of \( w \). This is justified by the asymptotic ARL analysis, which shows that ARL grows exponentially in $b$ as \( b \rightarrow \infty \). Thus, using the same threshold ensures that the procedures achieve the same ARL asymptotically. This setup is a standard setup when comparing procedures (see, e.g., \cite{tartakovsky2014sequential, cao2018multi}). While a more precise characterization allowing the threshold $b$ to depend on $w$ could, in principle, improve accuracy, it would entail a highly complex—if not intractable—case-by-case analysis, which we do not pursue in this work.  On the other hand, the oracle procedure can be viewed as the window-limited procedure \( T_{w} \) with \( w = \infty \). This allows us to extend the result and obtain  
\[
\EE_0[T_{w^\star}] - \inf_{w^\star \leq w \leq \infty} \EE_0[T_{w}] \leq \varepsilon.
\]
Furthermore, as shown in the proof of Theorem~1, choosing \( w = o(b) \) results in a constant performance gap, leading to  
\[
\EE_0[T_{w^\star}] - \inf_{w \geq 2} \EE_0[T_{w}] \leq \varepsilon,
\]
where the condition \( w \geq 2 \) is technical, since at least two sequential samples are needed to estimate the unbiased MMD. In this regard, Theorem \ref{thm:opT_{w}max} also implies that the optimal window length defined in \eqref{eq:w_star} also achieves an $\varepsilon$-optimal EDD, compared with that attained by the theoretically optimal window length.  
\end{remark}

\begin{remark}[Extension to weakly dependent data]  
In practice, the assumption of i.i.d. sequential data is often unrealistic. Our theoretical analysis can be extended to weakly dependent data. By leveraging recent advances in concentration inequalities for the maximum mean discrepancy (MMD) under dependence \citep{cherief2022finite}, we can show that Lemma~\ref{lma:EDD} remains valid: the EDD \(\mathbb{E}_0[T_w]\) achieves the same order as in \eqref{eq:first_edd_approx}, provided that the post-change observations \(\{Y_t\}_{t \in \NN^+}\) form a stationary process with summable generalized mixing coefficients under a given kernel. This assumption is satisfied, for example, by stationary autoregressive time series (see Assumption 2.1 and the accompanying discussion in \citep{cherief2022finite}). The proof follows similar reasoning to the i.i.d. case, with the key difference being the use of a concentration inequality tailored to weakly dependent data. For brevity, we omit the proof and instead provide numerical experiments in Section~\ref{sec:simu} to illustrate the robustness of our procedure under data dependence.
\end{remark}
\color{black}

\begin{remark}[Selection of block size under temporal dependence]
In the presence of temporal dependence, the block size $B$ serves as a robustness parameter that balances residual dependence and effective sample size: small $B$ increases variability due to within-block dependence, whereas overly large $B$ reduces the number of blocks and may degrade detection power (an numerical example illustrating this can be found in Appendix \ref{appendix:ar1_blocksize}). Consequently, no universally optimal $B$ exists. In practice, rather than explicitly estimating the long-run variance, we calibrate detection thresholds to control the average run length (ARL) under the null and choose the window length $w$ sufficiently large to include effective block sizes, yielding stable performance except at extreme choices of $B$.
\end{remark}

\section{Experiments} \label{sec:numerical}

In this section, we \textcolor{black}{present numerical experiments to show the effect of hyperparameters and compare our method extensively with other methods}.  Implementation is available online at \url{https://github.com/SongWei-GT/online_kernel_cusum}.

\subsection{Choice of hyperparameters}\label{appendix:hyperparameter}

We start by studying how hyperparameters affect the detection performance: the window length $w$, the choice of kernel and its bandwidth, and the number of pre-change blocks $N$; more explanations and results can be found in Appendix \ref{appendix:hyperparameter}.

\begin{figure}[!htp]
\centerline{
\includegraphics[width = \textwidth]{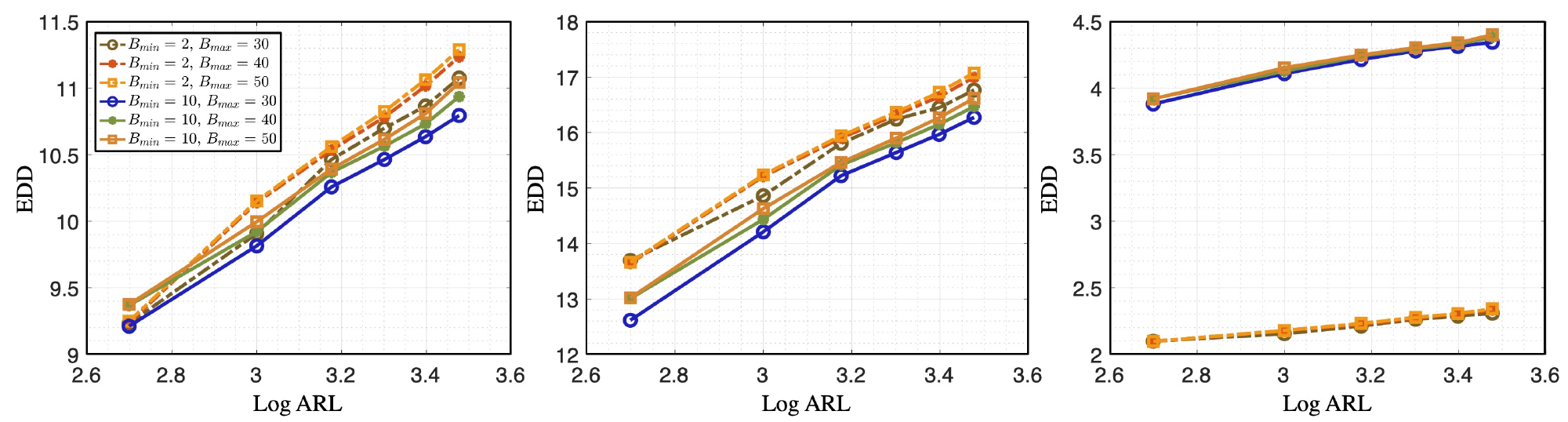}
}
\caption{Comparison of EDD for different search regions $[B_{\text{\rm min}}: B_{\text{\rm max}}]$; $w = B_{\rm max}$. The setting is: the pre-change distribution is ${\cN}(\mathbf{0}_{20}, I_{20})$, and the post-change distributions are, from left to right, ${\cN}(\mathbf{0}_{20},0.01 \ I_{20})$; ${\cN}(\mathbf{0}_{20},0.2 \ I_{20})$; ${\cN}(\mathbf{0}_{20},9 \ I_{20})$.
}
\label{fig:EDDARL_compare_bminbmax}
\end{figure}

 \vspace{.1in}
 \noindent{\it Effect of window length $w$.} Recall that we use \([2:w]\) as the search region to locate the change-point. For practical implementation, in addition to the right endpoint \(w\) (i.e., the window length), we also set the left endpoint of \(B \in [B_{\text{min}}: B_{\text{max}}]\), where \(B_{\text{max}}=w\). 
Fig. \ref{fig:EDDARL_compare_bminbmax} shows that \(B_{\text{max}}\) controls the success of detection—the procedure typically fails to raise an alarm within a reasonable time if the EDD of the oracle procedure \eqref{eq:detect_stat_noWL} exceeds \(B_{\text{max}}\). However, this does not suggest choosing an excessively large window length, as this incurs a higher computation and memory cost and a larger false alarm. Furthermore, EDD improvement plateaus when \(w \gtrsim \log(\text{ARL})\), which is numerically verified in the left panel of Figure~\ref{fig:EDDARL_compare_bminbmax}, where the EDD gain from increasing \(B_{\text{max}}\) from 40 to 50 is smaller than that from 30 to 40. 
On the other hand, \(B_{\text{min}}\) is important for quick detection—setting \(B_{\text{min}}\) larger than the EDD of the optimal procedure causes power loss (e.g., \(B_{\text{min}} = 10\) in the right panel). Typically, we choose \(B_{\text{min}} = 2\).

\begin{figure}[!htp]
\centerline{
\includegraphics[width = 0.7\textwidth]{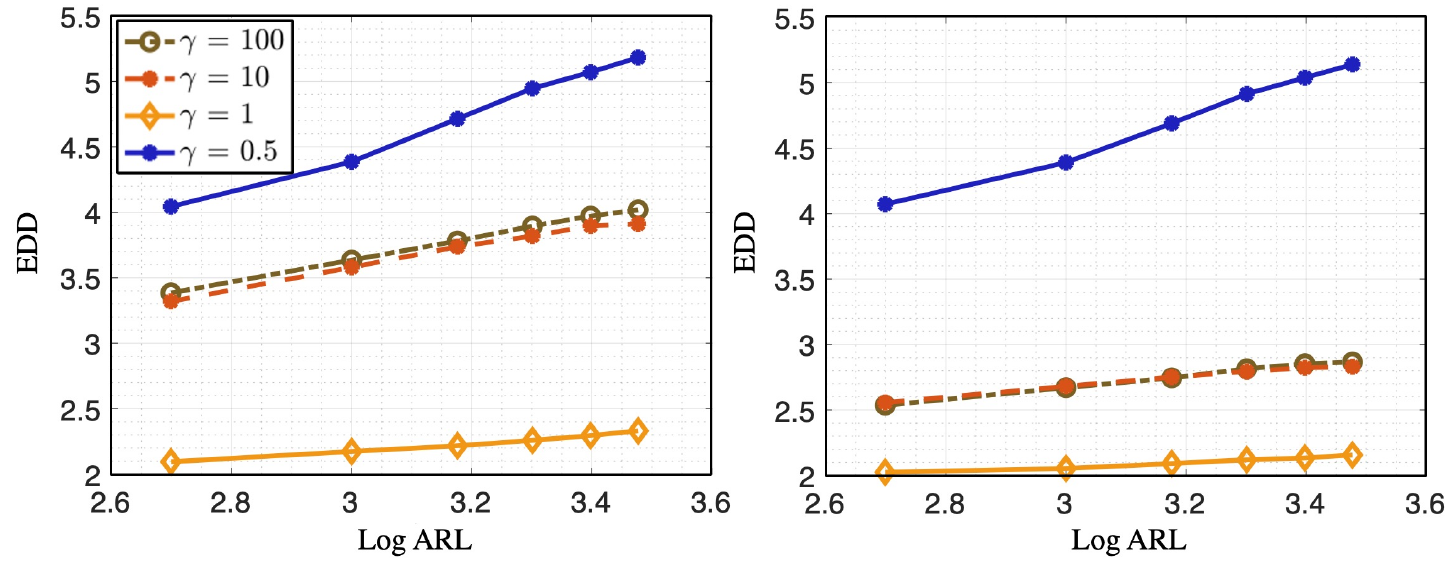}
}
\caption{Comparison of EDD for different kernel bandwidths of the Gaussian RBF kernel function. We consider different bandwidth choices ${{r}} = \gamma \texttt{med}_{{p}}$ where $\gamma \in \{0.5,1,10,100\}$, where $\texttt{med}_{{p}}$ is chosen via median heuristic. The change is from ${\cN}(\mathbf{0}_{20}, I_{20})$ to:  Left: ${\cN}(\mathbf{0}_{20},9 \ I_{20})$; Right: ${\cN}(\mathbf{1}_{20},9 \ I_{20})$.
}
\label{fig:EDDARL_compare_bandwidth}
\end{figure}

 \vspace{.1in}
 \noindent \textit{Effect of kernel bandwidth.}
\textcolor{black}{
In our experiments, for the kernel-based procedures, we choose Gaussian RBF kernel for all kernel methods because Gaussian RBF is a common choice when data are continuous \citep{li2019optimality}, and it is non-degenerate (see Section 4.3.3 in \citet{shashua2009introduction} for the expressions of $\lambda_j$ and $\varphi_j(\cdot)$ \eqref{eq:eigen_decomp} for Gaussian RBF kernel); the choice of kernel is further discussed in Appendix~\ref{appendix:hyperparameter}. }
Here, we study the effect of kernel bandwidth on the EDD. Here, we slightly abuse the notation \(\gamma\): the kernel bandwidth is set as \(r = \gamma\, \texttt{med}_{p}\), where \(\texttt{med}_{p}\) is the median of the \(\ell_2\) distance matrix among pre-change samples. Values \(\gamma < 1\) correspond to underestimation and \(\gamma > 1\) to overestimation of the bandwidth. As shown in Figure~\ref{fig:EDDARL_compare_bandwidth}, both under- and over-estimation lead to increased EDD, indicating that the median heuristic is an effective choice in our setting.

\begin{figure}[!htp]
\centerline{
\includegraphics[width = 0.5\textwidth]{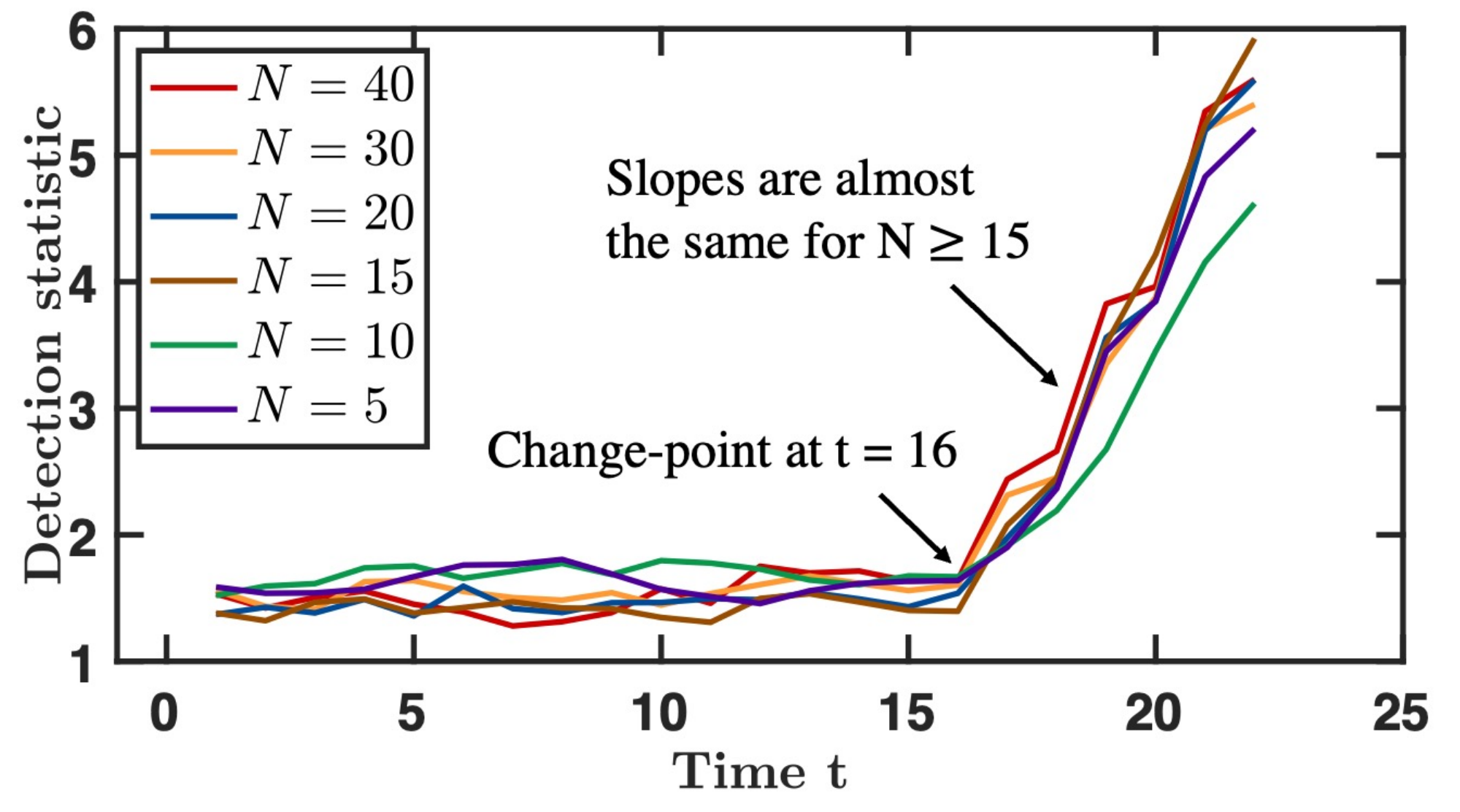}
}
\caption{Comparison of mean trajectories over 50 independent trials for different pre-change blocks' numbers $N$. The change is from ${\cN}(\mathbf{0}_{20}, I_{20})$ to Gaussian mixture with ${\cN}(\mathbf{0}_{20}, I_{20}) \ \text{\rm w.p. } 0.3$ and ${\cN}(\mathbf{0}_{20},4 \ I_{20}) \ \text{\rm w.p. } 0.7$ occurs at $t=16$. 
}
\label{fig:N_effect}
\end{figure}

 \vspace{.1in}
 \noindent \textit{Effect of number of number of pre-change blocks $N$.} 
Lastly, we examine the behavior of the detection statistic under different values of \(N\). Figure~\ref{fig:N_effect} shows the trajectory of the averaged detection statistic over 50 independent runs for various \(N\). The results indicate that performance is stable as long as \(N \geq 15\), and after that, the detection performance is not sensitive to the choice of $N$. In practice, \(N\) should be chosen large enough to avoid performance degradation.

\color{black}

\subsection{Simulations}\label{sec:simu}

We conduct a comprehensive comparison of our method with existing approaches, including the Scan \(B\)-procedure, Hotelling’s \(T^2\), multivariate exponentially weighted moving average (MEWMA) \citep{lowry1992multivariate}, W-GLR, window-limited CUSUM (W-CUSUM) \citep{xie2023window}, neural network (NN)-based detection methods \citep{hushchyn2020online,lee2023training}, and KCUSUM \citep{flynn2019change}, which uses a linear-time MMD statistic; detailed explanation of the methods are provided in Appendix~\ref{appendix:benchmarks}.

To ensure that each detection procedure meets an ARL constraint, we select the detection threshold $b$ for each detection procedure separately via $1000$ Monte Carlo trials. We then obtain the EDDs for each procedure as the average detection delays over another $1000$ Monte Carlo trials with $10000$ reference samples following ${p}$ and $50$ sequential samples following ${q}$. For our proposed procedure and Scan $B$-procedure, we choose $N=15$, $w = 50$, and use a Gaussian RBF kernel with bandwidth parameters selected via median heuristic. These choices are suggested by the experiments in \citet{li2019scan} and our simulation result in Appendix~\ref{appendix:hyperparameter}. 

\begin{figure}[!htp]
\centerline{
\includegraphics[width = .95\textwidth]{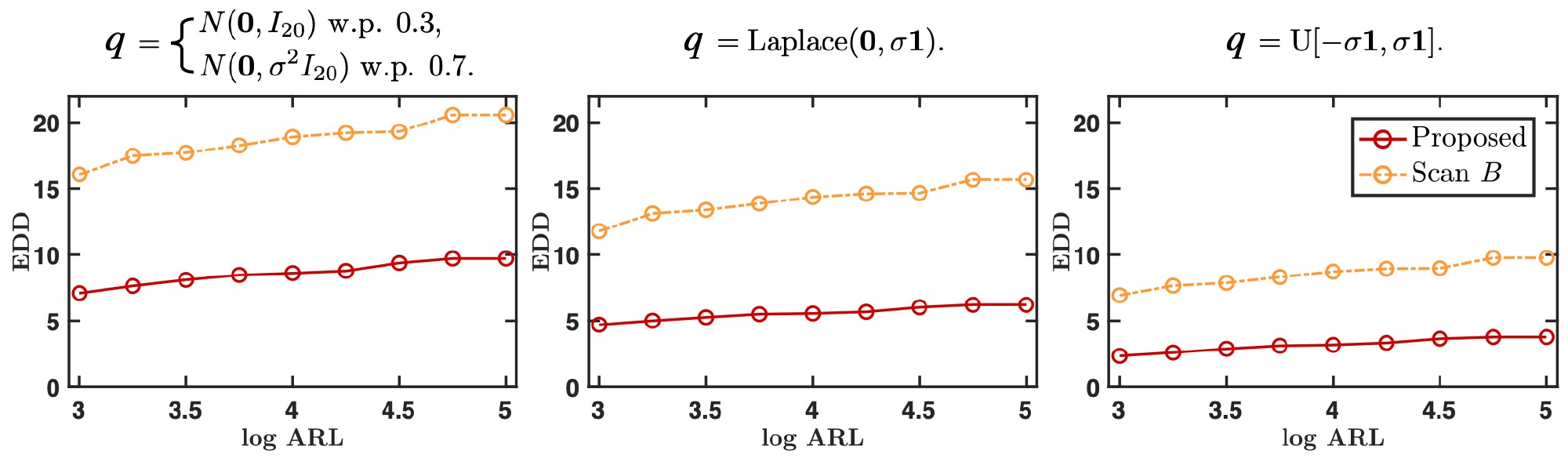}
}
\caption{Comparison of EDD for given ARL between our proposed (red) and the Scan $B$ (orange) detection procedures. In the x-axis, the ARL is in the base $10$ logarithm. The pre-change distribution is ${\cN}(\mathbf{0}_{20}, I_{20})$ and the post-change distribution $q$ is specified on the top of each figure (with $\sigma = 2$). 
}
\label{fig:EDDARL_compare_w_scanB}
\end{figure}

We first compare the proposed and Scan-B procedures in Figure~\ref{fig:EDDARL_compare_w_scanB} by plotting the EDD versus log-ARL, where we can observe quicker detection of our proposed procedure across  ARLs. 

Then we consider more settings of various post-change distributions, including changes from \(\cN(\mathbf{0}_{20}, I_{20})\) to Gaussian mixtures (Settings 1 and 2), Laplace (Setting 3), Exponential (Setting 4), and Uniform (Setting 5). We set \(M = 2500\), \(w = 80\), and \(N = 30\). More details of the detection procedures and the settings are in Appendix~\ref{appendix:more_comparison_settings}. We report the EDD for \(\kappa = 100\) (within 1000 sequential samples), as well as the number of false alarms (i.e., detection before \(\kappa\)) and failures (i.e., no detection by the end of the sequential data) over 1000  Monte Carlo trials in Tables~\ref{tab:main_comparison_results} and \ref{tab:additional_iid_comparison_results}. Table~\ref{tab:main_comparison_results} also includes the exact CUSUM \eqref{eq:cusum} as a benchmark, which performs best, as expected (under the assumption of knowing $p$ and $q$ precisely).

 Tables~\ref{tab:main_comparison_results} and \ref{tab:additional_iid_comparison_results} show that our proposed procedure achieves the best EDD in various settings while also having a number of successful detections very close to the best. We also see that although W-GLR has more successful detection and a similar EDD to our proposed method in Setting 1, its performance quickly deteriorates in higher dimensions (Setting 2), which can be explained by the fact that estimating the post-change parameters via maximum likelihood is more difficult in high dimensions. 
Although neural network-based methods, especially NN-CUSUM \citep{lee2022neural}, consistently detect changes (i.e., they have the highest number of successful trials), tuning their training hyperparameters to achieve smaller EDDs than kernel methods is challenging, and their performance degrades quickly in real problems with limited data (as shown in Settings 10 and 11 of Table~\ref{tab:additional_dependent_comparison_results}) (we provide additional results for different ARL in Appendix~\ref{appendix:more_comparison_results}; further validating the strong performance of our proposed procedure).

\begin{table}
\caption{\label{tab:main_comparison_results} Post-change data are i.i.d. Gaussian Mixture with different dimensions $d$: Comparison with other methods, in their EDD with ARL constraint $\gamma = 1000$.}


\centering
\begin{small}
\resizebox{\textwidth}{!}{
\begin{tabular}{cclcccccccccc}
\toprule
& & \multirow{2}{*}{Method} & Exact &  \multirow{2}{*}{Hotelling $T^2$} & \multirow{2}{*}{MEWMA} & \multirow{2}{*}{W-GLR} & \multirow{2}{*}{W-CUSUM} & NN & NN & \multirow{2}{*}{KCUSUM} & \multirow{2}{*}{Scan $B$} & \multirow{2}{*}{Proposed} \\
& & & CUSUM &  &  &  &  & Classifier & CUSUM & & \\
\midrule
\parbox[t]{2mm}{\multirow{4}{*}{\rotatebox[origin=c]{90}{Setting 1}}} & \parbox[t]{2mm}{\multirow{4}{*}{\rotatebox[origin=c]{90}{($d = 20$)}}} & EDD (std.) & $13.3_{(6.5)}$ & $125.9_{(83.0)}$ & $83.6_{(77.4)}$ & $30.9_{(15.0)}$ & $163.3_{(130.6)}$ & $97.8_{(82.6)}$ & $150.8_{(67.4)}$ & $296.3_{(209.8)}$ & $35.4_{(10.5)}$ & $28.6_{(14.4)}$ \\
& & Success & $840_{/1000}$ & $939_{/1000}$ & $815_{/1000}$ & $830_{/1000}$ & $870_{/1000}$ & $629_{/1000}$ & $897_{/1000}$ & $607_{/1000}$ & $794_{/1000}$ & $794_{/1000}$ \\
& & False Alarm & $160_{/1000}$ & $60_{/1000}$ & $185_{/1000}$ & $170_{/1000}$ & $125_{/1000}$ & $335_{/1000}$ & $103_{/1000}$ & $193_{/1000}$ & $206_{/1000}$ & $206_{/1000}$ \\
& & Failure & $0_{/1000}$ & $1_{/1000}$ & $0_{/1000}$ & $0_{/1000}$ & $5_{/1000}$ & $36_{/1000}$ & $0_{/1000}$ & $200_{/1000}$ & $0_{/1000}$ & $0_{/1000}$ \\
\cmidrule{3-13}
\parbox[t]{2mm}{\multirow{4}{*}{\rotatebox[origin=c]{90}{Setting 2}}} & \parbox[t]{2mm}{\multirow{4}{*}{\rotatebox[origin=c]{90}{($d = 50$)}}} & EDD (std.) & $2.0_{(1.5)}$ & $-$ & $407.3_{(238.6)}$ & $392.6_{(260.4)}$ & $55.0_{(150.8)}$ & $637.6_{(113.8)}$ & $328.6_{(168.6)}$ & $398.8_{(250.6)}$ & $49.6_{(13.4)}$ & $47.1_{(18.4)}$ \\
& & Success & $902_{/1000}$ & $0_{/1000}$ & $62_{/1000}$ & $84_{/1000}$ & $414_{/1000}$ & $396_{/1000}$ & $974_{/1000}$ & $251_{/1000}$ & $868_{/1000}$ & $869_{/1000}$ \\
& & False Alarm & $98_{/1000}$ & $7_{/1000}$ & $104_{/1000}$ & $100_{/1000}$ & $22_{/1000}$ & $0_{/1000}$ & $24_{/1000}$ & $92_{/1000}$ & $132_{/1000}$ & $131_{/1000}$ \\
& & Failure & $0_{/1000}$ & $993_{/1000}$ & $834_{/1000}$ & $816_{/1000}$ & $564_{/1000}$ & $604_{/1000}$ & $2_{/1000}$ & $657_{/1000}$ & $0_{/1000}$ & $0_{/1000}$ \\
\bottomrule
\end{tabular} }
\end{small}
\end{table}

\begin{table}
\caption{\label{tab:additional_iid_comparison_results} Post-change data are i.i.d. (under various distributions), EDD comparison for different procedures with ARL constraint $\gamma = 1000$. If there are fewer than $20$ successful trials, we will mark the EDD as ``$-$''. We highlight the best EDD in bold font and the second best in italic font.}
\vspace{0.025in}
\centering
\begin{small}
\resizebox{\textwidth}{!}{
\begin{tabular}{clccccccccc}
\toprule
& \multirow{2}{*}{Method} & \multirow{2}{*}{Hotelling $T^2$} & \multirow{2}{*}{MEWMA} & \multirow{2}{*}{W-GLR} & \multirow{2}{*}{W-CUSUM} & NN & NN & \multirow{2}{*}{KCUSUM} & \multirow{2}{*}{Scan $B$} & \multirow{2}{*}{Proposed} \\
& &  &  &  &  & Classifier & CUSUM &  & \\
\midrule
\parbox[t]{2mm}{\multirow{4}{*}{\rotatebox[origin=c]{90}{Setting 3}}} & EDD (std.) & $-$ & $428.6_{(240.7)}$ & $33.9_{(8.9)}$ & $\textit{21.1}_{(9.2)}$ & $430.1_{(218.5)}$ & $153.9_{(41.1)}$ & $390.5_{(232.7)}$ & $26.5_{(6.8)}$ & $\textbf{14.7}_{(4.1)}$ \\
 & Success & $0_{/1000}$ & $180_{/1000}$ & $911_{/1000}$ & $934_{/1000}$ & $686_{/1000}$ & $998_{/1000}$ & $404_{/1000}$ & $849_{/1000}$ & $891_{/1000}$ \\
 & False Alarm & $9_{/1000}$ & $101_{/1000}$ & $89_{/1000}$ & $66_{/1000}$ & $0_{/1000}$ & $2_{/1000}$ & $94_{/1000}$ & $151_{/1000}$ & $109_{/1000}$ \\
& Failure & $991_{/1000}$ & $719_{/1000}$ & $0_{/1000}$ & $0_{/1000}$ & $314_{/1000}$ & $0_{/1000}$ & $502_{/1000}$ & $0_{/1000}$ & $0_{/1000}$ \\
\cmidrule{2-11}
\parbox[t]{2mm}{\multirow{4}{*}{\rotatebox[origin=c]{90}{Setting 4}}} & EDD (std.) & $-$ & $374.3_{(241.9)}$ & $44.4_{(16.3)}$ & $\textit{31.8}_{(15.1)}$ & $452.6_{(215.9)}$ & $160.7_{(59.1)}$ & $384.6_{(245.6)}$ & $32.8_{(9.9)}$ & $\textbf{20.7}_{(8.2)}$ \\
& Success & $2_{/1000}$ & $483_{/1000}$ & $918_{/1000}$ & $922_{/1000}$ & $646_{/1000}$ & $990_{/1000}$ & $469_{/1000}$ & $870_{/1000}$ & $893_{/1000}$ \\
& False Alarm & $7_{/1000}$ & $101_{/1000}$ & $82_{/1000}$ & $78_{/1000}$ & $0_{/1000}$ & $10_{/1000}$ & $104_{/1000}$ & $130_{/1000}$ & $107_{/1000}$ \\
& Failure & $991_{/1000}$ & $416_{/1000}$ & $0_{/1000}$ & $0_{/1000}$ & $354_{/1000}$ & $0_{/1000}$ & $427_{/1000}$ & $0_{/1000}$ & $0_{/1000}$ \\
\cmidrule{2-11}
\parbox[t]{2mm}{\multirow{4}{*}{\rotatebox[origin=c]{90}{Setting 5}}} & EDD (std.) & $-$ & $40.3_{(34.6)}$ & $10.4_{(1.9)}$ & $\textit{9.7}_{(3.0)}$ & $363.6_{(219.0)}$ & $83.6_{(25.3)}$ & $214.8_{(188.7)}$ & $15.2_{(3.9)}$ & $\textbf{5.4}_{(1.1)}$ \\
& Success & $0_{/1000}$ & $900_{/1000}$ & $930_{/1000}$ & $938_{/1000}$ & $742_{/1000}$ & $996_{/1000}$ & $882_{/1000}$ & $845_{/1000}$ & $905_{/1000}$ \\
& False Alarm & $7_{/1000}$ & $100_{/1000}$ & $70_{/1000}$ & $62_{/1000}$ & $0_{/1000}$ & $4_{/1000}$ & $96_{/1000}$ & $155_{/1000}$ & $95_{/1000}$ \\
& Failure & $993_{/1000}$ & $0_{/1000}$ & $0_{/1000}$ & $0_{/1000}$ & $258_{/1000}$ & $0_{/1000}$ & $22_{/1000}$ & $0_{/1000}$ & $0_{/1000}$ \\
\bottomrule
\end{tabular} }
\end{small}
\end{table}

Finally, we also perform a simulation with dependent post-change data. Consider simulated examples where the distribution changes from \(\cN(\mathbf{0}_{20}, I_{20})\) to a 20-dimensional vector autoregressive (VAR) time series with lag one at \(\kappa = 100\). We test various VAR coefficient matrices: scaled identity (Setting 6), random diagonal (Setting 7), random symmetric (Setting 8), and random sparse (Setting 9). The results, shown in Table~\ref{tab:additional_dependent_comparison_results}, indicate that our proposed method achieves the best EDD while maintaining a high number of successful detections. Moreover, in Setting 9, where the post-change data is centered on removing any mean shift, both Hotelling’s \(T^2\) and Scan \(B\) (which perform similarly to our method in Settings 6, 7, and 8) become considerably less effective. In contrast, our proposed procedure achieves the best EDD in this challenging scenario.

\begin{table}
\caption{\label{tab:additional_dependent_comparison_results}  Dependent post-change data: EDD comparison of procedures with ARL constraint $\gamma = 1000$. If there are less than $20$ successful trials, we will mark the EDD as ``$-$''. We highlight the best EDD in bold font and the second best in italic font.}
\vspace{0.025in}
\centering
\begin{small}
\resizebox{\textwidth}{!}{
\begin{tabular}{cclccccccccc}
\toprule
& & \multirow{2}{*}{Method} & \multirow{2}{*}{Hotelling $T^2$} & \multirow{2}{*}{MEWMA} & \multirow{2}{*}{W-GLR} & \multirow{2}{*}{W-CUSUM} & NN & NN & \multirow{2}{*}{KCUSUM} & \multirow{2}{*}{Scan $B$} & \multirow{2}{*}{Proposed} \\
& & &  &  &  &  & Classifier & CUSUM &  & \\
\midrule
\parbox[t]{2mm}{\multirow{16}{*}{\rotatebox[origin=c]{90}{Simulated Examples (Settings $6-9$)}}} & \parbox[t]{2mm}{\multirow{4}{*}{\rotatebox[origin=c]{90}{Setting 6}}} & EDD (std.) & $368.4_{(224.2)}$ & $389.6_{(247.7)}$ & $391.0_{(259.1)}$ & $374.8_{(239.5)}$ & $616.8_{(102.6)}$ & $422.9_{(239.9)}$ & $374.8_{(246.3)}$ & $\textit{365.3}_{(249.1)}$ & $\textbf{350.6}_{(246.9)}$ \\
& & Success & $607_{/1000}$ & $533_{/1000}$ & $569_{/1000}$ & $568_{/1000}$ & $421_{/1000}$ & $605_{/1000}$ & $511_{/1000}$ & $482_{/1000}$ & $530_{/1000}$ \\
& & False Alarm & $8_{/1000}$ & $93_{/1000}$ & $58_{/1000}$ & $53_{/1000}$ & $0_{/1000}$ & $9_{/1000}$ & $113_{/1000}$ & $155_{/1000}$ & $98_{/1000}$ \\
& & Failure & $385_{/1000}$ & $374_{/1000}$ & $373_{/1000}$ & $379_{/1000}$ & $579_{/1000}$ & $386_{/1000}$ & $376_{/1000}$ & $363_{/1000}$ & $372_{/1000}$ \\
\cmidrule{3-12}
& \parbox[t]{2mm}{\multirow{4}{*}{\rotatebox[origin=c]{90}{Setting 7}}} & EDD (std.) & $365.3_{(229.1)}$ & $390.1_{(248.9)}$ & $395.4_{(256.8)}$ & $382.8_{(238.5)}$ & $620.7_{(112.0)}$ & $410.2_{(236.2)}$ & $375.3_{(248.7)}$ & $\textit{363.2}_{(237.8)}$ & $\textbf{352.3}_{(247.9)}$ \\
& & Success & $618_{/1000}$ & $538_{/1000}$ & $556_{/1000}$ & $565_{/1000}$ & $382_{/1000}$ & $593_{/1000}$ & $522_{/1000}$ & $466_{/1000}$ & $524_{/1000}$ \\
& & False Alarm & $10_{/1000}$ & $80_{/1000}$ & $69_{/1000}$ & $44_{/1000}$ & $0_{/1000}$ & $19_{/1000}$ & $98_{/1000}$ & $154_{/1000}$ & $100_{/1000}$ \\
& & Failure & $372_{/1000}$ & $382_{/1000}$ & $375_{/1000}$ & $391_{/1000}$ & $618_{/1000}$ & $388_{/1000}$ & $380_{/1000}$ & $380_{/1000}$ & $376_{/1000}$ \\
\cmidrule{3-12}
& \parbox[t]{2mm}{\multirow{4}{*}{\rotatebox[origin=c]{90}{Setting 8}}} & EDD (std.) & $372.6_{(225.8)}$ & $391.5_{(243.5)}$ & $384.9_{(256.6)}$ & $374.2_{(246.0)}$ & $619.5_{(115.5)}$ & $415.0_{(238.7)}$ & $365.8_{(246.3)}$ & $\textit{365.0}_{(242.2)}$ & $\textbf{348.7}_{(243.5)}$ \\
& & Success & $612_{/1000}$ & $526_{/1000}$ & $555_{/1000}$ & $565_{/1000}$ & $405_{/1000}$ & $574_{/1000}$ & $536_{/1000}$ & $482_{/1000}$ & $511_{/1000}$ \\
& & False Alarm & $8_{/1000}$ & $96_{/1000}$ & $66_{/1000}$ & $50_{/1000}$ & $0_{/1000}$ & $14_{/1000}$ & $103_{/1000}$ & $144_{/1000}$ & $102_{/1000}$ \\
& & Failure & $380_{/1000}$ & $378_{/1000}$ & $379_{/1000}$ & $385_{/1000}$ & $595_{/1000}$ & $412_{/1000}$ & $361_{/1000}$ & $374_{/1000}$ & $387_{/1000}$ \\
\cmidrule{3-12}
& \parbox[t]{2mm}{\multirow{4}{*}{\rotatebox[origin=c]{90}{Setting 9}}} & EDD (std.) & $395.2_{(251.8)}$ & $\textit{372.6}_{(238.1)}$ & $385.1_{(247.1)}$ & $379.2_{(251.9)}$ & $504.3_{(182.8)}$ & $406.5_{(224.7)}$ & $382.6_{(248.1)}$ & $377.1_{(256.5)}$ & $\textbf{344.1}_{(251.9)}$ \\
& & Success & $196_{/1000}$ & $403_{/1000}$ & $369_{/1000}$ & $413_{/1000}$ & $659_{/1000}$ & $482_{/1000}$ & $499_{/1000}$ & $234_{/1000}$ & $421_{/1000}$ \\
& & False Alarm & $9_{/1000}$ & $94_{/1000}$ & $83_{/1000}$ & $48_{/1000}$ & $0_{/1000}$ & $3_{/1000}$ & $112_{/1000}$ & $135_{/1000}$ & $125_{/1000}$ \\
& & Failure & $795_{/1000}$ & $503_{/1000}$ & $548_{/1000}$ & $539_{/1000}$ & $341_{/1000}$ & $515_{/1000}$ & $389_{/1000}$ & $631_{/1000}$ & $454_{/1000}$ \\
\cmidrule{3-12}
\parbox[t]{2mm}{\multirow{8}{*}{\rotatebox[origin=c]{90}{Real Examples}}} & \parbox[t]{2mm}{\multirow{4}{*}{\rotatebox[origin=c]{90}{setting 10}}} & EDD (std.) & $-$ & $-$ & $-$ & $-$ & $104.4_{(37.5)}$ & $96.6_{(57.3)}$ & $98.9_{(56.3)}$ & $\textit{95.7}_{(63.7)}$ & $\textbf{91.7}_{(59.1)}$ \\
& & Success & $14_{/1000}$ & $17_{/1000}$ & $0_{/1000}$ & $5_{/1000}$ & $400_{/1000}$ & $137_{/1000}$ & $145_{/1000}$ & $39_{/1000}$ & $112_{/1000}$ \\
& & False Alarm & $74_{/1000}$ & $126_{/1000}$ & $14_{/1000}$ & $79_{/1000}$ & $0_{/1000}$ & $97_{/1000}$ & $155_{/1000}$ & $153_{/1000}$ & $161_{/1000}$ \\
& & Failure & $912_{/1000}$ & $857_{/1000}$ & $986_{/1000}$ & $916_{/1000}$ & $600_{/1000}$ & $766_{/1000}$ & $700_{/1000}$ & $808_{/1000}$ & $727_{/1000}$ \\
\cmidrule{3-12}
& \parbox[t]{2mm}{\multirow{4}{*}{\rotatebox[origin=c]{90}{setting 11}}} & EDD (std.) & $\textbf{19.1}_{(11.5)}$ & $53.3_{(45.6)}$ & $63.1_{(49.1)}$ & $39.8_{(40.5)}$ & $103.3_{(34.0)}$ & $101.1_{(56.1)}$ & $86.1_{(56.4)}$ & $40.7_{(10.7)}$ & $\textit{24.3}_{(13.8)}$ \\
& & Success & $899_{/1000}$ & $852_{/1000}$ & $917_{/1000}$ & $86_{/1000}$ & $419_{/1000}$ & $150_{/1000}$ & $511_{/1000}$ & $854_{/1000}$ & $834_{/1000}$ \\
& & False Alarm & $101_{/1000}$ & $119_{/1000}$ & $22_{/1000}$ & $67_{/1000}$ & $0_{/1000}$ & $139_{/1000}$ & $157_{/1000}$ & $146_{/1000}$ & $166_{/1000}$ \\
& & Failure & $0_{/1000}$ & $29_{/1000}$ & $61_{/1000}$ & $847_{/1000}$ & $581_{/1000}$ & $711_{/1000}$ & $332_{/1000}$ & $0_{/1000}$ & $0_{/1000}$ \\
\bottomrule
\end{tabular} }
\end{small}
\end{table}

We examine the potential loss of detection power induced by block partitioning by comparing the proposed block CUSUM with a single-block (non-block) CUSUM implementation. 

\begin{figure}[h!]
    \centering
    \includegraphics[width=0.7\textwidth]{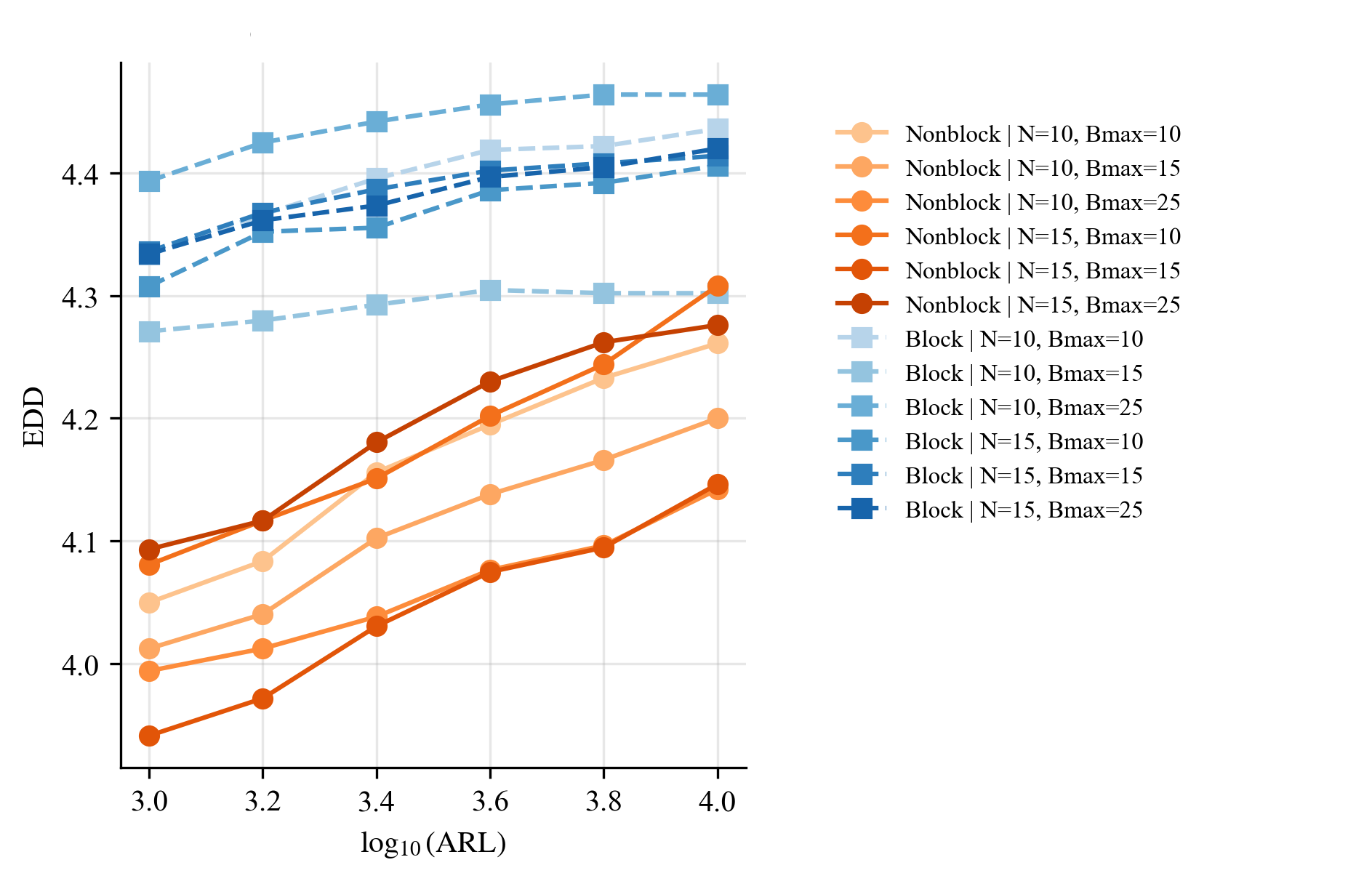}
\caption{Comparison between the proposed block CUSUM and the single-block (non-block) CUSUM detection statistics. The curves show the EDD versus $\log_{10}(\mathrm{ARL})$ for a $20$-dimensional Gaussian mixture shift from $\mathcal{N}(0, I_{20})$ to $0.3\mathcal{N}(0, I_{20}) + 0.7\mathcal{N}(0, 4 I_{20})$. Results are shown for different combinations of the number of blocks $N$ and maximum window size $B_{\max}$. The block CUSUM exhibits only a modest increase in detection delay relative to the non-block CUSUM across the full ARL range.}
    \label{fig:block_vs_nonblock}
\end{figure}

 \vspace{.1in}
 \noindent{\it Effect of block partitioning on detection power.} A natural concern with block-based implementations is the potential loss of detection power compared to using a single block of reference samples. Figure~\ref{fig:block_vs_nonblock} shows that this loss is minimal across a wide range of average run length (ARL) levels. This observation is consistent with results from offline two-sample testing using block MMD \citep{zaremba2013b}, where block-based statistics can achieve detection power comparable to full MMD while controlling the Type-I error. Heuristically, a similar phenomenon arises in the online setting considered here: in CUSUM-type procedures, detection delay is driven primarily by the accumulation of post-change information. As long as the block size is not chosen excessively small, the effective information available for detection remains comparable to that of a single-block implementation.

\subsection{Real-world data applications}

We now demonstrate the performance of our proposed procedure on several real-world data applications. Note that in real data, observations can be temporally dependent, thereby also illustrating that our method is effective for dependent data.

\begin{figure}[!htp]
\centerline{
\includegraphics[width = 0.4\textwidth]{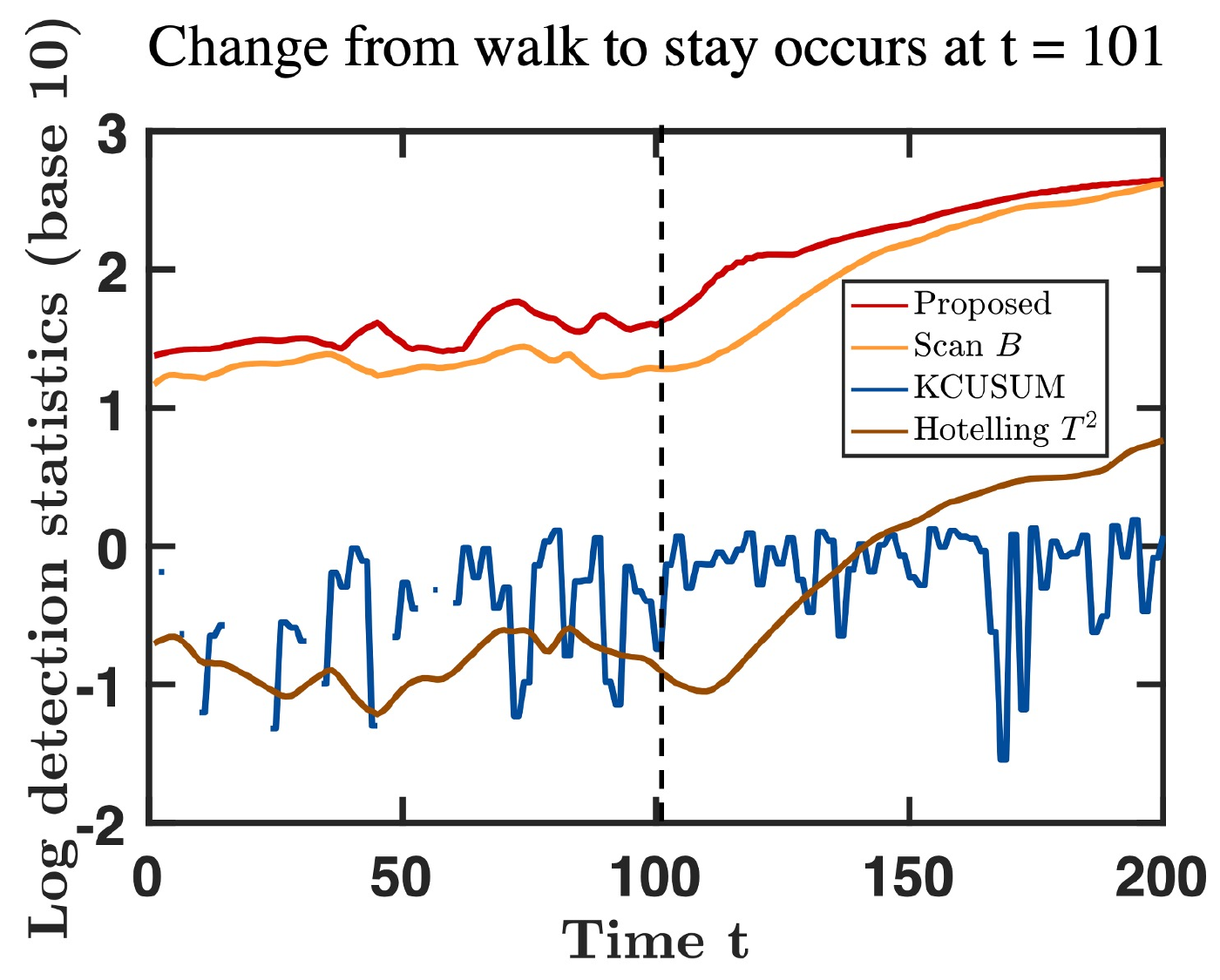}
}
\caption{The trajectories of different detection statistics. Human subject 101 activity changes from walking to staying at time $t = 101$. The recursive update of KCUSUM \eqref{eq:KCUSUM_update} yields some zero statistics, which correspond to the ``no value part'' on the blue line. 
}
\label{fig:realexp_detectstat_subj101}
\end{figure}

\vspace{.1in}
\noindent
{\it Human activity shift detection with HASC dataset.}
The Human Activity Sensing Consortium (HASC) dataset ({available at \url{http://hasc.jp/hc2011}})  contains $3$-dimensional measurements of human activities collected by portable 3D accelerometers. To compare the performance of different detection procedures, we consider the change from walking to staying of human subject 101 and plot the detection statistics in Figure~\ref{fig:realexp_detectstat_subj101}. The EDD and miss rate (out of 10 repeats) of each procedure are reported in Table~\ref{table:real_exp_101}. The detection threshold is chosen as the $80\%$ quantile of the largest detection statistics over all repeats under $H_0$ (i.e., the post-change activity is the same as the pre-change activity) to control the false alarm rate at $0.2$.

\begin{table}
\caption{Human activity data change detection. EDD (over successful detections) and miss (out of 10 repeats) of human activity change (from walking to other activities) detection for subject 101. The detection threshold is chosen as the $80 \%$ quantile of the maximum detection statistics under $H_0$ (i.e., the post-change activity is still walking). 
}
\centering
\begin{small}
\resizebox{.85\textwidth}{!}{%
\begin{tabular}{lcccccccccc}
\toprule[1pt]
Activity & \multicolumn{2}{c}{{Jog}} & \multicolumn{2}{c}{{Skip}} & \multicolumn{2}{c}{{Stay}} & \multicolumn{2}{c}{{Stair down}} & \multicolumn{2}{c}{{Stair up}}\\
 & EDD & Failure & EDD & Failure & EDD & Failure & EDD & Failure & EDD & Failure \\
\cmidrule(l){2-3} \cmidrule(l){4-5} \cmidrule(l){6-7} \cmidrule(l){8-9} \cmidrule(l){10-11}
Proposed & \textbf{37.7} & \textbf{0} & \textbf{50.6} & \textbf{0} & \textbf{13.0} & 8 & 517.2 & 4 & 227.1 & \textbf{2} \\ 
Scan $B$ & 73.7 & \textbf{0} & 82.4 & \textbf{0} & 72.2 & \textbf{0} & 329.7 & \textbf{3} & 207.3 & 7 \\ 
KCUSUM & 635.0 & 2 & 560.7 & 1 & 527.6 & \textbf{0} & 429.3 & 4 & $-$ & 10 \\ 
Hotelling $T^2$ & 77.6 & 5 & 242.8 & 5 & 26.5 & 8 & \textbf{114.0} & 5 & \textbf{44.7} & 7 \\ 
\bottomrule[1pt]
\end{tabular}
}
\end{small}
\label{table:real_exp_101}
\end{table}

As shown in Table~\ref{table:real_exp_101}, our procedure is robust and can successfully detect the change in almost all scenarios. In particular, our procedure achieves the quickest detection for changes from walking to jogging and skipping. 
We extend our experiment to the rest of the five human subjects and report the EDD and failure rates in Table~\ref{table:real_exp_therest} in Appendix~\ref{appendix:more_real_data_exp}, which further demonstrates the excellent performance of our procedure for most scenarios. We also provide more details of the HASC dataset and the choices of hyperparameters in Appendix~\ref{appendix:more_real_data_exp}.

\vspace{.1in}

\noindent \textit{Hand-written digits distribution shift detection with MNIST dataset.} The experiment considers detecting MNIST hand-written digits shift: we observe a sequence of digits that shift from one kind to another at some unknown time. The ambient dimension is 784 (28$\times$28-pixel image). The observations from pre- or post-change distribution are randomly sampled for a digit from $\{0,1,\dots,9\}$. 
The detection threshold is obtained with 500 independent experiments with $3000$ history data and $1000$ sequential observations, both from the same pre-change distribution. Then we perform another 500 independent experiments with $3000$ history data and $100$ sequential observations. To make the detection task even more challenging, each digit's data is centered by subtracting its coordinate-wise average vector, ensuring zero mean for all digits. The above experiment is then repeated on this centered MNIST dataset, and the results are presented in Table~\ref{table:MNIST}. In most cases, our proposed method outperforms the Scan \(B\)-procedure, achieving faster change detection. In the remaining cases where Scan \(B\) performs better, the two methods yield similar EDDs, indicating comparable performance. Overall, these findings verify the effectiveness of our proposed procedure in a high-dimensional setting.

\begin{table}
\caption{\label{table:MNIST}MNIST data, EDD for target ARL $\gamma = 1000$. In the 10-by-10 table, the $(i,j)$-th entry contains EDDs (left: our proposed, right: the Scan $B$) under the setting where the pre-change digit is $i$, and the post-change digit is $j$. In each entry, the smaller EDD is in a larger font, indicating the corresponding detection procedure is better under that specific setting; the entries where our proposed method outperforms Scan $B$ are highlighted in bold fonts.}
\centering
\begin{small}
\begin{sc}
\resizebox{.95\textwidth}{!}{%
\begin{tabular}{l|cccccccccc}
 & 0 & 1 & 2 & 3 & 4 & 5 & 6 & 7 & 8 & 9 \\
 \toprule[1pt]
0& -  & $_{31.04}30.31$ & $ \boldsymbol{3.44}_{24.32}$ & $ \boldsymbol{3.68}_{25.22}$ & $ \boldsymbol{6.44}_{40.17}$ & $ \boldsymbol{9.48}_{26.94}$ & $ \boldsymbol{9.28}_{30.83}$ & $ \boldsymbol{16.26}_{49.01}$ & $ \boldsymbol{3.78}_{21.73}$ & $ \boldsymbol{11.64}_{47.18}$ \\ 
1 & $ \boldsymbol{17.99}_{23.03}$& -  & $ \boldsymbol{19.18}_{23.81}$ & $ \boldsymbol{25.50}_{27.64}$ & $_{31.63}30.73$ & $ \boldsymbol{20.98}_{24.89}$ & $ \boldsymbol{25.88}_{27.88}$ & $_{37.25}33.56$ & $ \boldsymbol{25.10}_{27.37}$ & $_{34.23}32.10$ \\ 
2 & $ \boldsymbol{29.03}_{46.57}$ & $ \boldsymbol{24.14}_{27.52}$& -  & $ \boldsymbol{10.44}_{30.95}$ & $ \boldsymbol{12.47}_{46.50}$ & $ \boldsymbol{25.51}_{47.05}$ & $ \boldsymbol{28.40}_{47.78}$ & $ \boldsymbol{29.69}_{46.93}$ & $ \boldsymbol{8.72}_{29.42}$ & $ \boldsymbol{28.42}_{47.55}$ \\ 
3 & $ \boldsymbol{30.81}_{44.90}$ & $ \boldsymbol{27.12}_{29.56}$ & $ \boldsymbol{12.02}_{34.26}$& -  & $ \boldsymbol{10.80}_{35.16}$ & $ \boldsymbol{24.07}_{31.80}$ & $ \boldsymbol{26.40}_{45.41}$ & $ \boldsymbol{23.82}_{46.31}$ & $ \boldsymbol{7.87}_{19.91}$ & $ \boldsymbol{21.45}_{45.23}$ \\ 
4 & $ \boldsymbol{37.42}_{40.39}$ & $ \boldsymbol{30.85}_{31.82}$ & $ \boldsymbol{27.27}_{48.39}$ & $ \boldsymbol{15.90}_{44.92}$& -  & $ \boldsymbol{32.74}_{47.76}$ & $ \boldsymbol{28.47}_{47.58}$ & $ \boldsymbol{17.78}_{31.63}$ & $ \boldsymbol{10.71}_{33.48}$ & $ \boldsymbol{13.63}_{16.30}$ \\ 
5 & $ \boldsymbol{16.53}_{37.83}$ & $_{30.93}29.46$ & $ \boldsymbol{4.74}_{31.81}$ & $ \boldsymbol{4.67}_{13.90}$ & $ \boldsymbol{6.41}_{32.98}$& -  & $ \boldsymbol{12.13}_{34.06}$ & $ \boldsymbol{14.81}_{46.76}$ & $ \boldsymbol{2.64}_{10.94}$ & $ \boldsymbol{9.61}_{37.99}$ \\ 
6 & $ \boldsymbol{20.68}_{44.07}$ & $_{34.30}31.77$ & $ \boldsymbol{10.38}_{47.44}$ & $ \boldsymbol{6.32}_{37.23}$ & $ \boldsymbol{4.69}_{30.83}$ & $ \boldsymbol{18.14}_{42.48}$& -  & $ \boldsymbol{16.63}_{49.15}$ & $ \boldsymbol{4.66}_{30.58}$ & $ \boldsymbol{12.23}_{46.41}$ \\ 
7 & $_{39.89}35.51$ & $_{39.60}34.91$ & $_{41.33}40.69$ & $ \boldsymbol{19.34}_{47.91}$ & $ \boldsymbol{6.29}_{24.95}$ & $ \boldsymbol{37.31}_{41.23}$ & $ \boldsymbol{33.27}_{43.89}$& -  & $ \boldsymbol{16.33}_{46.44}$ & $ \boldsymbol{9.61}_{18.39}$ \\ 
8 & $ \boldsymbol{30.63}_{46.34}$ & $ \boldsymbol{26.33}_{29.79}$ & $ \boldsymbol{16.17}_{34.58}$ & $ \boldsymbol{11.24}_{22.23}$ & $ \boldsymbol{10.19}_{27.69}$ & $ \boldsymbol{23.96}_{33.44}$ & $ \boldsymbol{27.29}_{44.77}$ & $ \boldsymbol{23.42}_{47.34}$& -  & $ \boldsymbol{18.12}_{38.59}$ \\ 
9 & $_{39.92}38.31$ & $_{35.87}34.49$ & $ \boldsymbol{29.96}_{42.89}$ & $ \boldsymbol{12.08}_{45.15}$ & $ \boldsymbol{3.58}_{7.44}$ & $ \boldsymbol{28.91}_{45.68}$ & $ \boldsymbol{25.96}_{45.47}$ & $ \boldsymbol{9.77}_{19.13}$ & $ \boldsymbol{8.50}_{33.95}$& -  \\ 
\end{tabular}
}
\end{sc}
\end{small}
\end{table}

\vspace{.1in}
\color{black}
\noindent{\it Stock market shift detection.} Finally, we consider a financial data application using a semi-real limit order book stock market dataset generated by the Agent-Based Interactive Discrete Event Simulator (ABIDES) \citep{amrouni2021abides}. Within ABIDES, we can configure background traders to generate orders (or trades), producing various financial time series such as returns. The goal is to detect market shifts resulting from changes in the background trader configuration as early as possible, allowing traders to adapt their trading strategies to avoid losses.

By adding one or nine momentum traders \citep{vyetrenko2020get} to the background trader configuration (Settings 10 and 11 in Table~\ref{tab:additional_dependent_comparison_results}, respectively), we simulate minor and major market shifts that cause an unknown shift in the data distribution. Further details are provided in Appendix~\ref{appendix:more_real_data_exp_ABIDES}. We generate \(M = 800\) reference samples and 400 sequential observations of the market return time series, with \(w = 70\) and \(N = 10\). We perform 1000 Monte Carlo trials to determine the detection threshold satisfying the ARL constraint \(\gamma = 1000\), followed by another 1000 Monte Carlo trials (with the change occurring at \(\kappa = 200\)) to compute the EDD.

As shown in Table~\ref{tab:additional_dependent_comparison_results} (Settings 10 and 11), the major market shift (Setting 11) is quickly detected by parametric methods (Hotelling’s \(T^2\) and MEWMA), which, however, perform poorly under the minor market shift (Setting 10). Although neural networks can still detect minor market shifts (especially the classifier by \citealp{hushchyn2020online}), tuning and training them to achieve satisfactory EDD remains difficult. In contrast, our proposed online kernel CUSUM is robust, requires minimum training (except for tuning a handful of hyperparameters), and achieves the best EDD (Setting 10) or is very close to the best (Setting 11).

\color{black}
\section{Conclusion and Discussions}\label{sec:conclusion}

We presented an online kernel CUSUM procedure for change-point detection to overcome the limitations of existing Shewhart chart-type procedures, such as the Scan $B$-procedure, and boost the detection power from the linear-time kernel detection statistic, such as \citet{flynn2019change}.  We present accurate theoretical approximations to two standard performance metrics, the ARL and EDD, from which we establish the optimal window length to be logarithmic in ARL, analogous to the classical parametric results. We validate the effectiveness of our procedure through extensive numerical experiments, demonstrating its superior performance compared to various existing procedures. Our findings contribute to the non-parametric change-point detection literature, providing a practical and robust method for detecting changes in sequential data.

A future direction is to study the dependence of the performance on the data dimension and structure and see how to extend existing results for fixed sample two-sample tests: for instance, when the data comes from high-dimensional isotropic Gaussian distribution, the kernel method has a curse of dimensionality, as the population MMD drops to zero exponentially fast as the dimension increases \citep{ramdas2015decreasing,reddi2015high}; when data has an intrinsically low-dimensional structure such as manifold structures (many high-dimensional data including the commonly studied MNIST handwritten digits have this property), the MMD two-sample test has no curse-of-dimensionality \citep{cheng2021kernel}.

\section*{Acknowledgement}
This work is partially supported by an NSF CAREER CCF-1650913, and NSF DMS-2134037, CMMI-2015787, DMS-1938106, and DMS-1830210. The authors thank Lekun Wang for helping with additional experiments during the revision.


 \newcommand{\noop}[1]{}

\newpage

\appendix
\noindent
\textbf{\LARGE \centering Appendix of \mytitle}

\section{\color{black}Recursive implementation of detection statistic\color{black}}\label{appendix:moment}

\color{black}

\noindent \textit{Derivations for Algorithm~\ref{algo:kernel_CUSUM}.}
Recall the definitions of the Scan $B$-statistic \eqref{eq:scan-b} and $\rho$ \eqref{eq:cpq}, we have:
\begin{align*}
    {Z}_B(t) 
    & = \frac{\hat{\mathcal{D}}_B(t)}{\sqrt{\operatorname{Var}_{\infty}(\hat{\mathcal{D}}_B(t))}} = \rho \sqrt{B(B-1)} \hat{\mathcal{D}}_B(t) \\
     & = \frac{\rho}{N} \sqrt{B(B-1)} \sum_{n=1}^N {\hat{\mathcal{D}}(\mathbf{X}_B^{(n)},\mathbf{Y}_B(t))} \\
     & = \frac{\rho}{N} \sqrt{B(B-1)} \sum_{n=1}^N \frac{\sum_{i=1}^{B} \sum_{j \neq i}^{B} h\left(X^{(n)}_{i}, X^{(n)}_{j}, Y_{i}, Y_{j}\right)}{B(B-1)} \\
     & = \frac{\rho}{N\sqrt{B(B-1)}}  \sum_{n=1}^N \sum_{i=1}^{B} \sum_{j \neq i}^{B} h\left(X^{(n)}_{i}, X^{(n)}_{j}, Y_{i}, Y_{j}\right) \\
     & = \frac{\rho}{N\sqrt{B(B-1)}}  \sum_{n=1}^N \sum_{i=1}^{B} \sum_{j \neq i}^{B} k\left(X^{(n)}_{i}, X^{(n)}_{j}\right)+k\left(Y_{i}, Y_{j}\right)-k\left(X^{(n)}_{i}, Y_{j}\right)-k\left(X^{(n)}_{j}, Y_{i}\right) \\
     & = \frac{2\rho}{N\sqrt{B(B-1)}}  \sum_{n=1}^N \sum_{i=1}^{B} \sum_{j > i}^{B} k\left(X^{(n)}_{i}, X^{(n)}_{j}\right)+k\left(Y_{i}, Y_{j}\right)-k\left(X^{(n)}_{i}, Y_{j}\right)-k\left(X^{(n)}_{j}, Y_{i}\right) \\
     & = \underbrace{\frac{2\rho}{N} \frac{1}{\sqrt{B(B-1)}}}_{\text{\rm constant factor}}   \underbrace{\sum_{i=1}^{B} \sum_{j > i}^{B} \left(\sum_{n=1}^N k\left(X^{(n)}_{i}, X^{(n)}_{j}\right)+k\left(Y_{i}, Y_{j}\right)-k\left(X^{(n)}_{i}, Y_{j}\right)-k\left(X^{(n)}_{j}, Y_{i}\right)\right)}_{z_{B}}.
\end{align*}
Therefore, to calculate $z_{B}$ using $z_{B-1}$, we only need to calculate the following increment:
\[\Delta_{B} = z_{B} - z_{B-1} = \sum_{i=1}^{B} \sum_{n=1}^N k\left(X^{(n)}_{i}, X^{(n)}_{B}\right)+k\left(Y_{i}, Y_{B}\right)-k\left(X^{(n)}_{i}, Y_{B}\right)-k\left(X^{(n)}_{B}, Y_{i}\right).\]
Now, we have derived the ``constant factor'' as well as the increment to update ``$z$'' in Algorithm~\ref{algo:kernel_CUSUM}. 

Note  that, in the post-change block of Algorithm~\ref{algo:kernel_CUSUM}, the most recent sample has index \( i = w \) (rather than \( i = 1 \)). Therefore, to compute \( z_B \), we use the last (rather than the first) \( B \) samples in the post-change block. This distinction should not cause any confusion in understanding the updates in Algorithm~\ref{algo:kernel_CUSUM}, particularly given the graphical illustration in Figure~\ref{fig:online_update_illus}. This concludes the derivations for Algorithm~\ref{algo:kernel_CUSUM}.

\section{Moments of the Scan $B$-Statistic}\label{appendix:moment}

\begin{figure}[!htp]
\centerline{
\includegraphics[width = .95\textwidth]{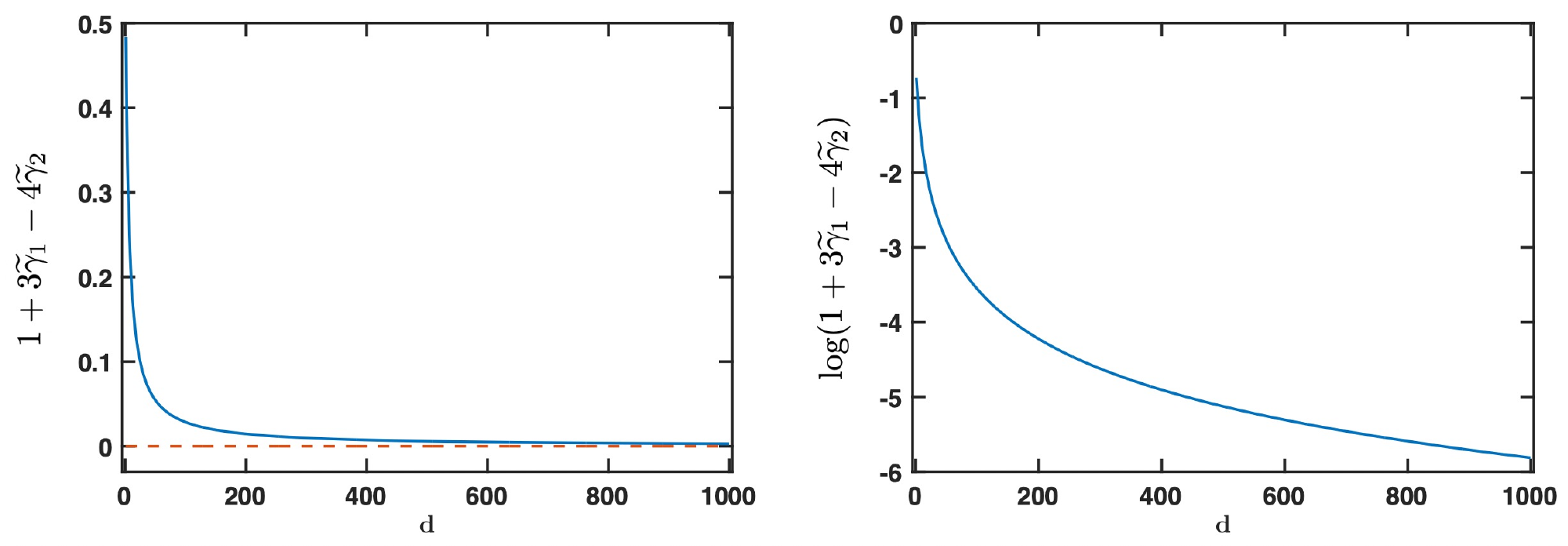}
}
\caption{Numerical evaluation of $1 + 3\tilde{{{r}}}_1 - 4\tilde{{{r}}}_2$ up to $d=1000$ when we take ${{r}}^2 = d \sigma^2$. We plot its raw value and its logarithm in the left and right panels, respectively. 
}
\label{fig:c1c2}
\end{figure}

\vspace{.1in}
\noindent \textit{Derivations in Remark~\ref{rmk:Neffect}.} For notational simplicity, we denote
\begin{align}\label{eq:constant_moments}
\begin{split}
 C_1 & = \mathbb{E}\left[h^{2}\left(X, X^{\prime}, Y, Y^{\prime}\right)\right], \\
 C_2 & = \operatorname{Cov}\left[h(X,X',Y,Y'), h\left(X^{\prime \prime}, X^{\prime \prime \prime}, Y, Y^{\prime}\right)\right],
\end{split}
\end{align}
where $X, X^{\prime}, X^{\prime \prime}, X^{\prime \prime \prime}, Y, Y^{\prime} $ are i.i.d. random variables following ${p}$.
By calculation, we have
\if\mycolumn1
\begin{align*}
    C_1 - C_2 =& \mathbb{E}[k^2(X,X')] - (\mathbb{E}[k(X,X')])^2 - 4(\mathbb{E}[k(X,X')k(X,Y)] - \mathbb{E}[k(X,X')k(X'',Y)]).
\end{align*}
\else
\begin{align*}
    C_1 - C_2 =& \mathbb{E}[k^2(X,X')] - (\mathbb{E}[k(X,X')])^2 \\
    &- 4(\mathbb{E}[k(X,X')k(X,Y)] - \mathbb{E}[k(X,X')k(X'',Y)]).
\end{align*}
\fi
When we choose a Gaussian RBF kernel 
with bandwidth parameter ${{r}} > 0$ and consider ${p} = {\cN}(\mu \mathbf{1}_{d}, \sigma^2 \ I_{d})$, we can evaluate the above quantity in closed-form.
For notational simplicity, we denote
\begin{align*}
        \tilde{{{r}}}_0 =& \left(\frac{{{r}}^2}{4\sigma^2 + {{r}}^2}\right)^{d/2},\\
    \tilde{{{r}}}_1=& \left(1-\frac{4\sigma^4}{(2\sigma^2 + {{r}}^2)^2}\right)^{d/2}, \\
    \tilde{{{r}}}_2=& \left(1-\frac{3\sigma^4}{(3\sigma^2 + {{r}}^2)(\sigma^2 + {{r}}^2)}\right)^{d/2}.
\end{align*}
Then, we have
\begin{align*}
     C_1 - C_2 = \tilde{{{r}}}_0 \left(1-\tilde{{{r}}}_1 - 4\tilde{{{r}}}_2 + 4\tilde{{{r}}}_1\right) = \tilde{{{r}}}_0 \left(1 + 3\tilde{{{r}}}_1 - 4\tilde{{{r}}}_2\right).
\end{align*}
We can see the problem boils down to evaluating $1 + 3\tilde{{{r}}}_1 - 4\tilde{{{r}}}_2$. When we use the median heuristic to select kernel bandwidth, we will have ${{r}}^2 \sim d\sigma^2$. 
We plug it in $1 + 3\tilde{{{r}}}_1 - 4\tilde{{{r}}}_2$ and we will have $1 + 3\tilde{{{r}}}_1 - 4\tilde{{{r}}}_2$ approximated as follows:
\begin{align*}
 1 + 3\left(1-\frac{3}{(3+d)(1+d)}\right)^{d/2} - 4 \left(1-\frac{4}{(2+d)^2}\right)^{d/2}.
\end{align*}
On one hand, we know $(1+1/d^2)^d \sim \exp\{1/d\} \rightarrow 1$ as $d \rightarrow \infty$. On the other hand, for relatively small $d$, we can calculate its value in Figure~\ref{fig:c1c2}.

As we can see in Figure~\ref{fig:c1c2}, the quantity $C_1 - C_2$ will be positive at least for $d$ up to $1000$. Since we also observe a monotonic decreasing trend, we conjecture that this quantity will always be positive.
In fact, this can be theoretically proved. By Taylor expansion with Lagrange remainder, we will have
\if\mycolumn1
\begin{align*}
    1 + 3\tilde{{{r}}}_1 - 4\tilde{{{r}}}_2
    \geq & 1+3 \exp\{-2d\sigma^4/{{r}}^4\}  - 4 \exp\{-3d\sigma^4/(2{{r}}^4)\}   - \frac{8d\sigma^6}{{{r}}^6}\exp\{{2d\sigma^6}/{{{r}}^6}\}.
\end{align*}
\else
\begin{align*}
    1 + 3\tilde{{{r}}}_1 - 4\tilde{{{r}}}_2
    \geq & 1+3 \exp\{-2d\sigma^4/{{r}}^4\} \\
    & - 4 \exp\{-3d\sigma^4/(2{{r}}^4)\}  \\
    & - \frac{8d\sigma^6}{{{r}}^6}\exp\{{2d\sigma^6}/{{{r}}^6}\}.
\end{align*}
\fi
Therefore, if we apply Taylor expansion and only keep the leading term, as long as
\begin{equation}\label{eq:gamma_condi}
    \sigma^2 =  \cO({{r}}^2/\sqrt{d}),
\end{equation}
we will have
\if\mycolumn1
\begin{align*}
    1+3 \exp\{-2d\sigma^4/{{r}}^4\} - 4 \exp\{-3d\sigma^4/(2{{r}}^4)\} 
    - \frac{8d\sigma^6}{{{r}}^6}\exp\{{2d\sigma^6}/{{{r}}^6}\} = \frac{3d\sigma^4}{2{{r}}^4} (1+o(1)).
\end{align*}
\else
\begin{align*}
    1+3 \exp\{-2d\sigma^4/{{r}}^4\} - 4 \exp\{-3d\sigma^4/(2{{r}}^4)\} & \\
    - \frac{8d\sigma^6}{{{r}}^6}\exp\{{2d\sigma^6}/{{{r}}^6}\} = &\frac{3d\sigma^4}{2{{r}}^4} (1+o(1)).
\end{align*}
\fi
As we can see, the median heuristic clearly satisfies condition~\eqref{eq:gamma_condi}. We can not only verify $1 + 3\tilde{{{r}}}_1 - 4\tilde{{{r}}}_2 \geq 0$ for large enough $d$, but also show the rate (at which it decays to zero) is $ \cO(1/d)$.

\begin{lemma}[Covariance structure of $\hat{\mathcal{D}}_B(t)$ under $H_0$]\label{lma:cov_H0}
Under $H_0$, for the un-normalized statistic $\hat{\mathcal{D}}_B(t)$ \eqref{eq:scan-b_unmoralized}, for block sizes $B_1, B_2 \geq 2$ and non-negative integer $s$, at time steps $t$ and $t+s$, the covariance is as follows:
\begin{align*}
    \operatorname{Cov}\left(\hat{\mathcal{D}}_{B_1}(t),\hat{\mathcal{D}}_{B_2}(t+s)\right)=C_2 \left(\begin{array}{c}
\ell \\
2
\end{array}\right)  \Bigg/ \left(\begin{array}{c}
B_1 \\
2
\end{array}\right) \left(\begin{array}{c}
B_2 \\
2
\end{array}\right),
\end{align*}where \begin{equation*}
  \ell =
    \begin{cases}
      0 &\text{, if } B_2 - s < 0;\\
      B_2 - s &\text{, if } 0 \leq B_2 - s < B_1;\\
      B_1 &\text{, otherwise},
    \end{cases}       
\end{equation*}
and $C_2$ is defined in \eqref{eq:constant_moments}.
\end{lemma}

\begin{proof}
Notice that $\ell$ represents the length of overlap between $\hat{\mathcal{D}}_{B_1}(t)$ and $\hat{\mathcal{D}}_{B_2}(t+s)$. By Lemma B.2 \citep{li2019scan} and notice that $\hat{\mathcal{D}}_{B_1}(t)$ and $\hat{\mathcal{D}}_{B_2}(t+s)$ have different normalizing factors, we complete the proof.

\end{proof}

\begin{lemma}[Lemma 6.1  \citep{li2019scan}]\label{lma:third_moment_H0}
Given block size $B \geq 2$ and the number of blocks $N$, under $H_0$, the third order moment of $\hat{\mathcal{D}}_B(t)$ \eqref{eq:scan-b_unmoralized} has an analytic expression as follows: 
\if\mycolumn1
\begin{equation*}\label{eq:third_moment}
    \begin{aligned}
&\mathbb{E}_{\infty}\left[(\hat{\mathcal{D}}_B(t))^3\right]\\
=& \ \frac{8(B-2)}{B^{2}(B-1)^{2}}\left\{\frac{1}{N^{2}} \mathbb{E}\left[h\left(X, X^{\prime}, Y, Y^{\prime}\right) h\left(X^{\prime}, X^{\prime \prime}, Y^{\prime}, Y^{\prime \prime}\right) h\left(X^{\prime \prime}, X, Y^{\prime \prime}, Y\right)\right]\right.\\
&\ +\frac{3(N-1)}{N^{2}} \mathbb{E}\left[h\left(X, X^{\prime}, Y, Y^{\prime}\right) h\left(X^{\prime}, X^{\prime \prime}, Y^{\prime}, Y^{\prime \prime}\right) h\left(X^{\prime \prime \prime}, X^{\prime \prime \prime \prime}, Y^{\prime \prime}, Y\right)\right] \\
&\ \left.+\frac{(N-1)(N-2)}{N^{2}} \mathbb{E}\left[h\left(X, X^{\prime}, Y, Y^{\prime}\right) h\left(X^{\prime \prime}, X^{\prime \prime \prime}, Y^{\prime}, Y^{\prime \prime}\right) h\left(X^{\prime \prime \prime \prime}, X^{\prime \prime \prime \prime \prime}, Y^{\prime \prime}, Y\right)\right]\right\} \\
&\  + \frac{4}{B^{2}(B-1)^{2}}\left\{\frac{1}{N^{2}} \mathbb{E}\left[h\left(X, X^{\prime}, Y, Y^{\prime}\right)^{3}\right]\right.\\
& \  +\frac{3(N-1)}{N^{2}} \mathbb{E}\left[h\left(X, X^{\prime}, Y, Y^{\prime}\right)^{2} h\left(X^{\prime \prime}, X^{\prime \prime \prime}, Y, Y^{\prime}\right)\right] \\
& \  \left.+\frac{(N-1)(N-2)}{N^{2}} \mathbb{E}\left[h\left(X, X^{\prime}, Y, Y^{\prime}\right) h\left(X^{\prime \prime}, X^{\prime \prime \prime}, Y, Y^{\prime}\right) h\left(X^{\prime \prime \prime \prime}, X^{\prime \prime \prime \prime \prime}, Y, Y^{\prime}\right)\right]\right\},
\end{aligned}
\end{equation*}
where $X, X^{\prime}, X^{\prime \prime}, X^{\prime \prime \prime}, X^{\prime \prime \prime \prime}, X^{\prime \prime \prime \prime \prime}, Y, Y^{\prime}, Y^{\prime \prime} $ are all i.i.d. random variables following ${p}$.
\else
\begin{equation*}\label{eq:third_moment}
    \begin{aligned}
\mathbb{E}_{\infty}\left[\hat{\mathcal{D}}_B^3(t)\right] = & \frac{8(B-2)}{B^{2}(B-1)^{2}}\left\{\frac{1}{N^{2}} \mathbb{E}\left[h_{12,12} h_{13,12} h_{13,13} \right]\right.\\
&+\frac{3(N-1)}{N^{2}} \mathbb{E}\left[h_{12,12} h_{13,12} h_{45,13} \right] \\
&\left.+\frac{(N-1)(N-2)}{N^{2}} \mathbb{E}\left[h_{12,12} h_{34,23} h_{56,13}\right]\right\} \\
+& \frac{4}{B^{2}(B-1)^{2}}\left\{\frac{1}{N^{2}} \mathbb{E}\left[h_{12,12}^{3}\right]\right.\\
&+\frac{3(N-1)}{N^{2}} \mathbb{E}\left[h_{12,12}^{2} h_{34,12}\right] \\
&\left.+\frac{(N-1)(N-2)}{N^{2}} \mathbb{E}\left[h_{12,12} h_{34,12} h_{56,12}\right]\right\},
\end{aligned}
\end{equation*}
where
\begin{align*}
    &h_{12,12} = h(X, X^{\prime}, Y, Y^{\prime}),
    &h_{13,23} = h(X^{\prime}, X^{\prime \prime}, Y^{\prime}, Y^{\prime \prime}), \\
    &h_{13,13} = h(X^{\prime \prime}, X, Y^{\prime \prime}, Y),
    &h_{45,13} = h(X^{\prime \prime \prime}, X^{\prime \prime \prime \prime}, Y^{\prime \prime}, Y), \\
    &h_{34,23} = h(X^{\prime \prime}, X^{\prime \prime \prime}, Y^{\prime}, Y^{\prime \prime}),
    &h_{56,13} = h(X^{\prime \prime \prime \prime}, X^{\prime \prime \prime \prime \prime}, Y^{\prime \prime}, Y), \\
    &h_{34,12} = h(X^{\prime \prime}, X^{\prime \prime \prime}, Y, Y^{\prime}),
    &h_{56,12} = h(X^{\prime \prime \prime \prime}, X^{\prime \prime \prime \prime \prime}, Y, Y^{\prime}),
\end{align*}
and 
$X, X^{\prime}, X^{\prime \prime}, X^{\prime \prime \prime}, X^{\prime \prime \prime \prime}, X^{\prime \prime \prime \prime \prime}, Y, Y^{\prime}, Y^{\prime \prime} $ are all i.i.d. random variables which follow ${p}$. 
\fi

The constant
$\mathbb{E}_{\infty} [Z_B^3(t)] = \mathbb{E}_{\infty} [(\hat{\mathcal{D}}_B(t))^3] \operatorname{Var}(\hat{\mathcal{D}}_B(t))^{-3/2}$ can be pre-calculated together with the constants in Lemma~\ref{lma:var_H0}.
\end{lemma}

Lastly, we evaluate the moment of Scan $B$-statistic \eqref{eq:scan-b} under $H_1$ as follows:

\begin{lemma}[Mean of $Z_B(t)$ under the alternative]\label{lma:mean_under_H1} 
Under $H_1$, assume change occurs at time step 0, at time step $t <  w $, the expectation of the Scan $B$-statistic \eqref{eq:scan-b} $\mu_B(t) = \mathbb{E}_{0} [Z_B(t)]$ has the following explicit expression:
\begin{align*}
    &\mu_B(t) = {\rho}\mathcal{D}({p},{q})  \sqrt{B(B-1)},\quad B \in [2:t],\\
    &\mu_B(t) = {\rho}\mathcal{D}({p},{q})  \frac{t(t-1)}{\sqrt{B(B-1)}},\quad B \in [t+1: w] ,
\end{align*}
where constant ${\rho}$ is defined in \eqref{eq:cpq}.
\end{lemma}

\begin{proof}
 We can observe that when a change occurs at $t=0$, the most recent $t$ samples will be post-change samples. Therefore, when $B > t$, the post-change block will also contain pre-change samples. Notice that $\mathcal{D}({p},{p}) = 0$ (see, e.g., Lemma 4 in \citet{gretton2012kernel}). Then it reduces to a simple evaluation of the expectation (it is simple because the expectation is linear). We omit further detailed derivations and conclude the proof. 
\end{proof}

\section{Proof of Lemma~\ref{lma:arl}}\label{appendix:thm1}

Under $H_0$, the event that the statistic crosses the threshold $b$ is rare. Therefore, as $b \rightarrow \infty$, 
for large enough $m$ such that
\begin{equation}\label{eq:ARL_condi}
    \frac{\log b}{w m}\rightarrow 0, \quad \frac{w m b}{e^{b^2/2}} \rightarrow 0, 
\end{equation}
the null distribution of $T_{w}$ can be approximated via the following exponential distribution with parameter $\lambda$:
\if\mycolumn1
\begin{equation}\label{eq:exp_approx}
\begin{split}
   \PP_{\infty} (T_{w} \leq m)  = \PP_{\infty} \left(\max_{t \in [m]} \max_{B \in [2:w]} Z_B(t) \geq b\right) \approx 1 - \exp\{-\lambda m\} \approx \lambda m.
\end{split}
\end{equation}
\else
\begin{equation}\label{eq:exp_approx}
\begin{split}
   \PP_{\infty} (T_{w} \leq m) &= \PP_{\infty} \left(\max_{t \in [m]} Z_t \geq b\right) \\
   & = \PP_{\infty} \left(\max_{t \in [m]} \max_{B \in [2:w]} Z_B(t) \geq b\right) \\
   &\approx 1 - \exp\{-\lambda m\} \approx \lambda m.
\end{split}
\end{equation}
\fi

The above result can be derived using the Poisson approximation technique (see, e.g., Theorem 1 \citep{arratia1989two} and 
Appendix D \citep{li2019scan}). Intuitively, the interarrival time of the first rare event will follow an exponential limiting distribution. More precise statements, as well as their detailed proof regarding the above exponential approximation, can be found in Appendix D in \citet{li2019scan}, and we omit further technical details here. 

One difference in our exponential approximation from that of \citet{li2019scan} is that they only require
\begin{equation}\label{eq:ARL_condi_li}
    \frac{\log b}{m}\rightarrow 0, \quad \frac{m b}{e^{b^2/2}} \rightarrow 0.
\end{equation}
By comparing condition~\eqref{eq:ARL_condi} and the above condition~\eqref{eq:ARL_condi_li}, we know that we cannot afford exponentially large $w$ with respect to $b$. Nevertheless, on one hand, window length $w$ is held constant in practice, which satisfies condition~\eqref{eq:ARL_condi} automatically; on the other hand, we will see in the following EDD analysis we only need $w \sim b$, which again satisfies condition~\eqref{eq:ARL_condi}.

\eqref{eq:exp_approx} tells us that the problem of approximating $\mathbb{E}_{\infty}[T_{w}]$ reduces to studying the extremes of random field $\{Z_B(t)\}$ with parameter $(t,B)$, which can be solved by determining the parameter $\lambda$ of the limiting exponential distribution. To achieve this goal, a popular method is described in Section~\ref{sec:arl_approx} (see steps (i), (ii), (iii) therein). 
A very clear and detailed instruction on how to prove the ARL approximation can be found in Appendices C and D \citep{li2019scan}. Here, we sketch the proof of Lemma~\ref{lma:arl} as follows:

\begin{proof}[Proof Sketch of Lemma~\ref{lma:arl}]\label{proof:thm_arl}
We give high-level ideas in each step described in Section~\ref{sec:arl_approx}. Moreover, we highlight the difference from the proof in \citet{li2019scan} as follows:
\begin{itemize}
    \item[(i)] We usually make use of log moment generating function $\psi_B(\theta)=\log \mathbb{E}_{\infty}[e^{\theta Z_{B}(t)}]$ and the new measure is defined as
$$d {P}_B=\exp \{\theta_B Z_{B}(t)-\psi_B(\theta_B)\} d \PP_{\infty}.$$
Note that the new measure $d {P}_B$ is within the exponential family.
An advantage of this is that we can easily manipulate the mean of $Z_B(t)$ under this new measure, which is $\Dot{\psi_B}(\theta_B)$, by properly choosing $\theta_B$. Specifically, we will choose $\theta_B$ such that $\Dot{\psi_B}(\theta_B) = b$.
\item[(ii)] We will apply the likelihood ratio identity. Before this, we first define some terms: the log-likelihood ratio is $$\ell_{B}(t) = \log(d {P}_B/d \PP_{\infty}) = \theta_B Z_{B}(t)-\psi_B(\theta_B).$$ 
The {\it global term} is $\tilde{\ell}_{B}(t)=\theta_B\left(Z_{B}(t)-b\right)$ 
and {\it local terms} are
\begin{align*}
    M_{B}&=\max _{t \in [m], \ s \in [2:w]} e^{\ell_{s}(t)-\ell_{B}(t)}, \\
    S_{B}&=\sum_{t = 1}^m \sum_{s \in [2:w]} e^{\ell_{s}(t)-\ell_{B}(t)}.
\end{align*}
For notational simplicity, we denote $$\Omega = [m]\times[2:w].$$ By likelihood ratio identity,  $\mathbb{E}[X ; A] = \mathbb{E}[X \mathds{1}_A]$, where $\mathds{1}_A = 1$ if event $A$ is true and zero otherwise, we have
\if\mycolumn1
\begin{equation*}
\begin{aligned}
 & \ \PP_{\infty}\left(\max_{(t,B) \in \Omega} Z_{B}(t)\geq b\right) \\
=& \ \mathbb{E}_{\infty}\left[1 ; \max_{(t,B) \in \Omega} Z_{B}(t)\geq b\right] =  \mathbb{E}_{\infty}\Bigg[\underbrace{\frac{\sum_{(t,B) \in \Omega} e^{\ell_{B}(t)}}{\sum_{(t,s) \in \Omega} e^{\ell_{s}(t)}}}_{=1} ; \max_{(t,B) \in \Omega} Z_{B}(t)\geq b\Bigg] \\
=& \sum_{(t,B) \in \Omega} \mathbb{E}_{\infty}\left[\frac{e^{\ell_{B}(t)}}{\sum_{(t,s)\in \Omega} e^{\ell_{s}(t)}} ; \max_{(t,B) \in \Omega} Z_{B}(t)\geq b\right] \\
=& \sum_{(t,B) \in \Omega} \mathbb{E}_{B}\left[\frac{1}{\sum_{(t,s)} e^{\ell_{s}(t)}} ; \max_{(t,B)\in \Omega} Z_{B}(t)\geq b\right] \\
=& \ m \sum_{B = 2}^w e^{\psi_B(\theta_B) - \theta_B b} \ \mathbb{E}_{B}\left[\frac{M_{B}}{S_{B}} e^{-\left(\tilde{\ell}_{B}+\log M_{B}\right)} ; \tilde{\ell}_{B}+\log M_{B} \geq 0\right].
\end{aligned}
\end{equation*}
\else
\begin{equation*}
\begin{aligned}
 &\PP_{\infty}\left\{\max_{(t,B) \in \Omega} Z_{B}(t)\geq b\right\}= \ \mathbb{E}_{\infty}\left[1 ; \max_{(t,B) \in \Omega} Z_{B}(t)\geq b\right]\\
=& \ \mathbb{E}_{\infty}\Bigg[\underbrace{\frac{\sum_{(t,B) \in \Omega} e^{\ell_{B}(t)}}{\sum_{(t,s) \in \Omega} e^{\ell_{s}(t)}}}_{=1} ; \max_{(t,B) \in \Omega} Z_{B}(t)\geq b\Bigg] \\
=& \sum_{(t,B) \in \Omega} \mathbb{E}_{\infty}\Bigg[\frac{e^{\ell_{B}(t)}}{\sum_{(t,s)\in \Omega} e^{\ell_{s}(t)}} ;  \max_{(t,B) \in \Omega} Z_{B}(t)\geq b\Bigg] \\
=& \sum_{(t,B)} {E}_{B}\left[\frac{1}{\sum_{(t,s)} e^{\ell_{s}(t)}} ; \max_{(t,B)} Z_{B}(t)\geq b\right] \\
=& \ m \sum_{B = 2}^w e^{\psi_B(\theta_B) - \theta_B b} \ {E}_{B}\Bigg[\frac{M_{B}}{S_{B}} e^{-\left(\tilde{\ell}_{B}+\log M_{B}\right)} ; \\
&\quad \quad \quad \quad \quad \quad \quad \quad \quad \quad \quad \quad \quad \quad  \tilde{\ell}_{B}+\log M_{B} \geq 0\Bigg].
\end{aligned}
\end{equation*}
\fi
\item[(iii)] We invoke the localization theorem (see, e.g., \citet{siegmund2010tail} and  \citet{yakir2013extremes}, for details) to show the asymptotic independence of local fields and global term in the equation above. 
Matching the equation in step (ii) above to \eqref{eq:exp_approx}, we know that
$$\lambda = \sum_{B = 2}^w e^{\psi_B(\theta_B) - \theta_B b} \ \mathbb{E}_{B}\left[\frac{M_{B}}{S_{B}} e^{-\left(\tilde{\ell}_{B}+\log M_{B}\right)} ; \tilde{\ell}_{B}+\log M_{B} \geq 0\right].$$
Therefore, the ARL can be approximated as $1/\lambda$. To evaluate the expectation in the above equation, we investigate the covariance structure of this random field as in Lemma~\ref{lma:cov_H0}. 
As suggested by \citet{xie2021sequentialb}, we can just consider taking $B_1=B_2=B$ in Lemma~\ref{lma:cov_H0} when evaluating the covariance, and doing this can still maintain the required level of accuracy. Therefore, the covariance structure is exactly the same as the one in \citet{li2019scan} under the online setting, and the rest of the proof directly follows the proof therein.
\end{itemize}

\end{proof}

In addition to the proof of the ARL approximation, we also show how to further simplify the expression when we only consider the first two moments' information as follows:

\vspace{.1in}
\noindent \textit{Derivation of \eqref{eq:ARL_Gaussian_assumption}.}
As $\mu \rightarrow \infty$, we have
$$    \nu(\mu) \approx \frac{(2 / \mu)(\Phi(\mu / 2)-0.5)}{(\mu / 2) \Phi(\mu / 2)+\phi(\mu / 2)} = \frac{2}{\mu^2} (1+o(1)).$$
Therefore, by plugging \eqref{eq:theta_B_Gaussian_assumption} into \eqref{eq:arl_approx}, we have $\mathbb{E}_{\infty}[T_{w}]$ expressed as follows:
\begin{align*}
      & \ \frac{\sqrt{2 \pi}}{b} \left\{\sum_{B = 2}^w e^{-b^2/2} \frac{\left(2 B-1\right)}{ B\left(B-1\right)}  \frac{2}{\left(b \sqrt{\frac{2\left(2 B-1\right)}{B\left(B-1\right)}}\right)^2}\right\}^{-1} \\
     = & \ \frac{\sqrt{2 \pi}}{b} e^{b^2/2} \left\{\sum_{B = 2}^w  \frac{\left(2 B-1\right)}{ B\left(B-1\right)}  \frac{{B\left(B-1\right)}}{b^2 {{\left(2 B-1\right)}}}\right\}^{-1} (1+o(1))  = \frac{\sqrt{2 \pi}b}{ w } e^{b^2/2} (1+o(1)).
\end{align*}

\color{black}
\section{Proof of Lemma~\ref{WEDD}}\label{appendix:WEDD}

Suppose that the change happens at $\kappa\geq 0$. Recall that the detection statistic for the detection procedure is defined as $Z_t = \max_{B\in [2:\min\{w,t\}]}Z_B(t)$.
Consider a detection statistic $Z_{t,\kappa}$ generated by {\it starting} the procedure at time $\tau$, which means using samples $Y_{\kappa+1}, \ldots, Y_{t}$ to form the first detection statistic and running the recursion from there. Clearly, one can see that $Z_{t,\kappa}\leq Z_t$, because the second term may possibly be the maximum taken over more terms. Thus, this suggests that the associated stopping time, $T_{w,t}$, defined similar to $T_w$ using $Z_{t,\kappa}$ replacing $Z_t$, will be ``greater'' than $T$. To be precise,
\begin{equation}
\mathbb E_{\kappa}[T_w-\kappa | T > \kappa, \mathcal F_\kappa] \leq \mathbb E[T_{w,\kappa}-\kappa | T > \kappa, \mathcal F_\kappa] = \mathbb E_0[T_{w}].
\end{equation}
The last equality is true due to the stationarity of the detection statistic in our setting, and that from time $\kappa+1$ on, the detection procedure $T_{w,\kappa}$ does not employ any information from $\mathcal F_\kappa$.

\color{black}

\section{Proof of Lemma~\ref{lma:EDD}}\label{appendix:thm2}
In this subsection, we will formally prove Lemma~\ref{lma:EDD}. This proof involves many technical details, but the high-level idea is simple. To help readers understand this simple idea, we first sketch this proof. For notational simplicity, we denote the detection statistic of our procedure as:
\begin{equation}\label{eq:proposed_stat}
    Z_t = \max_{B \in [2:w]} {Z}_B(t),
\end{equation}
where ${Z}_B(t)$ is the Scan $B$-statistic defined in \eqref{eq:scan-b}. 

In the following, to improve readability, we first sketch the proof of Lemma~\ref{lma:EDD} since it requires several technical lemmas. After that, we will prove all technical results rigorously.

\begin{proof}[Proof of Lemma~\ref{lma:EDD}]
On one hand, for large enough window length choice \eqref{eq:Bmax_choice}, the detection procedure will not stop too early nor too late. To be precise, for any $ w  \geq 7 b/({\rho}\mathcal{D}({p},{q})) $ and $t_1 = b/(4{\rho}\mathcal{D}({p},{q}) )$, as $b \rightarrow \infty$, we will have
\if\mycolumn1
\begin{align}
    &\PP_{0}\left(T_{w} < t_1\right) \leq 2 t_1^2 \left(\exp\left\{-\frac{Nb^2}{512K^2}\right\}+\exp\left\{-\frac{b^2}{128K^2}\right\}\right), \label{eq:pTb<t1} \\
\begin{split} \label{eq:pTb>t2}
    &\PP_{0}\left(T_{w} >  w \right) \leq \left(2 \exp\left\{-\frac{Nb^2}{512K^2}\right\} +  2\exp\left\{-\frac{ b^3}{128 {\rho}\mathcal{D}({p},{q}) K^2}\right\}\right)^{\frac{4b}{{\rho}\mathcal{D}({p},{q})}}. 
\end{split}
\end{align}
\else
\begin{align}
    &\PP_{0}\left(T_{w} < t_1\right) \leq 2 t_1^2 \left(\exp\left\{-\frac{Nb^2}{512K^2}\right\}+\exp\left\{-\frac{b^2}{128K^2}\right\}\right), \label{eq:pTb<t1} \\
\begin{split} \label{eq:pTb>t2}
    &\PP_{0}\left(T_{w} >  w \right) \\
    &\leq \left(2 \exp\left\{-\frac{Nb^2}{512K^2}\right\} +  2\exp\left\{-\frac{ b^3}{128 {\rho}\mathcal{D}({p},{q}) K^2}\right\}\right)^{4b/({\rho}\mathcal{D}({p},{q})) }. 
\end{split}
\end{align}
\fi

On the other hand, for any $t \in [t_1,  w ]$, we can show that for our detection statistic $Z_t = \max_{B \in [2:w] } Z_B(t)$, the maximum is attained at $B = t$ with high probability. To be precise, we denote $B_1(t) = {t}/{(8({\rho}\mathcal{D}({p},{q})  \vee 1))},$ and 
\begin{align} \label{event:max_at_t}
\begin{split}
     A_{1,t} &= \left\{\max_{2 \leq B \leq t \wedge w} Z_B(t) = \max_{B_1(t) \leq B \leq t} Z_B(t)\right\},\\
    A_{2,t} &= \left\{\max_{B_1(t) \leq B \leq t} Z_B(t) = Z_t(t)\right\},   
\end{split}
\end{align}
For any $t \in [t_1,  w ]$, as $b \rightarrow \infty$, we will have
\if\mycolumn1
\begin{align}
    &\PP_{0}\left(A_{t,1}^{\mathsf{c}}\right) \leq  \frac{b}{16({\rho}\mathcal{D}^2({p},{q}) \vee 1)} \Bigg(\exp \left\{-\frac{N b^2}{2^{15} ({\rho}\mathcal{D}^2({p},{q}) \vee 1) K^2}\right\} \label{eq:pAt1} \\
    & \quad \quad \quad \quad \quad \quad  \quad \quad \quad \quad \quad \quad \quad \quad + \exp\left\{-\frac{b^2}{2^{13} ({\rho}\mathcal{D}^2({p},{q}) \vee 1) K^2}\right\}\Bigg) (1+o(1)), \nonumber\\
    &\PP_{0}(A_{t,2}^{\mathsf{c}}) \leq 6 \Bigg( \exp\left\{-\frac{b {\rho}\mathcal{D}({p},{q}) }{2^{13} ({\rho}\mathcal{D}({p},{q})  \vee 1) K^2}\right\} + \exp\left\{-\frac{N{\rho}\mathcal{D}^2({p},{q})}{512 K^2}\right\}\Bigg) (1+o(1)), \label{eq:pAt2} 
\end{align}
\else
\begin{align}
    &\PP_{0}\left(A_{t,1}^{\mathsf{c}}\right) \leq  \frac{b}{16({\rho}\mathcal{D}^2({p},{q}) \vee 1)} \Bigg(\exp \left\{-\frac{N b^2}{2^{15} ({\rho}\mathcal{D}^2({p},{q}) \vee 1) K^2}\right\} \nonumber \\
    &\quad \quad \quad \quad   + \exp\left\{-\frac{b^2}{2^{13} ({\rho}\mathcal{D}^2({p},{q}) \vee 1) K^2}\right\}\Bigg) (1+o(1)), \label{eq:pAt1}\\
    &\PP_{0}(A_{t,2}^{\mathsf{c}}) \leq 6 \Bigg( \exp\left\{-\frac{b {\rho}\mathcal{D}({p},{q}) }{2^{13} ({\rho}\mathcal{D}({p},{q})  \vee 1) K^2}\right\} \nonumber\\
    &\quad \quad \quad \quad \quad \quad \quad \quad \  + \exp\left\{-\frac{N{\rho}\mathcal{D}^2({p},{q})}{512 K^2}\right\}\Bigg) (1+o(1)) \label{eq:pAt2},
\end{align}
\fi
where the superscript $^{\mathsf{c}}$ denotes the complement of a set.

Now, equations~\eqref{eq:pTb<t1} and \eqref{eq:pTb>t2} tell us $T_{w}$ takes value from $[t_1,  w ]$ with high probability, and equations~\eqref{eq:pAt1} and \eqref{eq:pAt2} show that at time $t \in [t_1,  w ]$, with high probability, the maximum over $B \in [2:w]$ is attained at $B = t$. These enable us to calculate the expectation of the detection statistic at the stopping time as follows: 
\if\mycolumn1
\begin{align}\label{eq:EZTb}
    &\Big|\mathbb{E}_{0}[Z_{T_{w}}] - {\rho}\mathcal{D}({p},{q})  \mathbb{E}_{0}[{T_{w}}]\Big|  \leq   \Bigg({\rho}\mathcal{D}({p},{q}) + 12K\exp\left\{-\frac{N{\rho}\mathcal{D}^2({p},{q})}{512 K^2}\right\}  \\
    & \quad \quad \quad \quad \quad \quad \quad \quad \quad \quad \quad \quad \quad \quad \quad  +  12K\exp\left\{-\frac{b {\rho}\mathcal{D}({p},{q}) }{2^{13} ({\rho}\mathcal{D}({p},{q})  \vee 1) K^2}\right\}\Bigg) (1+o(1)).   \nonumber
\end{align}
\else
\begin{align}\label{eq:EZTb}
\begin{split}
    &\Big|\mathbb{E}_{0}[Z_{T_{w}}] - {\rho}\mathcal{D}({p},{q})  \mathbb{E}_{0}[{T_{w}}]\Big|  \leq   \Bigg({\rho}\mathcal{D}({p},{q})  + 12K\exp\left\{-\frac{N{\rho}\mathcal{D}^2({p},{q})}{512 K^2}\right\} \\
    & \quad \quad \quad \quad \quad +  12K\exp\left\{-\frac{b {\rho}\mathcal{D}({p},{q}) }{2^{13} ({\rho}\mathcal{D}({p},{q})  \vee 1) K^2}\right\}\Bigg) (1+o(1)).  
\end{split}
\end{align}
\fi

Finally, since the overshoot, which is the occurrence of detection statistic exceeding its target threshold $b$, will be $o(1)$ given that $b \rightarrow \infty$, we will have the detection statistic at $t = T_{w}$ approximated as follows:
\begin{equation}\label{eq:wald}
    \mathbb{E}_{0}[Z_{T_{w}}] = b(1+o(1)).
\end{equation}
By equations~\eqref{eq:EZTb} and \eqref{eq:wald}, we complete the proof. Detailed statement and proof of equations~\eqref{eq:pTb<t1} and \eqref{eq:pTb>t2} can be found in Lemma~\ref{lma1}. Equations~\eqref{eq:pAt1} and \eqref{eq:pAt2} are proved in Lemma~\ref{lma2}. Equation~\eqref{eq:EZTb} is proved in Lemma~\ref{lma:expectation_max}. 
\end{proof}

We now present and prove the technical lemmas used in the proof sketch above.
We will start with an important concentration result for the Scan $B$-statistic. Recall its definition in \eqref{eq:scan-b} as follows: 
\begin{equation*}
    {Z}_B(t) = \frac{\sum_{i=1}^N {\hat{\mathcal{D}}(\mathbf{X}_B^{(i)},\mathbf{Y}_B(t))}}{N \sqrt{\operatorname{Var}_{\infty}(\hat{\mathcal{D}}_B(t))}}.
\end{equation*}
As we can see, the difficulty comes from the fact that the average in Scan $B$-statistic is taken over correlated random variables since we reuse the post-change block $\mathbf{Y}_B(t)$. Nevertheless, we can leverage Azuma's inequality for martingale and get the following concentration result for Scan $B$-statistic:

\begin{lemma}[Concentration inequality for Scan $B$-statistic]\label{lma:scanb_concentration} For any window length choice $ w  \geq 2$, at time step $t \leq  w $, for any block size $B_0 \in [2:w]$, the Scan $B$-statistic $Z_{B_0}(t)$ \eqref{eq:scan-b} satisfies the following concentration inequality: for any $z_1, z_2 > 0$ and $z = z_1 + z_2$, we have
\if\mycolumn1
\begin{align}\label{eq:scanB_concentration}
\begin{split}
        \PP_{0}(  |Z_{B_0}(t) &- \mu_{B_0}(t)|  \geq z) \leq 2\exp \left\{-\frac{N z_1^2}{32K^2 }\right\} + 2\exp \left\{-\frac{{B_0}  z_2^2}{16K^2}\right\}, 
\end{split}
\end{align}
\else
\begin{align}\label{eq:scanB_concentration}
\begin{split}
        \PP_{0}(  |Z_{B_0}(t) &- \mu_{B_0}(t)|  \geq z) \\
        &\leq 2\exp \left\{-\frac{N z_1^2}{32K^2 }\right\} + 2\exp \left\{-\frac{{B_0}  z_2^2}{16K^2}\right\}, 
\end{split}
\end{align}
\fi
where the expectation $\mu_{B_0}(t)$ has explicit formula as shown in Lemma~\ref{lma:mean_under_H1}.
\end{lemma}

\begin{proof}
For $z = z_1 + z_2 > 0$ (with $z_1, z_2 > 0$), event $\{|x-y|>z\}$ is a subset of event $\{|x-w| \geq z_1\} \cup \{|w-y| \geq z_2\}$, since if we take the complement we can easily show $\{|x-w| < z_1\} \cap \{|w-y| < z_2\} \subset \{|x-y| < z\}$. Let us define the following conditional expectation (which is a random variable), 
$$f(\mathbf{Y}_B(t)) = \mathbb{E}_{0}\left[\hat{\mathcal{D}}(\mathbf{X}_B^{(1)},\mathbf{Y}_B(t)) \mid \mathbf{Y}_B(t)\right].$$ 
We will have
\begin{equation}\label{eq:triangel_ineq_proof}
    \{|Z_{B_0}(t) - \mu_{B_0}(t)|  \geq z\} \subset \{|Z_B(t) - f(\mathbf{Y}_B(t))|\geq z_1\} \cap \{|f(\mathbf{Y}_B(t)) - \mu_B(t)| \geq z_2\}.
\end{equation}

On one hand, we can show $$\sum_{i=1}^N \hat{\mathcal{D}}(\mathbf{X}_B^{(i)},\mathbf{Y}_B(t)) - Nf(\mathbf{Y}_B(t))$$ is a martingale with its difference bounded within $[-4K,4K]$. Therefore, by Azuma's inequality, for any $z>0$, we have
\if\mycolumn1
\begin{align*}
    \PP_{0}\left(  |Z_B(t) - f(\mathbf{Y}_B(t))| \geq z\right) =& \ \PP_{0}\left(  \left|\sum_{i=1}^N \hat{\mathcal{D}}(\mathbf{X}_B^{(i)},\mathbf{Y}_B(t)) - N f(\mathbf{Y}_B(t)))\right| \geq N z\right) \\
    \leq & \  2\exp \left\{-\frac{(N z)^2}{32K^2 N}\right\} = 2\exp \left\{-\frac{N z^2}{32K^2 }\right\}. 
\end{align*}
\else
\begin{align*}
    &\PP_{0}\left(  |Z_B(t) - f(\mathbf{Y}_B(t))| \geq z\right) \\
    = &\PP_{0}\left(  \left|\sum_{i=1}^N \hat{\mathcal{D}}(\mathbf{X}_B^{(i)},\mathbf{Y}_B(t)) - N f(\mathbf{Y}_B(t)))\right| \geq N z\right) \\
    \leq & 2\exp \left\{-\frac{(N z)^2}{32K^2 N}\right\} = 2\exp \left\{-\frac{N z^2}{32K^2 }\right\}. 
\end{align*}
\fi

On the other hand, we can apply a similar concentration inequality on the empirical estimator of MMD (with sample size $B$), which can be derived from the large deviation bound on U-statistics (cf. \citet{hoeffding1994probability} or Theorem 10 in  \citet{gretton2012kernel}). To be precise, for any $z>0$, we have:
\begin{align*}
    \PP_{0}\left(  |f(\mathbf{Y}_B(t)) - \mu_B(t)| \geq z\right) \leq 2\exp \left\{-\frac{B  z^2}{16K^2}\right\}. 
\end{align*}

Finally, by \eqref{eq:triangel_ineq_proof} and the above two concentration inequalities, we can prove \eqref{eq:scanB_concentration}.
\end{proof}

\begin{lemma}\label{lma1}
Under $H_1$ where ${p} \not= {q}$, assume change occurs at time step 0. Under Assumptions~\ref{assumption:bounded_kernel} and~\ref{assumption:detection},
as $b\rightarrow \infty$, for any $N > 0$, we have:

\noindent
(1) for any $t_1 < b/(2{\rho}\mathcal{D}({p},{q}) )$, where constant ${\rho}$ is defined in \eqref{eq:cpq},
the following holds for the stopping time of  online kernel CUSUM $T_{w}$ \eqref{eq:stopping_time}:
\begin{align*}
    &\PP_{0}\left(T_{w} < t_1\right) \leq 2 t_1^2 \left(\exp\left\{-\frac{Nb^2}{512K^2}\right\}+\exp\left\{-\frac{b^2}{128K^2}\right\}\right).
\end{align*}

\noindent
(2) for any $t_2 \geq 4b/({\rho}\mathcal{D}({p},{q})) $, as long as we choose $ w  > 3b/({\rho}\mathcal{D}({p},{q})) $, the following holds for $T_{w}$:
\if\mycolumn1
\begin{align*}
    \PP_{0}\left(T_{w} > t_2\right)
    \leq & \left(2 \exp\left\{-\frac{Nb^2}{512K^2}\right\} +  2\exp\left\{-\frac{ b^3}{128 {\rho}\mathcal{D}({p},{q}) K^2}\right\}\right)^{t_2 - 3b/({\rho}\mathcal{D}({p},{q})) } \\
    \leq &\left(2 \exp\left\{-\frac{Nb^2}{512K^2}\right\} +  2\exp\left\{-\frac{ b^3}{128 {\rho}\mathcal{D}({p},{q}) K^2}\right\}\right)^{b/({\rho}\mathcal{D}({p},{q})) }.
\end{align*}
\else
\begin{align*}
    &\PP_{0}\left(T_{w} > t_2\right)\\
    \leq &\left(2 \exp\left\{-\frac{Nb^2}{512K^2}\right\} +  2\exp\left\{-\frac{ b^3}{128 {\rho}\mathcal{D}({p},{q}) K^2}\right\}\right)^{t_2 - 3b/({\rho}\mathcal{D}({p},{q})) } \\
    \leq &\left(2 \exp\left\{-\frac{Nb^2}{512K^2}\right\} +  2\exp\left\{-\frac{ b^3}{128 {\rho}\mathcal{D}({p},{q}) K^2}\right\}\right)^{b/({\rho}\mathcal{D}({p},{q})) }.
\end{align*}
\fi
\end{lemma}

\begin{proof} 
Let us prove (2) first. Intuitively, it is a rare event that $T_{w}$ exceeds $t_2$ for such large $t_2$. By definition, we have:
\begin{align*}
    \PP_{0}(T_{w} > t_2) &= \PP_{0}(\max_{2 \leq t \leq t_2} Z_t < b) = \prod_{t=2}^{t_2} \PP_{0}(Z_t < b)\\
    &= \prod_{t=2}^{t_2} \prod_{B=2}^{t \wedge w} \PP_{0}(Z_{B}(t) < b).
\end{align*}
Here, due to our choice of $t_2$, we have
\begin{align*}
    \PP_{0}(T_{w} >  w ) \leq \prod_{t=3b/({\rho}\mathcal{D}({p},{q})) }^{t_2} \PP_{0}(Z_{t \wedge w}(t) < b).
\end{align*}
Notice that, as long as $B > 3b/({\rho}\mathcal{D}({p},{q}))  > 4/3$, we have
\begin{equation*}
    \mathbb{E}_{0}[Z_{B}(t)] =  {\rho}\mathcal{D}({p},{q})  \sqrt{B(B-1)} \geq {\rho}\mathcal{D}({p},{q})  \frac{B}{2} > \frac{3b}{2},
\end{equation*}
where the last inequality comes from the fact that $\sqrt{x(x-1)}>x/2, \forall x > 4/3$. The condition $3b/({\rho}\mathcal{D}({p},{q}))  > 4/3$ can easily be satisfied as $b$ goes to infinity.
Finally, by Lemma~\ref{lma:scanb_concentration}, we have 
\if\mycolumn1
\begin{align*}
    &\PP_{0}\left(T_{w} > t_2\right) 
    \leq \left(2 \exp\left\{-\frac{Nb^2}{512K^2}\right\} +  2\exp\left\{-\frac{ b^3}{128 {\rho}\mathcal{D}({p},{q}) K^2}\right\}\right)^{t_2 - 3b/({\rho}\mathcal{D}({p},{q})) }.
\end{align*}
\else
\begin{align*}
    &\PP_{0}\left(T_{w} > t_2\right) \\
    \leq &\left(2 \exp\left\{-\frac{Nb^2}{512K^2}\right\} +  2\exp\left\{-\frac{ b^3}{128 {\rho}\mathcal{D}({p},{q}) K^2}\right\}\right)^{t_2 - 3b/({\rho}\mathcal{D}({p},{q})) }.
\end{align*}
\fi

Next, let us prove (1). We have:
\begin{align*}
    \PP_{0}(T_{w} < t_1) &= \PP_{0}(\max_{0< t < t_1} Z_t \geq b) \\
    &\leq \sum_{t=1}^{t_1} \PP_{0}(Z_t \geq b) \leq  \sum_{t=1}^{t_1} \sum_{B=2}^{t \wedge w} \PP_{0}(Z_{B}(t) \geq b).
\end{align*}
Similarly, notice that
\begin{equation*}
    \mathbb{E}_{0}[Z_{B}(t)] = {\rho}\mathcal{D}({p},{q})  \sqrt{B(B-1)} \leq {\rho}\mathcal{D}({p},{q})  t_1 < b/2,
\end{equation*}
where the last inequality is guaranteed by the choice of $t_1$. 
Again, by Lemma~\ref{lma:scanb_concentration}, we have 
\if\mycolumn1
\begin{align*}
    \PP_{0}(T_{w} < t_1) = & \ \PP_{0}(\max_{0< t < t_1} Z_t \geq b) 
    \leq \sum_{t=1}^{t_1} \PP_{0}(Z_t \geq b) 
    \leq  \sum_{t=1}^{t_1} \sum_{B=2}^{t \wedge w} \PP_{0}(Z_{B}(t) \geq b)\\
    \leq & \ t_1^2 \Bigg(2\exp \left\{-\frac{N (b/4)^2}{32K^2 }\right\} + 2\exp \left\{-\frac{2 (b/4)^2}{16K^2}\right\}\Bigg).
\end{align*}
\else
\begin{align*}
    &\PP_{0}(T_{w} < t_1) = \PP_{0}(\max_{0< t < t_1} Z_t \geq b) \\
    \leq &\sum_{t=1}^{t_1} \PP_{0}(Z_t \geq b) 
    \leq  \sum_{t=1}^{t_1} \sum_{B=2}^{t \wedge w} \PP_{0}(Z_{B}(t) \geq b)\\
    \leq & t_1^2 \Bigg(2\exp \left\{-\frac{N (b/4)^2}{32K^2 }\right\} + 2\exp \left\{-\frac{2 (b/4)^2}{16K^2}\right\}\Bigg).
\end{align*}
\fi
Now we complete the proof.

\end{proof}

\begin{lemma}\label{lma2}
Under $H_1$, where ${p} \not= {q}$ and we assume change occurs at time step 0, and Under Assumptions~\ref{assumption:bounded_kernel} and~\ref{assumption:detection},
we choose 
$$t_1 = \frac{b}{4{\rho}\mathcal{D}({p},{q})}, \quad  w  > \frac{3b}{{\rho}\mathcal{D}({p},{q})},$$ 
which satisfies the constraints in Lemma~\ref{lma1} and constant ${\rho}$ is defined in \eqref{eq:cpq}, we further take $$B_1(t) = \frac{t}{8({\rho}\mathcal{D}({p},{q})  \vee 1)},$$
for any $t \in [t_1, w ]$, recall that we denote 
\begin{align*} 
    A_{1,t} &= \left\{\max_{2 \leq B \leq t \wedge w} Z_B(t) = \max_{B_1(t) \leq B \leq t} Z_B(t)\right\}, \\
    A_{2,t} &= \left\{\max_{B_1(t) \leq B \leq t} Z_B(t) = Z_t(t)\right\},
\end{align*}
for any $N>0$, as long as $t_1 > 4/3$, the following holds
\if\mycolumn1
\begin{align*}
     \PP_{0}(A_{1,t}^{\mathsf{c}}) 
    \leq  & \ 2\left(\exp\left\{-\frac{N b^2}{2^{13} ({\rho}\mathcal{D}^2({p},{q}) \vee 1) K^2}\right\} + \exp\left\{-\frac{({\rho}\mathcal{D}^2({p},{q}) \wedge 1) b^3}{2^{14} {\rho}\mathcal{D}({p},{q}) ^3 K^2}\right\}\right) \\
     &\quad \quad + \frac{b}{16({\rho}\mathcal{D}^2({p},{q}) \vee 1)} \exp\left\{-\frac{N b^2}{2^{15} ({\rho}\mathcal{D}^2({p},{q}) \vee 1) K^2}\right\} 
     \\
     &\quad \quad + \frac{b}{16({\rho}\mathcal{D}^2({p},{q}) \vee 1)} \exp\left\{-\frac{b^2}{2^{13} ({\rho}\mathcal{D}^2({p},{q}) \vee 1) K^2}\right\}.
\end{align*}
\else
\begin{align*}
    & \PP_{0}(A_{1,t}^{\mathsf{c}}) \\
    \leq & 2\left(\exp\left\{-\frac{N b^2}{2^{13} ({\rho}\mathcal{D}^2({p},{q}) \vee 1) K^2}\right\} + \exp\left\{-\frac{({\rho}\mathcal{D}^2({p},{q}) \wedge 1) b^3}{2^{14} {\rho}\mathcal{D}({p},{q}) ^3 K^2}\right\}\right) \\
     + &\frac{b}{16({\rho}\mathcal{D}^2({p},{q}) \vee 1)} \Bigg(\exp\left\{-\frac{N b^2}{2^{15} ({\rho}\mathcal{D}^2({p},{q}) \vee 1) K^2}\right\} \\
     + &\exp\left\{-\frac{b^2}{2^{13} ({\rho}\mathcal{D}^2({p},{q}) \vee 1) K^2}\right\}\Bigg).
\end{align*}
\fi
Therefore, as $b \rightarrow \infty$, we have
\if\mycolumn1
\begin{align*}
    \PP_{0}\left(A_{t,1}^{\mathsf{c}}\right) \leq & \ \frac{b}{16({\rho}\mathcal{D}^2({p},{q}) \vee 1)} \Bigg(\exp  \left\{-\frac{N b^2}{2^{15} ({\rho}\mathcal{D}^2({p},{q}) \vee 1) K^2}\right\} 
     \\
    & \quad \quad \quad \quad \quad \quad \quad \quad \quad \quad + \exp\left\{-\frac{b^2}{2^{13} ({\rho}\mathcal{D}^2({p},{q}) \vee 1) K^2}\right\}\Bigg) (1+o(1)).
\end{align*}
\else
\begin{align*}
    \PP_{0}\left(A_{t,1}^{\mathsf{c}}\right) \leq & \frac{b}{16({\rho}\mathcal{D}^2({p},{q}) \vee 1)} \Bigg(\exp  \left\{-\frac{N b^2}{2^{15} ({\rho}\mathcal{D}^2({p},{q}) \vee 1) K^2}\right\} \nonumber \\
    &+ \exp\left\{-\frac{b^2}{2^{13} ({\rho}\mathcal{D}^2({p},{q}) \vee 1) K^2}\right\}\Bigg) (1+o(1)).
\end{align*}
\fi
Similarly, as $b \rightarrow \infty$, we have
\if\mycolumn1
\begin{align*}
    \PP_{0}(A_{t,2}^{\mathsf{c}}) \leq  6 \Bigg(&\exp\left\{-\frac{N{\rho}\mathcal{D}^2({p},{q})}{512 K^2}\right\} 
     +  \exp\left\{-\frac{b {\rho}\mathcal{D}({p},{q}) }{2^{13} ({\rho}\mathcal{D}({p},{q})  \vee 1)  K^2}\right\}\Bigg) (1+o(1)).
\end{align*}
\else
\begin{align*}
    \PP_{0}(A_{t,2}^{\mathsf{c}}) \leq  6 \Bigg(&\exp\left\{-\frac{N{\rho}\mathcal{D}^2({p},{q})}{512 K^2}\right\} \\
    & +  \exp\left\{-\frac{b {\rho}\mathcal{D}({p},{q}) }{2^{13} ({\rho}\mathcal{D}({p},{q})  \vee 1)  K^2}\right\}\Bigg) (1+o(1)).
\end{align*}
\fi
\end{lemma}

\begin{proof} 
Firstly, we will deal with $\PP_{0}( A_{t,1}^{\mathsf{c}})$. We let
\begin{equation}\label{eq:t3}
    t_3 = ({\rho}\mathcal{D}({p},{q})  \wedge 1) t / 4
\end{equation}
and denote
\begin{align*}
    \tilde{A}_{1,t} = \left\{ \max_{B_1(t) \leq B \leq t} Z_B(t) > t_3\right\}.
\end{align*}
Then, we have 
\begin{align*}
    \PP_{0}(\tilde{A}_{1,t}) \geq \PP_{0}(Z_t(t) > t_3) = 1 - \PP_{0}(Z_t(t) \leq t_3).
\end{align*}
Notice that for any $t \geq t_1 > 4/3$, we have
\begin{align*}
    \mathbb{E}_{0}[Z_t(t)] = {\rho}\mathcal{D}({p},{q})  \sqrt{t(t-1)} \geq {\rho}\mathcal{D}({p},{q})  \frac{t}{2} \geq 2t_3,
\end{align*}
where the first inequality comes from the fact that $\sqrt{x(x-1)}>x/2, \ \forall x>4/3$ and the second inequality comes from our choice of $t_3$ \eqref{eq:t3}.
Therefore, by Lemma~\ref{lma:scanb_concentration}, we have
\if\mycolumn1
\begin{align}
    \PP_{0}(\tilde{A}_{1,t}^{\mathsf{c}}) \leq & \ \PP_{0}(Z_t(t) \leq t_3) 
    \leq 2\left(\exp\left\{-\frac{N t_3^2}{32K^2}\right\} + \exp\left\{-\frac{t \  t_3^2}{16K^2}\right\}\right) \label{eq:bound_pA1tc} \\
    \leq & \ 2\left(\exp\left\{-\frac{N ({\rho}\mathcal{D}^2({p},{q}) \wedge 1) t^2}{512K^2}\right\} + \exp\left\{-\frac{({\rho}\mathcal{D}^2({p},{q}) \wedge 1) t^3}{256K^2}\right\}\right). \nonumber
\end{align}
\else
\begin{align}
    &\PP_{0}(\tilde{A}_{1,t}^{\mathsf{c}}) \leq \PP_{0}(Z_t(t) \leq t_3) \nonumber \\
    \leq &2\left(\exp\left\{-\frac{N t_3^2}{32K^2}\right\} + \exp\left\{-\frac{t \  t_3^2}{16K^2}\right\}\right) \label{eq:bound_pA1tc}\\
    \leq &2\left(\exp\left\{-\frac{N ({\rho}\mathcal{D}^2({p},{q}) \wedge 1) t^2}{512K^2}\right\} + \exp\left\{-\frac{({\rho}\mathcal{D}^2({p},{q}) \wedge 1) t^3}{256K^2}\right\}\right). \nonumber
\end{align}
\fi
Now, we have
\begin{align*}
    \PP_{0}(A_{1,t}^{\mathsf{c}}) &= \PP_{0}(A_{1,t}^{\mathsf{c}} \cap \tilde{A}_{1,t}^{\mathsf{c}})  + \PP_{0}(A_{1,t}^{\mathsf{c}} \cap \tilde{A}_{1,t}) \\
    &\leq \PP_{0}( \tilde{A}_{1,t}^{\mathsf{c}}) + \PP_{0}(A_{1,t}^{\mathsf{c}} \cap \tilde{A}_{1,t}) \\
    &\leq \PP_{0}( \tilde{A}_{1,t}^{\mathsf{c}}) + \PP_{0}\left(\max_{2\leq B < B_1(t)} Z_{B}(t) > t_3\right) \\
    &\leq \PP_{0}( \tilde{A}_{1,t}^{\mathsf{c}}) + \sum_{B=2}^{B_1(t)} \PP_{0}( Z_{B}(t) > t_3).
\end{align*}
Similarly, notice that $\mathbb{E}_{0}[Z_{B}(t)] \leq {\rho}\mathcal{D}({p},{q})  B_1(t) \leq t_3/2$, we have 
\if\mycolumn1
\begin{align*}
     \sum_{B=2}^{B_1(t)} \PP_{0}( Z_{B}(t) > t_3) 
    \leq& \ 2B_1(t) \left(\exp\left\{-\frac{N (t_3/2)^2}{32K^2}\right\} + \exp\left\{-\frac{2(t_3/2)^2}{16K^2}\right\}\right)\\
    \leq& \ \frac{t}{4({\rho}\mathcal{D}({p},{q})  \vee 1)} \Bigg(\exp\left\{-\frac{N ({\rho}\mathcal{D}^2({p},{q}) \wedge 1) t^2}{2048 K^2}\right\} \\
    &\quad \quad \quad \quad \quad \quad \quad\quad \ + \exp\left\{-\frac{({\rho}\mathcal{D}^2({p},{q}) \wedge 1) t^2}{512 K^2}\right\}\Bigg).
\end{align*}
\else
\begin{align*}
    & \sum_{B=2}^{B_1(t)} \PP_{0}( Z_{B}(t) > t_3) \\
    \leq&  2B_1(t) \left(\exp\left\{-\frac{N (t_3/2)^2}{32K^2}\right\} + \exp\left\{-\frac{2(t_3/2)^2}{16K^2}\right\}\right)\\
    \leq&  \frac{t}{4({\rho}\mathcal{D}({p},{q})  \vee 1)} \Bigg(\exp\left\{-\frac{N ({\rho}\mathcal{D}^2({p},{q}) \wedge 1) t^2}{2048 K^2}\right\} \\
    &\quad \quad \quad \quad \quad \quad \quad \quad \quad \quad \ + \exp\left\{-\frac{({\rho}\mathcal{D}^2({p},{q}) \wedge 1) t^2}{512 K^2}\right\}\Bigg).
\end{align*}
\fi
Notice that we already have bounded $\PP_{0}(\tilde{A}_{1,t}^{\mathsf{c}})$ in \eqref{eq:bound_pA1tc}.
Here, we can observe that the rate of the above probability upper bound vanishing to zero depends on $t \in [t_1, w ]$. The choice of $t_1$ in Lemma~\ref{lma1} only needs to satisfy $t_1 < b/(2{\rho}\mathcal{D}({p},{q}) )$. Therefore, we will choose $t_1 = b/(4{\rho}\mathcal{D}({p},{q}) )$ here. This gives us the desired upper bound on $\PP_{0}(A_{1,t}^{\mathsf{c}})$.

Next, we handle event $A_{t,2}$. Recall that we denote the expectation of $Z_{B}(t)$ as $\mu_{B} (t)$, which is evaluated in Lemma~\ref{lma:mean_under_H1}. Denote $\delta_2 = {\rho}\mathcal{D}({p},{q}) /2$, and we have
\begin{align*}
    \mu_B(t) - \mu_{B-1}(t) &= {\rho}\mathcal{D}({p},{q})  (\sqrt{B(B-1)} - \sqrt{(B-1)(B-2)}) \\
    &= \frac{2\sqrt{B-1}}{\sqrt{B} + \sqrt{B-2}}{\rho}\mathcal{D}({p},{q})  > {\rho}\mathcal{D}({p},{q}) .
\end{align*}
Here, the last inequality comes from the fact that $\frac{2\sqrt{B-1}}{\sqrt{B} + \sqrt{B-2}} > 1$, which can be verified by taking the square and rearranging terms.
Now, by definition, we have $$\{|Z_{B}(t) - \mu_B(t)| < \mu_t(t) - \mu_B(t) - \delta_2, \ B \in [B_1(t) ,t-1]\} \cap \{|Z_t(t) - \mu_t(t)| < \delta_2\} \subset A_{t,2}.$$ This gives us
\if\mycolumn1
\begin{align*}
    \PP_{0}(A_{t,2}^{\mathsf{c}}) 
    \leq &\sum_{B = B_1(t)}^{t-1} \PP_{0}(|Z_{B}(t) - \mu_B(t)| < \mu_t(t) - \mu_B(t) - \delta_2) + \PP_{0}(|Z_t(t) - \mu_t(t)| < \delta_2) \\
     \leq &\sum_{B = B_1(t)}^{t-1}  \PP_{0}(|Z_{B}(t) - \mu_B(t)| < \delta_2 (1+2(t-1-B)))   + \PP_{0}(|Z_t(t) - \mu_t(t)| < \delta_2) .
\end{align*}
\else
\begin{align*}
    &\PP_{0}(A_{t,2}^{\mathsf{c}}) \\
    \leq &\sum_{B = B_1(t)}^{t-1} \PP_{0}(|Z_{B}(t) - \mu_B(t)| < \mu_t(t) - \mu_B(t) - \delta_2) \\
    &    + \PP_{0}(|Z_t(t) - \mu_t(t)| < \delta_2) \\
     \leq &\sum_{B = B_1(t)}^{t-1}  \PP_{0}(|Z_{B}(t) - \mu_B(t)| < \delta_2 (1+2(t-1-B))) \\
     &   + \PP_{0}(|Z_t(t) - \mu_t(t)| < \delta_2) .
\end{align*}
\fi
Notice that the last inequality comes from the following derivation: $
    \mu_t(t) - \mu_{B}(t) - \delta_2 = \mu_t(t) - \mu_{B+1}(t) + (\mu_{B+1}(t) - \mu_{B}(t) - \delta_2).$
As we have shown above, $\mu_{B+1}(t) - \mu_{B}(t) > {\rho}\mathcal{D}({p},{q})  = 2\delta_2$, therefore we have 
\begin{align*}
    \mu_t(t) - \mu_{B}(t) - \delta_2 &> \mu_t(t) - \mu_{B+1}(t) +  \delta_2 \\
    &= 2 \delta_2 (\sqrt{t(t-1)} - \sqrt{B(B+1)}) +\delta_2.
\end{align*}
What remains to be proved is $\sqrt{t(t-1)} - \sqrt{B(B+1)} > t-1-B$. Again, this can be verified by taking square and re-arranging terms. We omit further details for this simple derivation here.
Now, by Lemma~\ref{lma:scanb_concentration}, we have
\if\mycolumn1
\begin{align*}
    \PP_{0}(A_{t,2}^{\mathsf{c}}) 
    & \leq 2 \sum_{B = B_1(t)}^{t-1} \Bigg(\exp\left\{-\frac{N[(t-B-1/2)\delta_2]^2}{32K^2}\right\}  + \exp\left\{-\frac{B[(t-B-1/2)\delta_2]^2}{16K^2}\right\} \Bigg)\\
    &\quad + 2 \exp\left\{-\frac{N (\delta_2/2)^2}{32K^2}\right\} + 2 \exp\left\{-\frac{t(\delta_2/2)^2}{16K^2}\right\}\\
    & \leq 2 \sum_{B = B_1(t)}^{t-1} \Bigg(\exp\left\{-\frac{N[(t-B-1/2)\delta_2]^2}{32K^2}\right\}  +  \exp\left\{-\frac{B_1(t)[(t-B-1/2)\delta_2]^2}{16K^2}\right\}\Bigg)\\
    &\quad + 2 \exp\left\{-\frac{N (\delta_2/2)^2}{32K^2}\right\} + 2 \exp\left\{-\frac{t(\delta_2/2)^2}{16K^2}\right\}.
\end{align*}
\else
\begin{align*}
    &\PP_{0}(A_{t,2}^{\mathsf{c}}) \\
    & \leq 2 \sum_{B = B_1(t)}^{t-1} \Bigg(\exp\left\{-\frac{N[(t-B-1/2)\delta_2]^2}{32K^2}\right\} \\
    &\quad \quad \quad \quad \quad \quad + \exp\left\{-\frac{B[(t-B-1/2)\delta_2]^2}{16K^2}\right\} \Bigg)\\
    &\quad + 2 \exp\left\{-\frac{N (\delta_2/2)^2}{32K^2}\right\} + 2 \exp\left\{-\frac{t(\delta_2/2)^2}{16K^2}\right\}\\
    & \leq 2 \sum_{B = B_1(t)}^{t-1} \Bigg(\exp\left\{-\frac{N[(t-B-1/2)\delta_2]^2}{32K^2}\right\} \\
    &\quad \quad \quad \quad \quad \quad +  \exp\left\{-\frac{B_1(t)[(t-B-1/2)\delta_2]^2}{16K^2}\right\}\Bigg)\\
    &\quad + 2 \exp\left\{-\frac{N (\delta_2/2)^2}{32K^2}\right\} + 2 \exp\left\{-\frac{t(\delta_2/2)^2}{16K^2}\right\}.
\end{align*}
\fi

By the summation of geometric series, we can bound the above summation. We omit further details on the derivation. 

We should remark that $B_1(t) = t/(8({\rho}\mathcal{D}({p},{q})  \vee 1)) \geq t_1/(8({\rho}\mathcal{D}({p},{q})  \vee 1))$ can not be overly small. That is to say, we cannot choose overly small $t_1$ in Lemma~\ref{lma1}, which happens to agree with what we have done for event $A_{t,1}$ above. Here, the reason is that the concentration of Scan $B$-statistic depends on not only the number of blocks $N$, but also the block size $B$, which is lower bounded by $B_1(t)$ in the above analysis. The above choice $t_1 = b/(4{\rho}\mathcal{D}({p},{q}) )$ suffices to give a good enough rate of concentration.
Now, as $N \rightarrow \infty$ and $b \rightarrow \infty$, we have 
\if\mycolumn1
\begin{align*}
    \PP_{0} & (A_{t,2}^{\mathsf{c}}) 
     \leq 6 \left(\exp\left\{-\frac{N\delta_2^2}{128 K^2}\right\} +  \exp\left\{-\frac{B_1(t) \delta_2^2}{64K^2}\right\}\right) (1+o(1)).
\end{align*}
\else
\begin{align*}
    \PP_{0} & (A_{t,2}^{\mathsf{c}}) \\
    & \leq 6 \left(\exp\left\{-\frac{N\delta_2^2}{128 K^2}\right\} +  \exp\left\{-\frac{B_1(t) \delta_2^2}{64K^2}\right\}\right) (1+o(1)).
\end{align*}
\fi
Moreover, by our choice $t_1 = b/(4{\rho}\mathcal{D}({p},{q}) )$, we can further bound the above term for all $t \in [t_1, w ]$ as follows:
\if\mycolumn1
\begin{align*}
    \PP_{0}(A_{t,2}^{\mathsf{c}}) \leq 6 \Bigg(&\exp\left\{-\frac{N{\rho}\mathcal{D}^2({p},{q})}{512 K^2}\right\} 
    +  \exp\left\{-\frac{b {\rho}\mathcal{D}({p},{q}) }{2^{13} ({\rho}\mathcal{D}({p},{q})  \vee 1) K^2}\right\}\Bigg) (1+o(1)).
\end{align*}
\else
\begin{align*}
    \PP_{0}(A_{t,2}^{\mathsf{c}}) \leq 6 \Bigg(&\exp\left\{-\frac{N{\rho}\mathcal{D}^2({p},{q})}{512 K^2}\right\} \\
    &+  \exp\left\{-\frac{b {\rho}\mathcal{D}({p},{q}) }{2^{13} ({\rho}\mathcal{D}({p},{q})  \vee 1) K^2}\right\}\Bigg) (1+o(1)).
\end{align*}
\fi
We complete the proof.
\end{proof}

\begin{lemma}\label{lma:expectation_max}
Under $H_1$, where ${p} \not= {q}$ and we assume change occurs at time step 0, and under Assumptions~\ref{assumption:bounded_kernel} and~\ref{assumption:detection}, we choose 
$$ w  \geq \frac{7b}{{\rho}\mathcal{D}({p},{q})},$$ 
where constant ${\rho}$ is defined in \eqref{eq:cpq}, as $b \rightarrow \infty$, for any $N > 0$, we have
\if\mycolumn1
\begin{align*}
    \big|\mathbb{E}_{0}[Z_{T_{w}}] - {\rho}\mathcal{D}({p},{q})  \mathbb{E}_{0}[{T_{w}}]\big| \leq   \Bigg(&{\rho}\mathcal{D}({p},{q})  + 12K\exp\left\{-\frac{N{\rho}\mathcal{D}^2({p},{q})}{512 K^2}\right\} 
    \\
    & +  12K\exp\left\{-\frac{b {\rho}\mathcal{D}({p},{q}) }{2^{13} ({\rho}\mathcal{D}({p},{q})  \vee 1) K^2}\right\}\Bigg) (1+o(1)). 
\end{align*}
\else
\begin{align*}
    \big|\mathbb{E}_{0}[Z_{T_{w}}] - &{\rho}\mathcal{D}({p},{q})  \mathbb{E}_{0}[{T_{w}}]\big|  \leq   \Bigg({\rho}\mathcal{D}({p},{q})  + 12K\exp\left\{-\frac{N{\rho}\mathcal{D}^2({p},{q})}{512 K^2}\right\} \\
    &+  12K\exp\left\{-\frac{b {\rho}\mathcal{D}({p},{q}) }{2^{13} ({\rho}\mathcal{D}({p},{q})  \vee 1) K^2}\right\}\Bigg) (1+o(1)). 
\end{align*}
\fi
\end{lemma}

\begin{proof} 
\textit{Step 1:} We choose $ w  = 4b/({\rho}\mathcal{D}({p},{q})) $ as in Lemma~\ref{lma2}. For any $t \in [t_1, w ]$, recall the definitions of events $A_{1,t}$ and $A_{2,t}$ in \eqref{event:max_at_t}, we will have
\if\mycolumn1
\begin{align*}
    \mathbb{E}_{0} \left[\max_{2 \leq B \leq B_{\rm max}} Z_B(t)\right] &= \mathbb{E}_{0} \left[\max_{B_1 \leq B \leq B_2} Z_B(t); A_1\right] + \mathbb{E}_{0} \left[\max_{2 \leq B \leq B_{\rm max}} Z_B(t); A_1^{\mathsf{c}}\right]\\
    &= \mathbb{E}_{0} \left[\max_{B_1 \leq B \leq B_2} Z_B(t)\right] + \mathbb{E}_{0} \left[\max_{2 \leq B \leq B_{\rm max}} Z_B(t); A_1^{\mathsf{c}}\right]\\
    &= \mathbb{E}_{0} \left[\max_{B_1 \leq B \leq B_2} Z_B(t) - Z_t(t)\right] + \mathbb{E}_{0} \left[Z_t(t)\right] + \mathbb{E}_{0} \left[\max_{2 \leq B \leq B_{\rm max}} Z_B(t); A_1^{\mathsf{c}}\right]\\
     &= {\rho}\mathcal{D}({p},{q})  \sqrt{t(t-1)} \\
     &\quad + \mathbb{E}_{0} \left[\max_{B_1 \leq B \leq B_2} Z_B(t) - Z_t(t);A_2^{\mathsf{c}}\right]  + \mathbb{E}_{0} \left[\max_{2 \leq B \leq B_{\rm max}} Z_B(t); A_1^{\mathsf{c}}\right]\\
     &=  {\rho}\mathcal{D}({p},{q})  t + {\rho}\mathcal{D}({p},{q})  (\sqrt{t(t-1)} - t) \\
     &\quad +\mathbb{E}_{0} \left[\max_{B_1 \leq B \leq B_2} Z_B(t) - Z_t(t);A_2^{\mathsf{c}}\right]  + \mathbb{E}_{0} \left[\max_{2 \leq B \leq B_{\rm max}} Z_B(t); A_1^{\mathsf{c}}\right].
\end{align*}
\else
\begin{align*}
&\mathbb{E}_{0} \left[\max_{B \in [2:w] } Z_B(t)\right]\\
     = &\mathbb{E}_{0} \left[\max_{B \in [B_1:B_2]} Z_B(t); A_1\right] + \mathbb{E}_{0} \left[\max_{B \in [2:w] } Z_B(t); A_1^{\mathsf{c}}\right]\\
    = &\mathbb{E}_{0} \left[\max_{B \in [B_1:B_2]} Z_B(t)\right] + \mathbb{E}_{0} \left[\max_{B \in [2:w] } Z_B(t); A_1^{\mathsf{c}}\right]\\
    = &\mathbb{E}_{0} \left[Z_t(t)\right] + \mathbb{E}_{0} \left[\max_{B \in [B_1:B_2]} Z_B(t) - Z_t(t)\right] \\
    & + \mathbb{E}_{0} \left[\max_{B \in [2:w] } Z_B(t); A_1^{\mathsf{c}}\right]\\
     =&  {\rho}\mathcal{D}({p},{q})  t + {\rho}\mathcal{D}({p},{q})  (\sqrt{t(t-1)} - t) + \mathbb{E}_{0} \left[\max_{B \in [2:w] } Z_B(t); A_1^{\mathsf{c}}\right]\\
      & +\mathbb{E}_{0} \left[\max_{B \in [B_1:B_2]} Z_B(t) - Z_t(t);A_2^{\mathsf{c}}\right].
\end{align*}
\fi

As we have shown in Lemma~\ref{lma2},
we can easily bound $\PP_{0}( A_1^{\mathsf{c}}), \PP_{0}( A_2^{\mathsf{c}})$. 
We have, $\forall t \in [t_1, w ]$, 
\if\mycolumn1
\begin{align}\label{eq:1}
\begin{split}
        \Bigg|\mathbb{E}_{0} \left[\max_{B \in [2:w] } Z_B(t)\right]& -  {\rho}\mathcal{D}({p},{q})  t \Bigg| 
        \leq    {\rho}\mathcal{D}({p},{q})   + 2K(\PP_{0}( A_1^{\mathsf{c}}) + \PP_{0}( A_2^{\mathsf{c}})).
\end{split}
\end{align}
\else
\begin{align}\label{eq:1}
\begin{split}
        \Bigg|\mathbb{E}_{0} \left[\max_{B \in [2:w] } Z_B(t)\right]& -  {\rho}\mathcal{D}({p},{q})  t \Bigg| \\
        \leq  &  {\rho}\mathcal{D}({p},{q})   + 2K(\PP_{0}( A_1^{\mathsf{c}}) + \PP_{0}( A_2^{\mathsf{c}})).
\end{split}
\end{align}
\fi

\vspace{.1in}
\noindent
\textit{Step 2:} 
We are ready to evaluate $\mathbb{E}_{0}[Z_{T_{w}}]$. Notice that $-2K \leq Z_B(t) \leq 2K$, the partition theorem gives us
\begin{align*}
    \mathbb{E}_{0}[Z_{T_{w}}] & = \mathbb{E}_{0}[Z_{T_{w}}|T_{w} < t_1] \PP_{0}(T_{w} < t_1) \\
    &\quad +  \mathbb{E}_{0}[Z_{T_{w}}|T_{w} >  w ] \PP_{0}(T_{w} >  w ) \\
    &\quad +  \mathbb{E}_{0}[Z_{T_{w}}|t_1 \leq T_{w} \leq  w ] \PP_{0}(t_1 \leq T_{w} \leq  w ).
\end{align*}
Combining everything above, we have 
\if\mycolumn1
\begin{align*}
    |\mathbb{E}_{0}[Z_{T_{w}}] - \mathbb{E}_{0}[Z_{T_{w}}|t_1
    \leq T_{w} \leq  w ]| 
    \leq & 2K (\PP_{0}(T_{w} >  w ) + \PP_{0}(T_{w} < t_1)
     + 1- \PP_{0}(t_1 \leq T_{w} \leq  w ))\\
     =&4K (\PP_{0}(T_{w} >  w ) + \PP_{0}(T_{w} < t_1)
     ).
\end{align*}
\else
\begin{align*}
    &|\mathbb{E}_{0}[Z_{T_{w}}] - \mathbb{E}_{0}[Z_{T_{w}}|t_1
    \leq T_{w} \leq  w ]| \\
    \leq & 2K (\PP_{0}(T_{w} >  w ) + \PP_{0}(T_{w} < t_1)
     + 1- \PP_{0}(t_1 \leq T_{w} \leq  w ))\\
     =&4K (\PP_{0}(T_{w} >  w ) + \PP_{0}(T_{w} < t_1)
     ).
\end{align*}
\fi
Recall that in Lemma~\ref{lma1}, we have
bounded both probabilities. 

In the above equation, $\mathbb{E}_{0}[Z_{T_{w}}|t_1 \leq T_{w} \leq  w ]$ can be evaluated by replacing $t$ with $[{T_{w}}|t_1 \leq T_{w} \leq  w ]$ and taking expectation with respect to $[{T_{w}}|t_1 \leq T_{w} \leq  w ]$. Combining everything gives us
\if\mycolumn1
\begin{align*}
    \big|\mathbb{E}_{0}[Z_{T_{w}}] - & {\rho}\mathcal{D}({p},{q})  \mathbb{E}_{0}[{T_{w}}|t_1 \leq T_{w} \leq  w ]\big| \\
    & \leq  {\rho}\mathcal{D}({p},{q})  + 2K\PP_{0}( A_1^{\mathsf{c}}) + 2K \PP_{0}( A_2^{\mathsf{c}}) + 4K \PP_{0}(T_{w} >  w )
    + 4K\PP_{0}(T_{w} < t_1).
\end{align*}
\else
\begin{align*}
    &\left|\mathbb{E}_{0}[Z_{T_{w}}] - {\rho}\mathcal{D}({p},{q})  \mathbb{E}_{0}[{T_{w}}|t_1 \leq T_{w} \leq  w ]\right| \\
    \leq & {\rho}\mathcal{D}({p},{q})  + 2K\PP_{0}( A_1^{\mathsf{c}}) + 2K \PP_{0}( A_2^{\mathsf{c}}) + 4K \PP_{0}(T_{w} >  w ) \\
    &+ 4K\PP_{0}(T_{w} < t_1)
     .
\end{align*}
\fi

\vspace{.1in}
\noindent
\textit{Step 3:} 
Now, we only need to bound the difference between $\mathbb{E}_{0}[{T_{w}}|t_1 \leq T_{w} \leq  w ]$ and $\mathbb{E}_{0}[T_{w}]$. Again, we apply the partition theorem and will get
\begin{align*}
    \mathbb{E}_{0}[{T_{w}}] & = \mathbb{E}_{0}[{T_{w}}|T_{w} < t_1] \PP_{0}(T_{w} < t_1) \\
    &\quad +  \mathbb{E}_{0}[{T_{w}}|T_{w} >  w ] \PP_{0}(T_{w} >  w ) \\
    &\quad +  \mathbb{E}_{0}[{T_{w}}|t_1 \leq T_{w} \leq  w ] \PP_{0}(t_1 \leq T_{w} \leq  w ).
\end{align*}
On one hand, it is easy to show that 
\begin{align*}
    0 \leq \mathbb{E}_{0}[{T_{w}}|T_{w} < t_1] \PP_{0}(T_{w} < t_1) \leq t_1 \PP_{0}(T_{w} < t_1).
\end{align*}
On the other hand, we have 
\begin{align*}
    0 \leq \mathbb{E}_{0}[{T_{w}}|T_{w} >  w ] = \sum_{\tau =  w }^\infty \PP_{0}(T_{w} > \tau) + w\PP_{0}\left(T_{w} >  w \right).
\end{align*}
Here, for $\tau \geq  w  = 4b / {\rho}\mathcal{D}({p},{q}) $, Lemma~\ref{lma1} and its proof therein give us
\if\mycolumn1
\begin{align*}
    &\PP_{0}\left(T_{w} > \tau\right)  \leq \left(2 \exp\left\{-\frac{Nb^2}{512K^2}\right\} +  2\exp\left\{-\frac{ b^3}{128 {\rho}\mathcal{D}({p},{q}) K^2}\right\}\right)^{\tau - 3b/({\rho}\mathcal{D}({p},{q})) }.
\end{align*}
\else
\begin{align*}
    &\PP_{0}\left(T_{w} > \tau\right) \\
    \leq &\left(2 \exp\left\{-\frac{Nb^2}{512K^2}\right\} +  2\exp\left\{-\frac{ b^3}{128 {\rho}\mathcal{D}({p},{q}) K^2}\right\}\right)^{\tau - 3b/({\rho}\mathcal{D}({p},{q})) }.
\end{align*}
\fi

Therefore, we have
\if\mycolumn1
\begin{align*} 
    & \sum_{\tau =  w }^\infty \PP_{0}(T_{w} > \tau) 
    \leq  \frac{\left(2 \exp\left\{-\frac{Nb^2}{512K^2}\right\} +  2\exp\left\{-\frac{ b^3}{128 {\rho}\mathcal{D}({p},{q}) K^2}\right\}\right)^{w - 3b/({\rho}\mathcal{D}({p},{q})) }}{1-\left(2 \exp\left\{-\frac{Nb^2}{512K^2}\right\} +  2\exp\left\{-\frac{ b^3}{128 {\rho}\mathcal{D}({p},{q}) K^2}\right\}\right)}.
\end{align*}
\else
\begin{align*} 
    & \sum_{\tau =  w }^\infty \PP_{0}(T_{w} > \tau) \\
    \leq & \frac{\left(2 \exp\left\{-\frac{Nb^2}{512K^2}\right\} +  2\exp\left\{-\frac{ b^3}{128 {\rho}\mathcal{D}({p},{q}) K^2}\right\}\right)^{w - 3b/({\rho}\mathcal{D}({p},{q})) }}{1-\left(2 \exp\left\{-\frac{Nb^2}{512K^2}\right\} +  2\exp\left\{-\frac{ b^3}{128 {\rho}\mathcal{D}({p},{q}) K^2}\right\}\right)}.
\end{align*}
\fi
Under the technical condition
\begin{equation}\label{eq:b_tech_condi}
    b \geq (4{\rho}\mathcal{D}({p},{q}) /9 \vee 32\sqrt{2 \log 2} K),
\end{equation}
we are able to show
\if\mycolumn1
\begin{align*}
    2 \exp\left\{-\frac{Nb^2}{512K^2}\right\} +  2\exp\left\{-\frac{ b^3}{128 {\rho}\mathcal{D}({p},{q}) K^2}\right\} \leq & 4 \exp\left\{-\frac{b^2}{128K^2} \left(\frac{N}{4} \wedge \frac{b}{{\rho}\mathcal{D}({p},{q}) }\right)\right\} \\
     \leq & \exp\left\{-\frac{b^2}{256K^2} \left(\frac{N}{4} \wedge \frac{b}{{\rho}\mathcal{D}({p},{q}) }\right)\right\} < 1/2.
\end{align*}
\else
\begin{align*}
    &2 \exp\left\{-\frac{Nb^2}{512K^2}\right\} +  2\exp\left\{-\frac{ b^3}{128 {\rho}\mathcal{D}({p},{q}) K^2}\right\} \\
    \leq &4 \exp\left\{-\frac{b^2}{128K^2} \left(\frac{N}{4} \wedge \frac{b}{{\rho}\mathcal{D}({p},{q}) }\right)\right\} \\
     \leq & \exp\left\{-\frac{b^2}{256K^2} \left(\frac{N}{4} \wedge \frac{b}{{\rho}\mathcal{D}({p},{q}) }\right)\right\} < 1/2.
\end{align*}
\fi
Thus, we have
\if\mycolumn1
\begin{align*}
     &\sum_{\tau =  w }^\infty \PP_{0}(T_{w} > \tau)  
     \leq 2\exp\left\{-\frac{b^2}{256K^2} \left(\frac{N}{4} \wedge \frac{b}{{\rho}\mathcal{D}({p},{q}) }\right) (w-3b/({\rho}\mathcal{D}({p},{q})) )\right\}.
\end{align*}
\else
\begin{align*}
     &\sum_{\tau =  w }^\infty \PP_{0}(T_{w} > \tau)  \\
     \leq& 2\exp\left\{-\frac{b^2}{256K^2} \left(\frac{N}{4} \wedge \frac{b}{{\rho}\mathcal{D}({p},{q}) }\right) (w-3b/({\rho}\mathcal{D}({p},{q})) )\right\}.
\end{align*}
\fi
As $b \rightarrow \infty$, the technical condition \eqref{eq:b_tech_condi} automatically holds (since we choose $w \geq 7b/({\rho}\mathcal{D}({p},{q})) $). Therefore, we have
\if\mycolumn1
\begin{align}
    w\PP_{0}\left(T_{w} >  w \right) + \sum_{\tau =  w }^\infty \PP_{0}(T_{w} > \tau) \leq&  w\exp\left\{-\frac{b^2}{256K^2} \left(\frac{N}{4} \wedge \frac{b}{{\rho}\mathcal{D}({p},{q}) }\right)(w-3b/({\rho}\mathcal{D}({p},{q})) )\right\} \nonumber  \\
     +& 2\exp\left\{-\frac{b^2}{128K^2} \left(\frac{N}{4} \wedge \frac{b}{{\rho}\mathcal{D}({p},{q}) }\right) (w-3b/({\rho}\mathcal{D}({p},{q})) )\right\}. \label{eq:wp+sump_bound}
\end{align}
\else
\begin{align}
    & w\PP_{0}\left(T_{w} >  w \right) + \sum_{\tau =  w }^\infty \PP_{0}(T_{w} > \tau) \nonumber \\
    \leq&  w\exp\left\{-\frac{b^2}{256K^2} \left(\frac{N}{4} \wedge \frac{b}{{\rho}\mathcal{D}({p},{q}) }\right)(w-3b/({\rho}\mathcal{D}({p},{q})) )\right\} \nonumber  \\
     +& 2\exp\left\{-\frac{b^2}{128K^2} \left(\frac{N}{4} \wedge \frac{b}{{\rho}\mathcal{D}({p},{q}) }\right) (w-3b/({\rho}\mathcal{D}({p},{q})) )\right\}. \label{eq:wp+sump_bound}
\end{align}
\fi
We notice that for any $a_1, a_2 > 0$ and $a_1 - a_2 \geq \log 2$, we will have
$$x e^{-a_1 x} < e^{-a_2 x}, \quad \forall x>1,$$
since $x < e^{(a_1 - a_2) x} \leq e^{x \log 2} = 2^x$.
Under the technical condition \eqref{eq:b_tech_condi}, we have
\begin{align*}
    \frac{1}{2}\frac{b^2}{256K^2} \left(\frac{N}{4} \wedge \frac{b}{{\rho}\mathcal{D}({p},{q}) }\right) \geq \frac{1}{2}\frac{32^2 \times 2 \times \log 2 K^2}{256K^2} \frac{1}{4} = \log 2.
\end{align*}
That is to say, for $w > 1$ (which holds for large enough $b$ and our choice $w \geq 7b/({\rho}\mathcal{D}({p},{q})) $), we will have 
\if\mycolumn1
\begin{align*}
    0 &\leq w\PP_{0}\left(T_{w} >  w \right) + \sum_{\tau =  w }^\infty \PP_{0}(T_{w} > \tau) \\
    &\leq  \exp\left\{-\frac{b^2}{512K^2} \left(\frac{N}{4} \wedge \frac{b}{{\rho}\mathcal{D}({p},{q}) }\right)(w-6b/({\rho}\mathcal{D}({p},{q})) )\right\} \\
    &\quad + 2\exp\left\{-\frac{b^2}{128K^2} \left(\frac{N}{4} \wedge \frac{b}{{\rho}\mathcal{D}({p},{q}) }\right) (w-3b/({\rho}\mathcal{D}({p},{q})) )\right\} \\
    &\leq 3\exp\left\{-\frac{b^2}{512K^2} \left(\frac{N}{4} \wedge \frac{b}{{\rho}\mathcal{D}({p},{q}) }\right)(w-6b/({\rho}\mathcal{D}({p},{q})) )\right\}.
    \\
    &\leq 3\exp\left\{-\frac{b^3}{512 {\rho}\mathcal{D}({p},{q})  K^2} \left(\frac{N}{4} \wedge \frac{b}{{\rho}\mathcal{D}({p},{q}) }\right)\right\},
\end{align*}
\else
\begin{align*}
    0 &\leq w\PP_{0}\left(T_{w} >  w \right) + \sum_{\tau =  w }^\infty \PP_{0}(T_{w} > \tau) \\
    &\leq  \exp\left\{-\frac{b^2}{512K^2} \left(\frac{N}{4} \wedge \frac{b}{{\rho}\mathcal{D}({p},{q}) }\right)(w-6b/({\rho}\mathcal{D}({p},{q})) )\right\} \\
    &\quad + 2\exp\left\{-\frac{b^2}{128K^2} \left(\frac{N}{4} \wedge \frac{b}{{\rho}\mathcal{D}({p},{q}) }\right) (w-3b/({\rho}\mathcal{D}({p},{q})) )\right\} \\
    &\leq 3\exp\left\{-\frac{b^2}{512K^2} \left(\frac{N}{4} \wedge \frac{b}{{\rho}\mathcal{D}({p},{q}) }\right)(w-6b/({\rho}\mathcal{D}({p},{q})) )\right\}.
    \\
    &\leq 3\exp\left\{-\frac{b^3}{512 {\rho}\mathcal{D}({p},{q})  K^2} \left(\frac{N}{4} \wedge \frac{b}{{\rho}\mathcal{D}({p},{q}) }\right)\right\},
\end{align*}
\fi
where the last inequality comes from our choice $w \geq 7b/({\rho}\mathcal{D}({p},{q})) $.

\vspace{.1in}
\noindent
\textit{Step 4:} 
Finally, combining the above derivations, we have
\if\mycolumn1
\begin{align*}
     |\mathbb{E}_{0}[{T_{w}}] - & \mathbb{E}_{0}[{T_{w}}|t_1 \leq T_{w} \leq  w ]| \\
    & \leq  
    t_1 \PP_{0}(T_{w} < t_1) +  w  (1-\PP_{0}(t_1 \leq T_{w} \leq  w ))
    + \mathbb{E}_{0}[{T_{w}}|T_{w} >  w ] \PP_{0}(T_{w} >  w ).
\end{align*}
\else
\begin{align*}
    & |\mathbb{E}_{0}[{T_{w}}] -  \mathbb{E}_{0}[{T_{w}}|t_1 \leq T_{w} \leq  w ]|\\
    \leq & 
    t_1 \PP_{0}(T_{w} < t_1) +  w  (1-\PP_{0}(t_1 \leq T_{w} \leq  w )) \\
    &+ \mathbb{E}_{0}[{T_{w}}|T_{w} >  w ] \PP_{0}(T_{w} >  w ).
\end{align*}
\fi
This gives us
\if\mycolumn1
\begin{align*}
    \left|\mathbb{E}_{0}[Z_{T_{w}}] - {\rho}\mathcal{D}({p},{q})  \mathbb{E}_{0}[{T_{w}}]\right|
    \leq & 2K(\PP_{0}( A_1^{\mathsf{c}}) + \PP_{0}( A_2^{\mathsf{c}})) + 4K (\PP_{0}(T_{w} >  w ) + \PP_{0}(T_{w} < t_1))\\
    &+ {\rho}\mathcal{D}({p},{q}) \Big(1 + t_1 \PP_{0}(T_{w} < t_1) +  w  (1-\PP_{0}(t_1 \leq T_{w} \leq  w )) \\
    &\quad \quad \quad \quad \quad \quad \quad \quad \quad \quad \quad \ \ +  \mathbb{E}_{0}[{T_{w}}|T_{w} >  w ] \PP_{0}(T_{w} >  w )\Big).
\end{align*}
\else
\begin{align*}
    &\left|\mathbb{E}_{0}[Z_{T_{w}}] - {\rho}\mathcal{D}({p},{q})  \mathbb{E}_{0}[{T_{w}}]\right| \\
    \leq & 2K(\PP_{0}( A_1^{\mathsf{c}}) + \PP_{0}( A_2^{\mathsf{c}})) + 4K (\PP_{0}(T_{w} >  w ) + \PP_{0}(T_{w} < t_1))\\
    + &{\rho}\mathcal{D}({p},{q}) \Big(1 + t_1 \PP_{0}(T_{w} < t_1) +  w  (1-\PP_{0}(t_1 \leq T_{w} \leq  w )) \\
    + & \mathbb{E}_{0}[{T_{w}}|T_{w} >  w ] \PP_{0}(T_{w} >  w )\Big).
\end{align*}
\fi
As $b \rightarrow \infty$, we only consider the leading term and will get 
\if\mycolumn1
\begin{align*}
    &\big|\mathbb{E}_{0}[Z_{T_{w}}] - {\rho}\mathcal{D}({p},{q})  \mathbb{E}_{0}[{T_{w}}]\big|  \leq   \Bigg({\rho}\mathcal{D}({p},{q})  + 12K\exp\left\{-\frac{N{\rho}\mathcal{D}^2({p},{q})}{512 K^2}\right\} \\
    &\quad \quad \quad \quad \quad \quad \quad \quad \quad \quad \quad \quad \quad \quad \quad \   +  12K\exp\left\{-\frac{b {\rho}\mathcal{D}({p},{q}) }{2^{13} ({\rho}\mathcal{D}({p},{q})  \vee 1) K^2}\right\}\Bigg) (1+o(1)). 
\end{align*}
\else
\begin{align*}
    &\big|\mathbb{E}_{0}[Z_{T_{w}}] - {\rho}\mathcal{D}({p},{q})  \mathbb{E}_{0}[{T_{w}}]\big|  \leq   \Bigg({\rho}\mathcal{D}({p},{q})  + 12K\exp\left\{-\frac{N{\rho}\mathcal{D}^2({p},{q})}{512 K^2}\right\} \\
    &\quad \quad \quad \quad \quad  +  12K\exp\left\{-\frac{b {\rho}\mathcal{D}({p},{q}) }{2^{13} ({\rho}\mathcal{D}({p},{q})  \vee 1) K^2}\right\}\Bigg) (1+o(1)). 
\end{align*}
\fi
Now, we complete the proof.
\end{proof}

\section{Proof of Theorem~\ref{thm:opT_{w}max}}\label{appendix:opT_{w}max}

For notational simplicity, let us denote the detection statistic in the oracle detection procedure as follows:
\begin{equation}\label{eq:detect_stat_noWL}
    Z_t^{\rm o} = \max_{B \in [2:t]} {Z}_B(t),
\end{equation}
where ${Z}_B(t)$ is the Scan $B$-statistic defined in \eqref{eq:scan-b}.

\begin{proof} It is important to recognize that we need to specify the choice of window length $w$ when we control ARL by properly choosing the threshold $b$. Thus, the question boils down to the ARL approximation for given $b$ and $w$. We will first show that it requires $w \sim b$ to guarantee $\varepsilon$-performance loss, and therefore, we need to show how to control ARL given the relationship $w \sim b$. Luckily, we can verify that the ARL approximation in Lemma~\ref{lma:arl} allows the window length $w$ to go to infinity at any polynomial rate with respect to $b$ (which includes the case $w \sim  b$). To begin with, we prove that we need $w \sim b$ when there is a change at time zero to meet the $\varepsilon$-performance loss constraint.

\vspace{.1in}
\noindent
\textit{Under $H_1$.} We assume change occurs at time step 0. Notice that the difference between our procedure and the oracle procedure only exists when $T_{w} > w$, otherwise, at time $t \leq w$, the scan region of our procedure would be $B \in [2: w \wedge t]$, which is exactly the same as that of the oracle procedure. That is, we can upper bound the performance loss as follows:
\begin{align*}
    0 \leq \mathbb{E}_0[T_{w}] - \mathbb{E}_0[T_{\rm o}] &\leq \mathbb{E}_{0}[{T_{w}}|T_{w} >  w ]  \PP_{0}\left(T_{w} >  w \right) \\
    &=  w\PP_{0}\left(T_{w} >  w \right) + \sum_{\tau =  w }^\infty \PP_{0}(T_{w} > \tau).
\end{align*}

We re-use the derivations to bound $w\PP_{0}\left(T_{w} >  w \right)$ and $\sum_{\tau =  w }^\infty \PP_{0}(T_{w} > \tau)$ in the \textit{step 3} in the proof of Lemma~\ref{lma:expectation_max}. Notice that, as $\gamma \rightarrow \infty$, by choosing $b \sim \log \gamma$, the technical conditions in that proof hold automatically. Thus, we have
\if\mycolumn1
\begin{align}
    0 \leq \mathbb{E}_0[T_{w}] - \mathbb{E}_0[T_{\rm o}] 
    &\leq  \exp\left\{-\frac{b^2}{512K^2} \left(\frac{N}{4} \wedge \frac{b}{{\rho}\mathcal{D}({p},{q}) }\right)\left(w -  \frac{6b}{{\rho}\mathcal{D}({p},{q})} \right)\right\} \nonumber \\
    &  + 2\exp\left\{-\frac{b^2}{128K^2} \left(\frac{N}{4} \wedge \frac{b}{{\rho}\mathcal{D}({p},{q}) }\right) \left(w -  \frac{3b}{{\rho}\mathcal{D}({p},{q})} \right)\right\} \nonumber \\
    &\leq 3\exp\left\{-\frac{b^2}{512K^2} \left(\frac{N}{4} \wedge \frac{b}{{\rho}\mathcal{D}({p},{q}) }\right)\left(w -  \frac{6b}{{\rho}\mathcal{D}({p},{q})} \right)\right\}. \label{eq:diff_bound}
\end{align}
\else
\begin{align}
    0 &\leq \mathbb{E}_0[T_{w}] - \mathbb{E}_0[T_{\rm o}] \nonumber \\
    &\leq  \exp\left\{-\frac{b^2}{512K^2} \left(\frac{N}{4} \wedge \frac{b}{{\rho}\mathcal{D}({p},{q}) }\right)\left((w -  \frac{6b}{{\rho}\mathcal{D}({p},{q})} \right)\right\} \nonumber \\
    &\quad + 2\exp\left\{-\frac{b^2}{128K^2} \left(\frac{N}{4} \wedge \frac{b}{{\rho}\mathcal{D}({p},{q}) }\right) (w-3b/({\rho}\mathcal{D}({p},{q})) )\right\} \nonumber \\
    &\leq 3\exp\left\{-\frac{b^2}{512K^2} \left(\frac{N}{4} \wedge \frac{b}{{\rho}\mathcal{D}({p},{q}) }\right)(w-6b/({\rho}\mathcal{D}({p},{q})) )\right\}. \label{eq:diff_bound}
\end{align}
\fi

Therefore, for any tolerance $\varepsilon > 0$, we only need the RHS of the above inequality to be smaller than $\varepsilon$. The sufficient condition to guarantee $\varepsilon$-performance loss is
$$w \geq \frac{6b}{{\rho}\mathcal{D}({p},{q}) } + \frac{512K^2 \log {3}/{\varepsilon}}{b^2 (N/4\wedge {b}/{({\rho}\mathcal{D}({p},{q})) })}.$$

However, as we just mentioned, this is just a sufficient condition, meaning that the RHS of the above equation is just an upper bound of the smallest possible window length, i.e., $w^\star$. We denote it by $\bar w$, i.e.,
$$\bar w = \frac{6b}{{\rho}\mathcal{D}({p},{q}) } + \frac{512K^2 \log {3}/{\varepsilon}}{b^2 (N/4\wedge {b}/{({\rho}\mathcal{D}({p},{q})) })}.$$
Now we know that $w^\star \leq \bar w$.
Clearly, this $\bar w$ is not exactly $w^\star$ since: (I) we manually choose $w > 3b/({\rho}\mathcal{D}({p},{q})) $ to help bound $\PP_{0}\left(T_{w} >  w \right)$ in Lemma~\ref{lma1} and (II) the performance loss bound of $\mathbb{E}_0[T_{w}] - \mathbb{E}_0[T_{\rm o}]$ in \eqref{eq:diff_bound} is just an upper bound of the original performance loss bound in \eqref{eq:wp+sump_bound}.

Nevertheless, we will show $w^\star \sim \bar w$, meaning that we characterize the optimal memory-efficient window length up to some constant factor. We will handle (II) first by showing that the upper bound is tight in the sense that the resulting $w$ to satisfy the $\varepsilon$-performance loss remains unchanged. Notice that we have already manually choose $w$ to be $ \cO(b)$, by the upper bound on the performance loss bound \eqref{eq:wp+sump_bound}, we have
\begin{align*}
    w\PP_{0}&\left(T_{w} >  w \right) +  \sum_{\tau =  w }^\infty \PP_{0}(T_{w} > \tau) \\
    \leq & w\exp\left\{-\frac{b^2}{256K^2} \left(\frac{N}{4} \wedge \frac{b}{{\rho}\mathcal{D}({p},{q}) }\right)(w-3b/({\rho}\mathcal{D}({p},{q})) )\right\} (1+o(1)).
\end{align*}
For notational simplicity, we denote 
$$\tilde b = \frac{b^2}{256K^2} \left(\frac{N}{4} \wedge \frac{b}{{\rho}\mathcal{D}({p},{q}) }\right).$$
Then, by the derivation in the \textit{step 3} in the proof of Lemma~\ref{lma:expectation_max}, we have
\if\mycolumn1
\begin{align*}
    \exp\left\{-\tilde b(w-3b/({\rho}\mathcal{D}({p},{q})) )\right\} & \leq w\exp\left\{-\tilde b(w-3b/({\rho}\mathcal{D}({p},{q})) )\right\} \\
    & \leq \exp\left\{-\frac{\tilde b}{2}(w-6b/({\rho}\mathcal{D}({p},{q})) )\right\}.
\end{align*}
\else
\begin{align*}
    \exp\left\{-\tilde b(w-3b/({\rho}\mathcal{D}({p},{q})) )\right\} & \leq w\exp\left\{-\tilde b(w-3b/({\rho}\mathcal{D}({p},{q})) )\right\} \\
    & \leq \exp\left\{-\frac{\tilde b}{2}(w-6b/({\rho}\mathcal{D}({p},{q})) )\right\}.
\end{align*}
\fi
As one can see, if we want LHS to meet the $\varepsilon$-performance loss constraint, the resulting window length $w$ will have the same order as $\bar w$, meaning that our performance bound is tight in terms of the order of the resulting window length to meet the $\varepsilon$-performance loss constraint.

Next, we handle (I). We will show that choosing $w = o(b)$ cannot guarantee $\varepsilon$-performance loss for any fixed $\varepsilon > 0$, given $b \rightarrow \infty$. We need to go back to the proof of Lemma~\ref{lma1} where we bound the probability $\PP_{0}\left(T_{w} >  w \right)$. Now, we choose $w = o(b)$ and will have
\begin{align*}
    \PP_{0}(T_{w} > w) = \PP_{0}\left(\max_{t \in [2:w]} Z_t < b\right) = \prod_{t=2}^{w} \PP_{0}(Z_t < b) = \prod_{t=2}^{w} \prod_{B=2}^{t \wedge w} \PP_{0}(Z_{B}(t) < b).
\end{align*}
Notice that, in Lemma~\ref{lma:mean_under_H1}, we calculated the expectation of the detection statistic, which is of order block size $B$. That is to say, for $B \leq w = o(b)$, we have
\if\mycolumn1
\begin{align*}
    \PP_{0}(Z_{B}(t) < b)  =& 1 - \PP_{0}(Z_{B}(t) \geq b) \\
     \geq & 1-2\exp \left\{-\frac{N (b-B)^2}{128K^2 }\right\} - 2\exp \left\{-\frac{{B}  (b-B)^2}{64K^2}\right\} \\
     \geq & 1-2\exp \left\{-\frac{N (b-w)^2}{128K^2 }\right\} - 2\exp \left\{-\frac{2  (b-w)^2}{64K^2}\right\}.
\end{align*}
\else
\begin{align*}
    &\PP_{0}(Z_{B}(t) < b)  = 1 - \PP_{0}(Z_{B}(t) \geq b) \\
     \geq & 1-2\exp \left\{-\frac{N (b-B)^2}{128K^2 }\right\} - 2\exp \left\{-\frac{{B}  (b-B)^2}{64K^2}\right\} \\
     \geq & 1-2\exp \left\{-\frac{N (b-w)^2}{128K^2 }\right\} - 2\exp \left\{-\frac{2  (b-w)^2}{64K^2}\right\}.
\end{align*}
\fi
Therefore, we have
\if\mycolumn1
\begin{align*}
    & \PP_{0}(T_{w} > w) 
    \geq  \left(1-2\exp \left\{-\frac{N (b-w)^2}{128K^2 }\right\} - 2\exp \left\{-\frac{2  (b-w)^2}{64K^2}\right\}\right)^{w^2}.
\end{align*}
\else
\begin{align*}
    & \PP_{0}(T_{w} > w) \\
    \geq & \left(1-2\exp \left\{-\frac{N (b-w)^2}{128K^2 }\right\} - 2\exp \left\{-\frac{2  (b-w)^2}{64K^2}\right\}\right)^{w^2}.
\end{align*}
\fi
Notice that $(1-o(1/n))^n \rightarrow 1$ as $n \rightarrow \infty$. Clearly, as $b \rightarrow \infty$, for $w = o(b)$ choice, we have
$$2\exp \left\{-\frac{N (b-w)^2}{128K^2 }\right\} + 2\exp \left\{-\frac{2  (b-w)^2}{64K^2}\right\} = o(1/w^2).$$
That is to say, $\PP_{0}(T_{w} > w) \rightarrow 1$ as $b \rightarrow \infty$. Recall that the performance loss bound has leading term $w\PP_{0}(T_{w} > w)$, which cannot shrink to zero (or even diverge) as $b \rightarrow \infty$. Thus, $w = o(b)$ can never satisfy the $\varepsilon$-performance loss constraint, and therefore, our upper bound $w^\star \leq \bar w$ is tight in terms of the order.

\vspace{0.1in}

\noindent
\textit{Under $H_0$.} We choose detection threshold $b$ to control the ARL, given that we choose $w \sim b$. The analytic approximation in Lemma~\ref{lma:arl} can partially answer the question --- we are able to give the relationship between $b$ and ARL to make sure our procedure is a constant ARL procedure in $\cC_{\gamma}$ \eqref{eq:constant_arl_procedure}.
Luckily, our choice $w \sim b$ satisfies the technical condition in \eqref{eq:ARL_condi}, and therefore, we can apply this ARL approximation here.
Even though it is difficult to obtain such closed-from ARL approximation for the oracle procedure, we are able to leverage the concentration results for Scan $B$-statistic (Lemma~\ref{lma:scanb_concentration} in Appendix~\ref{appendix:thm2}) to show it is also a constant ARL procedure in $\cC_{\gamma}$. 

For a given ARL, in our window-limited procedure \eqref{eq:stopping_time}, we approximate the true ARL $\mathbb{E}_{\infty}[T_{w}]$ using Lemma~\ref{lma:arl}. The given ARL constraint is
\begin{equation}\label{eq:condi_ARL}
    \mathbb{E}_{\infty}[T_{w}] \geq \gamma.
\end{equation}

For simplicity, we start with the coarse approximation under the Gaussian assumption (i.e., we only use the first and second-order moments' information). 
By \eqref{eq:ARL_Gaussian_assumption} as well as $ w  \sim b$, we have:
\begin{equation}\label{Bmax_cond1}
    \begin{split}
        \mathbb{E}_{\infty}[T_{w}] \sim e^{b^2/2}.
    \end{split}
\end{equation}
To satisfy condition~\eqref{eq:condi_ARL}, we need 
\begin{equation}\label{eq:b_choice_1}
    b = a \sqrt{\log \gamma}, \quad a \geq \sqrt{2}.
\end{equation}

In follow-up discussion of Theorem 4.2 \citep{li2019scan}, they explicitly claimed $\mathbb{E}_{\infty}[T_{w}] = \cO(e^{b^2})$ and proposed to consider $b = \cO(\sqrt{\log \gamma})$ in order to meet the ARL constraint \eqref{eq:condi_ARL}, which agree with \eqref{Bmax_cond1} and \eqref{eq:b_choice_1}, respectively. However, we may notice that $b = \cO(\sqrt{\log \gamma})$ might violate the lower bound in \citet{lorden1971procedures}, which is known as the best achievable EDD.
Recall that the KL divergence is a fundamental information-theoretic quantity that characterizes the hardness of the hypothesis testing problem. The Lorden's lower bound of EDD is on the order of $\log \gamma$ divided by the KL divergence, which will be larger than what we get above asymptotically since the KL divergence is treated as constant in our setting. 
This contradiction comes from the very coarse ARL approximation (see Figure~\ref{fig:arl_approx}), and  if we incorporate the first $2n$-th order moments' information in the ARL approximation (Lemma~\ref{lma:arl}), we will have
$$\psi_B(\theta) \approx \sum_{i=1}^{2n} \frac{\mathbb{E}_{\infty} [Z_B^i(t)]}{i!} \theta^i.$$
Note that solving $\dot{\psi_B}(\theta_B) = b$ analytically is very challenging, but we can fix $n$ and obtain
$$\theta_B \sim b^{1/(2n-1)},$$ 
when we consider the $b \rightarrow \infty$ limit, given that the $2n$-th moment of the Scan $B$-statistic will exist and not vanish to zero (Assumption~\ref{assumption:bounded_kernel}).
Plugging this into Lemma~\ref{lma:arl} gives us
\begin{equation}\label{eq:ARL_approx_n}
        \mathbb{E}_{\infty}[T_{w}]  \sim \frac{e^{b^{2n/(2n-1)}}}{b^{\frac{4n-4}{2n-1}}}.
\end{equation}
This yields a more strict condition on the threshold to be the ARL constraint \eqref{eq:condi_ARL}, i.e., 
\begin{equation}\label{eq:b_choice_2}
    b \sim ({\log \gamma})^{1-\frac{1}{2n}}.
\end{equation}
We should remark that this is also numerically verified by Figure~\ref{fig:arl_approx} in our experiment --- considering higher order moments will lead to better approximation, whereas failure to do so will result in ``underestimating'' threshold $b$. As one can see, the choice of detection threshold in \eqref{eq:b_choice_1} will not satisfy the above requirement \eqref{eq:b_choice_2}. Indeed, \eqref{eq:b_choice_1} corresponds to $n=1$ case in \eqref{eq:b_choice_1}. Thus, the threshold choice \eqref{eq:b_choice_1} will fail to meet the given ARL constraint asymptotically.

Now, both the numerical evidence and our theoretical analysis point out that the ARL approximation should involve all moments' information to meet the given ARL constraint. 
In addition, recall that we also want to obtain the smallest possible window length $ w $, which is in the same order as the detection threshold $b$ by our previous analysis.
Thus, we give the optimal order of $b$ as follows:
$$b \sim {\log \gamma}.$$

Although our current analysis cannot verify the conjecture that $\mathbb{E}_{\infty}[T_{w}] = \cO(e^{c_3b})$ for some constant $c_3 > 0$, \eqref{eq:ARL_approx_n} can tell us as $b \rightarrow \infty$
$$e^{c_3b}/\mathbb{E}_{\infty}[T_{w}] \rightarrow 0.$$
This tells us that our procedure will keep running for a very long time when there is no change, and the run length will be much larger than the detection threshold $b$. At time step $ t >  w $, we will have
$$\max_{2\leq B \leq t} Z_B(t) = Z_t^{\rm o} \geq Z_t = \max_{2\leq B \leq  w } Z_B(t).$$
This implies the oracle procedure will stop earlier since it has larger detection statistics (for the same detection threshold). Here, approximating the ARL for the oracle procedure is not trivial, and therefore, the detection threshold for both procedures is selected based on Lemma~\ref{lma:arl}. By the above analysis, for $b$ selected as $$b \sim {\log \gamma},$$
we know that 
\begin{align*}
    \mathbb{E}_{\infty}[T_{w}] &\geq \gamma,\\
    \mathbb{E}_{\infty}[T_{w}] &\geq \mathbb{E}_{\infty}[T_{\rm o}].
\end{align*}
In the following, we will show $\mathbb{E}_{\infty}[T_{\rm o}] \geq \gamma$. By our concentration result in Lemma~\ref{lma:scanb_concentration}, we have
\if\mycolumn1
\begin{align*}
    \PP_{\infty}(T_{\rm o} \leq t_0) = & \ \PP_{\infty}(\max_{1 \leq t \leq t_0} Z_t^{\rm o} > b) 
    = \PP_{\infty}(\max_{1 \leq t \leq t_0} \max_{2 \leq B \leq t} Z_B(t) > b) \\
     = & \ \PP_{\infty}(\cup_{1 \leq t \leq t_0} \cup_{2 \leq B \leq t} \{Z_B(t) > b\}) \\
     \leq &\sum_{1 \leq t \leq t_0} \sum_{2 \leq B \leq t} \PP_{\infty}(Z_B(t) > b) = \cO\left( t_0^2 \exp\left\{-\frac{2 (b/2)^2}{16K^2}\right\}\right).
\end{align*}
\else
\begin{align*}
    &\PP_{\infty}(T_{\rm o} \leq t_0) = \PP_{\infty}(\max_{1 \leq t \leq t_0} Z_t^{\rm o} > b) \\
    = &\PP_{\infty}(\max_{1 \leq t \leq t_0} \max_{2 \leq B \leq t} Z_B(t) > b) \\
     = &\PP_{\infty}(\cup_{1 \leq t \leq t_0} \cup_{2 \leq B \leq t} \{Z_B(t) > b\}) \\
     \leq &\sum_{1 \leq t \leq t_0} \sum_{2 \leq B \leq t} \PP_{\infty}(Z_B(t) > b) = \cO\left( t_0^2 \exp\left\{-\frac{2 (b/2)^2}{16K^2}\right\}\right).
\end{align*}
\fi
This gives us $\PP_{\infty}(T_{\rm o} \leq \gamma)$ drops exponentially fast to zero with increasing $b$.
That is to say, with probability almost one, the oracle procedure will not stop before $t_0 =  \cO(\gamma)$, which implies $\mathbb{E}_{\infty}[T_{\rm o}] \geq \gamma$. 
Now, we complete the proof.

\end{proof}

\section{Additional Experimental Results}\label{appendix:hyperparameter}

In this last section, we present additional information about the experiments and results.

\subsection{Choice of hyperparameters}

\subsubsection{Choice of kernel}

We start by giving a brief review of the ``kernel learning'' literature, aiming to choose the ``right kernel'' for better data representation, thereby improving testing power in kernel two-sample tests \citep{gretton2012kernel} and predictive (or classification) accuracy in kernel ridge regression \citep{chen2023kernel} (or kernel support vector machine \citep{lanckriet2004learning, zien2007multiclass}).
In the literature, there are simple heuristics such as \emph{1-(1)}, the Gaussian RBF kernel coupled with the median heuristic \citep{gretton2012kernel}, and \emph{1-(2)}, maximizing the empirical kernel statistic \citep{fukumizu2009kernel}. More principled approaches typically consider \emph{2-(1)} a linear combination of characteristic kernels $\{k_\ell\}_{\ell = 1}^{L}$ \citep{lanckriet2004learning, zien2007multiclass, gretton2012optimal}, i.e., $k_{\rm lin}(x,x') = \sum \beta_\ell k_\ell(x,x')$, or \emph{2-(2)} parameterizing the kernel with the scaling parameter $\Sigma$ in $k_\Sigma(x,x') = \phi((x-x')^\T \Sigma (x-x'))$ \citep{chen2023kernel}. In \emph{2-(2)}, when $\Sigma$ maps the feature to a subspace (e.g., a diagonal matrix $\Sigma$), this approach can be viewed as feature selection or dimension reduction to remove redundancy \citep{fukumizu2004dimensionality, fukumizu2009kernel}.
However, whether it is a two-sample test problem or a regression/classification problem, the kernel learning objective to learn $\{\beta_\ell\}_{\ell = 1}^{L}$ or $\Sigma$ involves samples from another distribution or class (in our setting, samples from $q$). For example, \citet{gretton2012optimal} used asymptotic normality to select the threshold to guarantee the test level and then optimized (or rather, maximized) the corresponding test's power with respect to the weights $\{\beta_\ell\}_{\ell = 1}^{L}$ of the kernel $k_{\rm lin}$. Apparently, empirically evaluating the test power requires samples from $q$.

In our problem, since we only have data from $p$ and have not seen samples from $q$ yet (i.e., the anomaly/change we aim to detect), we recommend choosing a kernel capable of detecting change from all possibilities. Thus, a ``characteristic'' kernel is a desirable and safe choice satisfying our Assumption (A2), and we will use the Gaussian RBF kernel as recommended by \citet{li2019optimality}.

Before we move on to the next study, we want to clarify that this work mainly focuses on developing a CUSUM-type procedure to boost the detection power for kernels satisfying (A2). We leave the (optimal) kernel design in both theory and practice (when there are only samples from one distribution or class) to future studies. Nevertheless, we mention a possible direction called adversarial kernel learning \citep{li2017mmd}: \citet{li2017mmd} parameterized the kernel as $k_{\rm adv}(x,x') = k(f(x),f(x'))$, where $f$ is an injective mapping (such as a neural network) --- This can be viewed as an extension of \emph{2-(2)}. They adopted the min-max formulation in generative adversarial nets (GANs) \citep{goodfellow2014generative} to learn $f$ that maximizes the empirical kernel statistic with $k_{\rm adv}$ as the discriminator (given only samples from $p$), which closely resembles the idea in \emph{1-(2)}.

\subsubsection{Choice of window length $w$}

\begin{figure}[!htp]
\centerline{
\includegraphics[width = .85\textwidth]{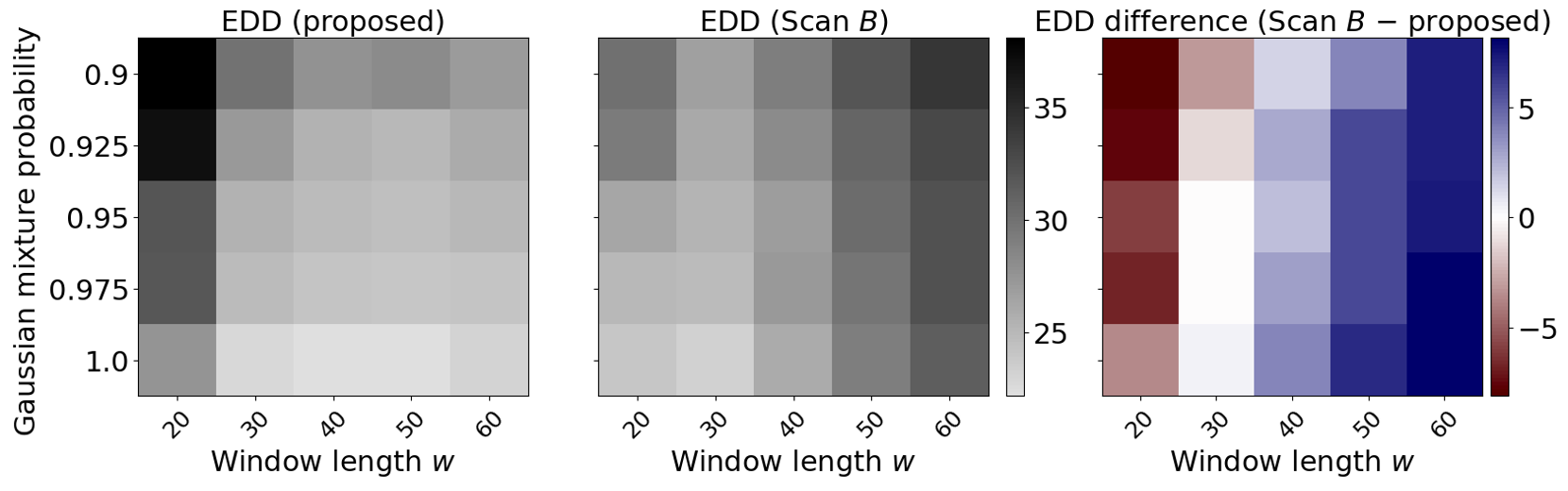}
}
\caption{EDD of the proposed procedure and the Scan $B$-procedure for different window length $w$, where the pre-change block number is fixed to be $N = 30$. The change from ${\cN}(\mathbf{0}_{20}, I_{20})$ to Gaussian Mixture: $(1-p_{\rm GM}) {\cN}(\mathbf{0}_{20}, I_{20}) + p_{\rm GM} {\cN}(1/4 \times \mathbf{1}_{20}, I_{20})$ occurs at $\kappa = 100$.
}
\label{fig:edd_w}
\end{figure}

\color{black} In addition to the leftmost panel in Figure~\ref{fig:EDDARL_compare_bminbmax}, we present further evidence in Figure~\ref{fig:edd_w} to numerically verify (ii). In both results, we can see keep increasing $w$ does not lead to EDD improvement when the window length $w$ is already sufficiently large. \color{black}

\subsection{Effect of number of pre-change blocks $N$}

\begin{figure}[!htp]
\centerline{
\includegraphics[width = \textwidth]{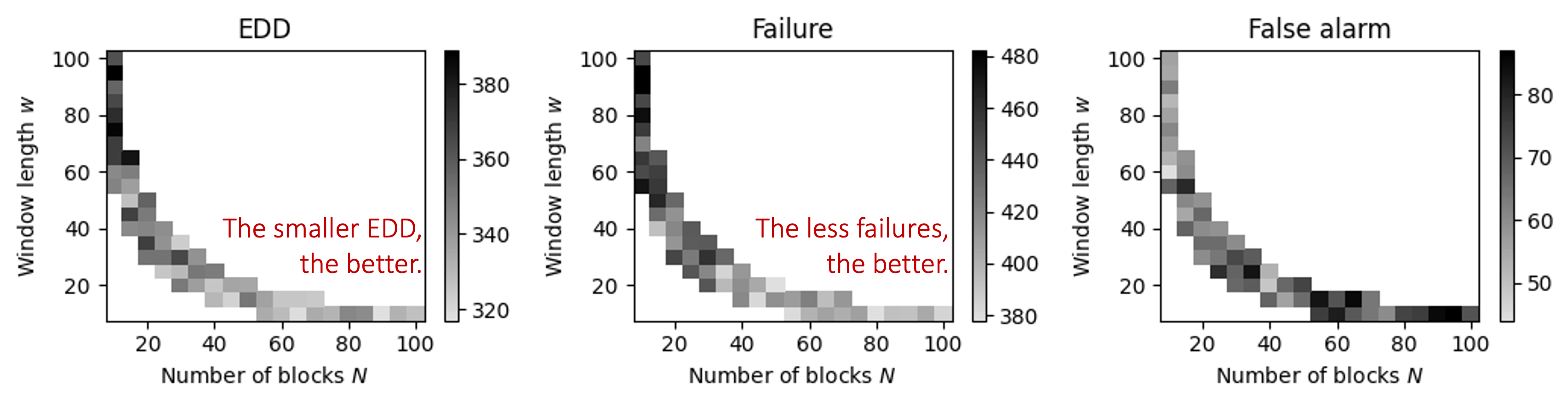}}
\caption{Numerical study on performance affected by hyperparameters $(w,N)$ pairs. We only carry out experiments for $M/2 \leq N  w \leq M$, and the color ``white'' means we do not experiment for the corresponding $(w,N)$.
}
\label{fig:AS_heatmap}
\end{figure}

We also observe that the choice of $N$ may depend on $w$. Consider a change from a 20-dimensional standard Gaussian distribution to a vector autoregressive time series with lag one at $\kappa = 31$ time point. We report EDD over successful trials, failures (no detection within 1000 sequential samples), and false alarms (detection within the first 30 samples) in Figure~\ref{fig:AS_heatmap}, which shows the results over a $(w, N)$-grid, with $N \in \{5,10,\dots,100\}$, $w \in \{5,10,\dots,100\}$ around $M/2 \leq Nw \leq M$ (exclude values outside this range due to inefficient data usage). 
As observed in Figure~\ref{fig:AS_heatmap}, smaller EDD and fewer failures can be achieved by choosing $N$ to be proportional to $w$. 

\section{Supporting results under temporal dependence}\label{appendix:ar1_blocksize}

To illustrate the trade-off in block size selection under temporal dependence, we conducted simulations using a one-dimensional AR(1) process with a change in dependence structure. Under the null hypothesis $H_0$, the data follow
\[
x_t = 0.3\, x_{t-1} + e_t, \qquad e_t \sim \mathcal{N}(0, 1 - 0.3^2),
\]
while under the alternative hypothesis $H_1$,
\[
x_t = 0.7\, x_{t-1} + e_t, \qquad e_t \sim \mathcal{N}(0, 1 - 0.7^2).
\]
The change is assumed to occur at the start of monitoring. We examined block sizes $B \in \{2,4,8,16,32,64\}$, with the number of blocks given by $\lfloor 1000 / B \rfloor$. Detection thresholds were calibrated to achieve an average run length (ARL) of 5000 under $H_0$, using $N_0 = 1000$ reference samples and $N_1 = 10000$ monitoring samples over 200 Monte Carlo runs. Detection performance under $H_1$ was evaluated using 500 Monte Carlo runs.

The results, summarized in Table~\ref{tab:ar1_blocksize}, where the \# failures indicates the number of cases where change is not detected, show that detection performance is best (bold entry) for intermediate block sizes, while both very small and very large block sizes lead to noticeable degradation. This empirical behavior is consistent with the discussion in the main text and supports tuning block sizes over a moderate range under temporal dependence, rather than relying on explicit long-run variance estimation.

\begin{table}
\caption{\label{tab:ar1_blocksize}Effect of block size $B$ on detection performance under an AR(1) dependence change; ARL calibrated to 5000 under $H_0$.}
\begin{tabular}{cccc}
\hline
block size $B$ & threshold & \# failures & EDD \\
\hline
2  & 1.3755 & 109 & 352.13 \\
4  & 0.9853 & 10  & 226.10  \\
8  & 0.5467 & 0   & 136.73  \\
16 & 0.2668 & 0   & {\bfseries 129.12}  \\
32 & 0.1288 & 2   & 182.08  \\
64 & 0.0594 & 7   & 232.63  \\\hline
\end{tabular}
\end{table}

\section{Comparison with other procedures}\label{appendix:add_exp}
In this last section, we present additional experimental configurations and results.

\subsection{Procedures for comparison}\label{appendix:benchmarks}

Before we present additional results on EDD against ARL comparison, let us briefly introduce the detection statistics of the aforementioned benchmark procedures as follows:

\paragraph{Hotelling $T^2$ statistic:}  
At time step $t$, for hypothetical change-point $\kappa < t$, we split the sample points into two parts: $U = (X_1,X_2,\dots,X_M,Y_1,\dots,Y_{\kappa-1}), V = (Y_{\kappa},\dots,Y_t).$ For notational simplicity, we denote the elements in $U$ and $V$ by $U_1,\dots,U_{M + \kappa - 1}$ and $V_1,\dots,V_{t - \kappa + 1}$, respectively. 
We define
$$
T^{2}_t(\kappa)=\frac{(M + \kappa - 1)(t - \kappa + 1)}{M + t}\left(\bar{U}-\bar{V}\right)^{\T} \widehat{\Sigma}^{-1}\left(\bar{U}-\bar{V}\right),
$$where $\bar{U} = \sum_{i=1}^{M + \kappa -1} U_i/(M + \kappa - 1)$, $\bar{V} =  \sum_{i=1}^{t-\kappa+1} V_i/(t - \kappa + 1)$ and $\widehat{\Sigma}$ is the pooled covariance matrix:
\if\mycolumn1
\begin{align*}
    \widehat{\Sigma}=(M + t-2)^{-1}\Bigg(&\sum_{i=1}^{M+\kappa-1}\left(U_{i}-\bar{U}\right)\left(U_{i}-\bar{U}\right)^{\T} +\sum_{i=1}^{t-\kappa+1}\left(V_{i}-\bar{V}\right)\left(V_{i}-\bar{V}\right)^{\T}\Bigg).
\end{align*}
\else
\begin{align*}
    \widehat{\Sigma}=(M + t-2)^{-1}\Bigg(&\sum_{i=1}^{M + \kappa -1}\left(U_{i}-\bar{U}\right)\left(U_{i}-\bar{U}\right)^{\T} \\
    &+\sum_{i=1}^{t-r+1}\left(V_{i}-\bar{V}\right)\left(V_{i}-\bar{V}\right)^{\T}\Bigg).
\end{align*}
\fi
The Hotelling $T^2$ detection statistic is defined as follows:
\begin{equation*}
    S^{{\rm HT}^2}_t = \max_{1\leq \kappa \leq t-1} T^{2}_t(\kappa).
\end{equation*}

\paragraph{Multivariate exponentially weighted moving average
(MEWMA) statistic:}  MEWMA computes
$$S_0^{\text{\rm MEWMA}} = 0, \quad S_t^{\text{\rm MEWMA}} = r_{\rm decay} Y_t + (1-r_{\rm decay}) S_{t-1}^{\text{\rm MEWMA}},$$
where $0 < r_{\rm decay} < 1$ is the decay rate. The detection statistic takes the form
$$(S_t^{\text{\rm MEWMA}})^\T \widehat{\Sigma}_t^{-1} S_t^{\text{\rm MEWMA}},$$ where $$\widehat{\Sigma}_t = \frac{r_{\rm decay}}{2 - r_{\rm decay}}(1-(1-r_{\rm decay})^{2t})\widehat{\Sigma}_0.$$
Here, $\widehat{\Sigma}_0$ is the covariance matrix of the reference samples.

\paragraph{Window-limited GLR statistic:} The general form of W-GLR is reviewed in Section~\ref{sec:para_methods}, and there is no recursion for calculating its detection statistics. This results in a high computational cost. In our experiments, we assume a Gaussian mean shift, and therefore, the detection statistic can be calculated as:
\begin{equation*}
S_t^{\text{\rm W-GLR}}=\max _{(t-w)^+ \leq \kappa<t} \frac{\left(\sum_{j=\kappa+1}^t\left(Y_j-\hat{\mu}\right)\right)^{\T} \widehat{\Sigma}^{-1}\left(\sum_{j=\kappa+1}^t\left(Y_j-\hat{\mu}\right)\right)}{t-\kappa},
\end{equation*}
where $\hat{\mu}$ and $\widehat{\Sigma}$ are estimated mean and covariance from reference data.

\paragraph{Window-limited CUSUM statistic:} \citet{xie2023window} assumed a fully known pre-change density $p$. Similar to W-GLR, the post-change density has form \(q(\cdot;\theta)\) where parameter $\theta \in \Theta$ is unknown. In W-CUSUM, this parameter is estimated using the most recent $w$ sequential data. In our experiments, $p$ is the density of a standard Gaussian distribution, and $q$ is also a Gaussian distribution (but with unknown mean and covariance). We use the maximum likelihood estimator for parameter $\theta$:
\begin{equation*}
\hat{\theta}_t=\underset{\theta \in \Theta}{\arg \max } \sum_{i=0}^{w-1} \log q\left(Y_{t-i}, \theta\right).
\end{equation*}
The W-CUSUM statistic calculation has the following recursive form:
\begin{equation*}
S_0^{\text{\rm W-CUSUM}} = 0, \quad S_t^{\text{\rm W-CUSUM}}= \left(S_{t-1}^{\text{\rm W-CUSUM}}+\log \frac{q\left(Y_{t}, \hat{\theta}_{t-1}\right)}{p(Y_t)}\right)^+.
\end{equation*}

\paragraph{Neural Network based methods:} We refer readers to \citep{hushchyn2020online,lee2023training} for the detection statistics of the NN classifier and NN CUSUM, respectively. In our implementation, we take stride size $10$, training window length $30$ and test window size $30$. The batch size is also $30$.

\paragraph{KCUSUM statistic:} \citet{flynn2019change} replaced the GLR statistic in the classic CUSUM recursive update \eqref{eq:cusum} with the linear-time MMD statistic \citep{gretton2012kernel} and proposed the KCUSUM procedure. 
\if\mycolumn1
The KCUSUM detection statistic has the following recursive update rule:
\begin{align}\label{eq:KCUSUM_update}
    S^{\rm K}_0 = 0, \quad S^{\rm K}_t=\left\{\begin{array}{ll}
\left(S^{\rm K}_{t-1} + h\left(X_{1,t}, X_{2,t}, Y_{t-1}, Y_{t}\right) - {\delta}_{\rm KCUSUM} \right)^+ & \text{ if $t$ is even}, \\
S^{\rm K}_{t-1} & \text{ if $t$ is odd},
\end{array}\right. 
\end{align}
\else
The KCUSUM detection statistic has the following recursive update rule: if $t$ is even, then
\begin{align}
    & S^{\rm K}_t \label{eq:KCUSUM_update} \\
    = & 0 \vee (S^{\rm K}_{t-1} + h\left(x_{1,t}, x_{2,t}, y_{t-1}, y_{t}\right) - \delta),  \nonumber
\end{align}
if $t$ is odd, then
\begin{align*}
    S^{\rm K}_t=  S^{\rm K}_{t-1},
\end{align*}
and $S^{\rm KCUSUM}(0) = 0$. 
\fi
where $h(\cdot,\cdot,\cdot,\cdot)$ is defined in \eqref{eq:kernel_h} and can be viewed as a linear-time MMD statistic with sample size $n = 2$. The hyperparameter ${\delta}_{\rm KCUSUM} > 0$ makes sure a negative drift and is chosen to be ${\delta}_{\rm KCUSUM} = 1/50$ as suggested by the original work. 

\paragraph{Scan $B$-statistic:} For a fixed block size $B_0 \geq 2$, the Scan $B$-statistic $Z_{B_0}(t)$ is defined in \eqref{eq:scan-b}. Here, we choose $B_0 = w$, where $ w $ is the window length of our procedure.

\subsection{Experiment settings}\label{appendix:more_comparison_settings}

In the simulation results presented in Section~\ref{sec:exp-add-procedure}, the settings in Tables~\ref{tab:main_comparison_results}, \ref{tab:additional_iid_comparison_results}, and \ref{tab:additional_dependent_comparison_results} are as follows:

\paragraph{Setting 1:} The change from ${\cN}(\mathbf{0}_{20}, I_{20})$ to Gaussian Mixture:  $7/8 \times {\cN}(1/4 \times \mathbf{1}_{20}, I_{20}) + 1/8 \times {\cN}(\mathbf{0}_{20}, I_{20})$ occurs at $\kappa = 100$.

\paragraph{Setting 2:} The change from ${\cN}(\mathbf{0}_{50}, I_{50})$ to Gaussian Mixture:  $1/2 \times {\cN}(\mathbf{0}_{50}, 1/3 \times I_{50}) + 1/2 \times {\cN}(\mathbf{0}_{50}, I_{50})$ occurs at $\kappa = 100$.

\paragraph{Setting 3:} The change from ${\cN}(\mathbf{0}_{20}, I_{20})$ to $20$-dimensional Laplace distribution (each coordinate is independent ${\rm Lap}(1/2,1/4)$) occurs at $\kappa = 100$.

\paragraph{Setting 4:} the change from ${\cN}(\mathbf{0}_{20}, I_{20})$ to $20$-dimensional Exponential distribution (each coordinate is independent ${\rm Exp}(-1,4/5)$) occurs at $\kappa = 100$.

\paragraph{Setting 5:} The change from ${\cN}(\mathbf{0}_{20}, I_{20})$ to $20$-dimensional Uniform distribution (each coordinate is independent ${\rm U}(1/2 - 1,1/2 + 1)$) occurs at $\kappa = 100$.

\paragraph{Setting 6: Identity matrix} The change from ${\cN}(\mathbf{0}_{20}, I_{20})$ to $20$-dimensional vector autoregressive time series (with Lag 1) occurs at $\kappa = 100$. To be more precise, the post-change data is generated as $Y_t = AY_{t-1} + \epsilon_t$, where $\epsilon_t$ is $20$-dimensional Gaussian noise (with mean zero and variance $1/4$ for each coordinate). The autoregressive coefficient matrix $A = 10^{-3} \times I_{20}$. 

\paragraph{Setting 7: Random diagonal matrix.}
We first randomly generate a $20$-dimensional standard Gaussian vector to create a diagonal matrix. Next, we scale it using the maximum of its absolute entries such that its entries take values from $[-1,1]$. Finally, denote this matrix by $\tilde A$, and we take $A = 10^{-3} \times \tilde A$.

\paragraph{Setting 8: Random symmetric matrix.}
We first generate a $20$-by-$20$ standard Gaussian matrix and then scale it by the maximum of its absolute entries such that its entries take values from $[-1,1]$. Denote this matrix by $\tilde A$, we take $A = 10^{-3} \times (\tilde A + \tilde A^\T)/2$. To ensure the resulting matrix leads to a stationary VAR time series, we check its largest absolute eigenvalue, and if it is greater than one, we will do $A \leftarrow A/1.1$.

\paragraph{Setting 9: Random sparse matrix.}
We first generate a $20$-by-$20$ standard Gaussian matrix and then remove the entries whose absolute value is smaller than $1.8$ to create a sparse matrix. Next, we scale it by the maximum of its absolute entries such that its entries take values from $[-1,1]$. Denote this matrix by $\tilde A$, we take $A = 3/8 \times 10^{-3} \times (\tilde A + \tilde A^\T)/2$. Additionally, to create an even more difficult case, we scale the entire post-change sequence to remove the mean shift compared to the reference data.

\subsection{Additional EDD Comparison Results}\label{appendix:more_comparison_results}

To further demonstrate the good performance of our procedure, we additionally conduct more experiments for (i) change from Gaussian to Gaussian mixture as well as change from Gaussian distribution to (ii) Laplace and (iii) uniform distributions with different ARL constraints. 
 We start with presenting the EDD against the ARL plot for kernel methods and one parametric approach procedure, in which the absence of a dot indicates that the corresponding procedure fails to detect the change before the end of the simulated trajectory $t = 50$. The experiment configurations are the same as those of Figure~\ref{fig:EDDARL_compare_w_scanB}. Our method achieves the quickest detection for the same ARL but also the most robust, capable of detecting even small changes that KCUSUM and parametric Hotelling $T^2$ procedures fail to detect under the considered settings.
In contrast, although KCUSUM occasionally achieves good detection performance, it tends to fail when the change is small or the ARL is large, indicating a lack of robustness.

Next, \color{black} we present results under more scenarios in Tables~\ref{table:EDD_Gmix}, \ref{table:EDD_lap} and \ref{table:EDD_uniform}, respectively.
From these tables, we can observe that our proposed online kernel CUSUM achieves the quickest detection among all benchmarks, even under the settings where the changes are too small for parametric procedures to detect. 
In addition to the aforementioned observations, we can see Hotelling's $T^2$ procedure performs pretty well when the mean shift is significant but fails easily otherwise. 
Under those small mean shift circumstances, kernel methods can easily achieve better performance than the Hotelling $T^2$ procedure; see the first two rows in the table for evidence. 
This observation agrees with our intuition since the Hotelling $T^2$ procedure is a parametric method that is designed to detect the mean shift.

\begin{table}
\caption{\label{table:EDD_Gmix}EDDs for given ARLs under more settings. The change is from standard Gaussian distribution  ${\cN}(\mathbf{0}_{20}, I_{20})$ to Gaussian mixture distribution. 
Non-value ``$-$'' indicates the detection procedure fails to detect the change. Smallest EDDs are highlighted in bold fonts. 
}
\centering
\begin{small}

\resizebox{.9\textwidth}{!}{%
\begin{tabular}{lccccccccccccccccc}
\multicolumn{16}{c}{\large{Setting (i): the change from ${\cN}(\mathbf{0}_{20}, I_{20})$ to
Gaussian mixture $0.3 {\cN}(\mathbf{0}_{20}, I_{20}) + 0.7 {\cN}(\mu \mathbf{1}_{20},\sigma^2  I_{20})$ occurs at $\kappa = 50$.}} \\ 
\toprule[1pt]
& \multicolumn{3}{c}{{$\boldsymbol{(\mu = 0.1, \ \sigma^2 = 0.1)}$}} & \multicolumn{3}{c}{$\boldsymbol{(\mu = 0.1, \ \sigma^2 = 0.3)}$} & \multicolumn{3}{c}{$\boldsymbol{(\mu = 0.1, \ \sigma^2 = 1)}$} & \multicolumn{3}{c}{$\boldsymbol{(\mu = 0.1, \ \sigma^2 = 4)}$} & \multicolumn{3}{c}{$\boldsymbol{(\mu = 0.1, \ \sigma^2 = 9)}$}\\
ARL & 500 & 1000 & 2000 & 500 & 1000 & 2000 & 500 & 1000 & 2000 & 500 & 1000 & 2000 & 500 & 1000 & 2000 \\
\cmidrule(l){2-4} \cmidrule(l){5-7} \cmidrule(l){8-10} \cmidrule(l){11-13} \cmidrule(l){14-16}
Proposed & \textbf{19.2} & \textbf{19.55} & \textbf{21.57} & $-$ & $-$ & $-$ & $-$ & $-$ & $-$ & \textbf{6.99} & \textbf{7.08} & \textbf{7.66} & \textbf{3.47} & \textbf{3.49} & \textbf{3.6} \\ 
Scan $B$ & 28.1 & 28.7 & 30.83 & $-$ & $-$ & $-$ & $-$ & $-$ & $-$ & 15.48 & 15.91 & 17.36 & 9.63 & 9.94 & 10.98 \\ 
KCUSUM & $-$ & $-$ & $-$ & $-$ & $-$ & $-$ & $-$ & $-$ & $-$ & $-$ & $-$ & $-$ & 6.03 & 6.89 & $-$ \\ 
Hotelling $T^2$ & $-$ & $-$ & $-$ & $-$ & $-$ & $-$ & $-$ & $-$ & $-$ & $-$ & $-$ & $-$ & 10.04 & 10.46 & 12.17 \\ 
\cmidrule(l){1-16}

& \multicolumn{3}{c}{{$\boldsymbol{(\mu = 0.3, \ \sigma^2 = 0.1)}$}} & \multicolumn{3}{c}{$\boldsymbol{(\mu = 0.3, \ \sigma^2 = 0.3)}$} & \multicolumn{3}{c}{$\boldsymbol{(\mu = 0.3, \ \sigma^2 = 1)}$} & \multicolumn{3}{c}{$\boldsymbol{(\mu = 0.3, \ \sigma^2 = 4)}$} & \multicolumn{3}{c}{$\boldsymbol{(\mu = 0.3, \ \sigma^2 = 9)}$}\\
ARL & 500 & 1000 & 2000 & 500 & 1000 & 2000 & 500 & 1000 & 2000 & 500 & 1000 & 2000 & 500 & 1000 & 2000 \\
\cmidrule(l){2-4} \cmidrule(l){5-7} \cmidrule(l){8-10} \cmidrule(l){11-13} \cmidrule(l){14-16}
Proposed & \textbf{12.55} & \textbf{12.77} & \textbf{14.19} & \textbf{17.26} & \textbf{17.53} & \textbf{19.46} & $-$ & $-$ & $-$ & \textbf{6.69} & \textbf{6.77} & \textbf{7.33} & \textbf{3.46} & \textbf{3.47} & \textbf{3.61} \\ 
Scan $B$ & 22.36 & 22.93 & 24.84 & 24.87 & 25.5 & 27.61 & $-$ & $-$ & $-$ & 14.85 & 15.28 & 16.69 & 9.53 & 9.85 & 10.91 \\ 
KCUSUM & $-$ & $-$ & $-$ & $-$ & $-$ & $-$ & $-$ & $-$ & $-$ & $-$ & $-$ & $-$ & 6.04 & 6.71 & 8.18 \\ 
Hotelling $T^2$ & $-$ & $-$ & $-$ & $-$ & $-$ & $-$ & $-$ & $-$ & $-$ & 17.34 & 17.88 & $-$ & 8.88 & 9.28 & 10.68 \\ 
\cmidrule(l){1-16}

& \multicolumn{3}{c}{{$\boldsymbol{(\mu = 1, \ \sigma^2 = 0.1)}$}} & \multicolumn{3}{c}{$\boldsymbol{(\mu = 1, \ \sigma^2 = 0.3)}$} & \multicolumn{3}{c}{$\boldsymbol{(\mu = 1, \ \sigma^2 = 1)}$} & \multicolumn{3}{c}{$\boldsymbol{(\mu = 1, \ \sigma^2 = 4)}$} & \multicolumn{3}{c}{$\boldsymbol{(\mu = 1, \ \sigma^2 = 9)}$}\\
ARL & 500 & 1000 & 2000 & 500 & 1000 & 2000 & 500 & 1000 & 2000 & 500 & 1000 & 2000 & 500 & 1000 & 2000 \\
\cmidrule(l){2-4} \cmidrule(l){5-7} \cmidrule(l){8-10} \cmidrule(l){11-13} \cmidrule(l){14-16}
Proposed & \textbf{3.33} & \textbf{3.34} & \textbf{3.45} & \textbf{3.63} & \textbf{3.67} & \textbf{4.05} & \textbf{4.79} & \textbf{4.85} & \textbf{5.26} & \textbf{4.65} & \textbf{4.7} & \textbf{5.15} & \textbf{3.36} & \textbf{3.37} & \textbf{3.46} \\ 
Scan $B$ & 9.64 & 10.01 & 11.11 & 10.11 & 10.46 & 11.69 & 11.39 & 11.81 & 13.23 & 11.16 & 11.56 & 12.94 & 9.07 & 9.36 & 10.41 \\ 
KCUSUM & $-$ & $-$ & $-$ & $-$ & $-$ & $-$ & $-$ & $-$ & $-$ & $-$ & $-$ & $-$ & 5.16 & 5.85 & 6.73 \\ 
Hotelling $T^2$ & 9.25 & 9.56 & 10.53 & 9.12 & 9.36 & 10.39 & 8.72 & 8.96 & 9.91 & 7.13 & 7.42 & 8.34 & 5.52 & 5.71 & 6.43 \\ 
\cmidrule(l){1-16}

& \multicolumn{3}{c}{{$\boldsymbol{(\mu = 1.5, \ \sigma^2 = 0.1)}$}} & \multicolumn{3}{c}{$\boldsymbol{(\mu = 1.5, \ \sigma^2 = 0.3)}$} & \multicolumn{3}{c}{$\boldsymbol{(\mu = 1.5, \ \sigma^2 = 1)}$} & \multicolumn{3}{c}{$\boldsymbol{(\mu = 1.5, \ \sigma^2 = 4)}$} & \multicolumn{3}{c}{$\boldsymbol{(\mu = 1.5, \ \sigma^2 = 9)}$}\\
ARL & 500 & 1000 & 2000 & 500 & 1000 & 2000 & 500 & 1000 & 2000 & 500 & 1000 & 2000 & 500 & 1000 & 2000 \\
\cmidrule(l){2-4} \cmidrule(l){5-7} \cmidrule(l){8-10} \cmidrule(l){11-13} \cmidrule(l){14-16}
Proposed & \textbf{3.12} & \textbf{3.13} & \textbf{3.18} & \textbf{3.16} & \textbf{3.17} & \textbf{3.21} & \textbf{3.3} & \textbf{3.3} & \textbf{3.39} & \textbf{3.62} & \textbf{3.64} & \textbf{3.86} & \textbf{3.32} & \textbf{3.32} & \textbf{3.38} \\ 
Scan $B$ & 7.18 & 7.31 & 8.24 & 7.33 & 7.48 & 8.5 & 8.06 & 8.39 & 9.35 & 9.27 & 9.62 & 10.72 & 8.75 & 8.98 & 10.01 \\ 
KCUSUM & 3.99 & 4.05 & 4.08 & 4.1 & 4.18 & 4.28 & 4.95 & 5.3 & 5.86 & 7.05 & 8.37 & $-$ & 4.78 & 5.07 & 5.62 \\ 
Hotelling $T^2$ & 5.74 & 5.86 & 6.59 & 5.72 & 5.87 & 6.51 & 5.63 & 5.8 & 6.34 & 5.11 & 5.24 & 5.77 & 4.37 & 4.49 & 4.98 \\ 
\cmidrule(l){1-16}

& \multicolumn{3}{c}{{$\boldsymbol{(\mu = 2, \ \sigma^2 = 0.1)}$}} & \multicolumn{3}{c}{$\boldsymbol{(\mu = 2, \ \sigma^2 = 0.3)}$} & \multicolumn{3}{c}{$\boldsymbol{(\mu = 2, \ \sigma^2 = 1)}$} & \multicolumn{3}{c}{$\boldsymbol{(\mu = 2, \ \sigma^2 = 4)}$} & \multicolumn{3}{c}{$\boldsymbol{(\mu = 2, \ \sigma^2 = 9)}$}\\
ARL & 500 & 1000 & 2000 & 500 & 1000 & 2000 & 500 & 1000 & 2000 & 500 & 1000 & 2000 & 500 & 1000 & 2000 \\
\cmidrule(l){2-4} \cmidrule(l){5-7} \cmidrule(l){8-10} \cmidrule(l){11-13} \cmidrule(l){14-16}
Proposed & \textbf{2.89} & \textbf{2.89} & \textbf{2.97} & \textbf{2.95} & \textbf{2.96} & \textbf{3.03} & \textbf{3.11} & \textbf{3.12} & \textbf{3.16} & \textbf{3.31} & \textbf{3.33} & \textbf{3.38} & \textbf{3.28} & \textbf{3.29} & \textbf{3.33} \\ 
Scan $B$ & 5.94 & 6.44 & 7.11 & 6.39 & 6.74 & 7.19 & 7.02 & 7.17 & 7.87 & 8.32 & 8.62 & 9.49 & 8.53 & 8.71 & 9.68 \\ 
KCUSUM & 3.98 & 3.98 & 4.01 & 3.98 & 4 & 4.01 & 4.03 & 4.03 & 4.08 & 4.77 & 5.07 & 5.56 & 4.4 & 4.59 & 4.89 \\ 
Hotelling $T^2$ & 4.31 & 4.38 & 5.17 & 4.34 & 4.44 & 5.08 & 4.37 & 4.45 & 4.92 & 4.14 & 4.24 & 4.6 & 3.72 & 3.82 & 4.16 \\ 
\bottomrule[1pt]
\end{tabular}
}

\end{small}

\end{table}

\begin{table}
\caption{\label{table:EDD_lap}EDDs for given ARLs under more settings. The change is from standard Gaussian distribution  ${\cN}(\mathbf{0}_{20}, I_{20})$ to Laplace distribution. The best results are highlighted in bold fonts.
}
\centering
\begin{small}
\resizebox{.9\textwidth}{!}{%
\begin{tabular}{lccccccccccccccccc}
\multicolumn{16}{c}{\large{Setting (ii): the change from ${\cN}(\mathbf{0}_{20}, I_{20})$ to
${\rm Laplace}(\mu \mathbf{1}_{20}, b \mathbf{1}_{20})$ occurs at $\kappa = 50$.}} \\ 
\toprule[1pt]
& \multicolumn{3}{c}{{$\boldsymbol{(\mu = 0.1, \ b^2 =  0.1)}$}} & \multicolumn{3}{c}{$\boldsymbol{(\mu = 0.1, \ b^2 =  0.3)}$} & \multicolumn{3}{c}{$\boldsymbol{(\mu = 0.1, \ b^2 =  1)}$} & \multicolumn{3}{c}{$\boldsymbol{(\mu = 0.1, \ b^2 =  4)}$} & \multicolumn{3}{c}{$\boldsymbol{(\mu = 0.1, \ b^2 =  9)}$}\\
ARL & 500 & 1000 & 2000 & 500 & 1000 & 2000 & 500 & 1000 & 2000 & 500 & 1000 & 2000 & 500 & 1000 & 2000 \\
\cmidrule(l){2-4} \cmidrule(l){5-7} \cmidrule(l){8-10} \cmidrule(l){11-13} \cmidrule(l){14-16}
Proposed & \textbf{10.16} & \textbf{10.33} & \textbf{11.33} & \textbf{15.91} & \textbf{16.17} & \textbf{17.73} & $-$ & $-$ & $-$ & \textbf{4.64} & \textbf{4.69} & \textbf{5.01} & \textbf{2.37} & \textbf{2.38} & \textbf{2.54} \\ 
Scan $B$ & 21.75 & 22.1 & 23.64 & 25.73 & 26.44 & 28.69 & $-$ & $-$ & $-$ & 11.25 & 11.6 & 12.99 & 6.82 & 7.03 & 7.7 \\ 
KCUSUM & $-$ & $-$ & $-$ & $-$ & $-$ & $-$ & $-$ & $-$ & $-$ & $-$ & $-$ & $-$ & 3.67 & 4.19 & 5.06 \\ 
Hotelling $T^2$ & $-$ & $-$ & $-$ & $-$ & $-$ & $-$ & $-$ & $-$ & $-$ & $-$ & $-$ & $-$ & 7.03 & 7.35 & 8.47 \\ 

\cmidrule(l){1-16}

& \multicolumn{3}{c}{{$\boldsymbol{(\mu = 0.3, \ b^2 =  0.1)}$}} & \multicolumn{3}{c}{$\boldsymbol{(\mu = 0.3, \ b^2 =  0.3)}$} & \multicolumn{3}{c}{$\boldsymbol{(\mu = 0.3, \ b^2 =  1)}$} & \multicolumn{3}{c}{$\boldsymbol{(\mu = 0.3, \ b^2 =  4)}$} & \multicolumn{3}{c}{$\boldsymbol{(\mu = 0.3, \ b^2 =  9)}$}\\
ARL & 500 & 1000 & 2000 & 500 & 1000 & 2000 & 500 & 1000 & 2000 & 500 & 1000 & 2000 & 500 & 1000 & 2000 \\
\cmidrule(l){2-4} \cmidrule(l){5-7} \cmidrule(l){8-10} \cmidrule(l){11-13} \cmidrule(l){14-16}
Proposed & \textbf{6.43} & \textbf{6.51} & \textbf{7.23} & \textbf{8.92} & \textbf{9.04} & \textbf{9.97} & \textbf{17.9} & \textbf{18.26} & \textbf{20.22} & \textbf{4.4} & \textbf{4.46} & \textbf{4.76} & \textbf{2.34} & \textbf{2.36} & \textbf{2.5} \\ 
Scan $B$ & 15.88 & 16.21 & 17.92 & 18.01 & 18.56 & 20.48 & 23.82 & 24.42 & 26.67 & 10.53 & 10.9 & 12.22 & 6.72 & 6.96 & 7.59 \\ 
KCUSUM & $-$ & $-$ & $-$ & $-$ & $-$ & $-$ & $-$ & $-$ & $-$ & $-$ & $-$ & $-$ & 3.62 & 4.12 & 4.83 \\ 
Hotelling $T^2$ & 24.95 & 25.4 & 27.12 & 24.11 & 24.56 & 26.1 & 20.89 & 21.38 & 23.08 & 11.83 & 12.31 & 13.99 & 6.12 & 6.43 & 7.31 \\ 

\cmidrule(l){1-16}

& \multicolumn{3}{c}{{$\boldsymbol{(\mu = 0.5, \ b^2 =  0.1)}$}} & \multicolumn{3}{c}{$\boldsymbol{(\mu = 0.5, \ b^2 =  0.3)}$} & \multicolumn{3}{c}{$\boldsymbol{(\mu = 0.5, \ b^2 =  1)}$} & \multicolumn{3}{c}{$\boldsymbol{(\mu = 0.5, \ b^2 =  4)}$} & \multicolumn{3}{c}{$\boldsymbol{(\mu = 0.5, \ b^2 =  9)}$}\\
ARL & 500 & 1000 & 2000 & 500 & 1000 & 2000 & 500 & 1000 & 2000 & 500 & 1000 & 2000 & 500 & 1000 & 2000 \\
\cmidrule(l){2-4} \cmidrule(l){5-7} \cmidrule(l){8-10} \cmidrule(l){11-13} \cmidrule(l){14-16}
Proposed & \textbf{4.15} & \textbf{4.18} & \textbf{4.6} & \textbf{5.08} & \textbf{5.17} & \textbf{5.65} & \textbf{7.51} & \textbf{7.61} & \textbf{8.32} & \textbf{4.05} & \textbf{4.07} & \textbf{4.35} & \textbf{2.3} & \textbf{2.31} & \textbf{2.45} \\ 
Scan $B$ & 12.07 & 12.56 & 13.49 & 13.06 & 13.41 & 14.49 & 15.1 & 15.51 & 16.97 & 9.58 & 9.98 & 11.19 & 6.62 & 6.86 & 7.48 \\ 
KCUSUM & $-$ & $-$ & $-$ & $-$ & $-$ & $-$ & $-$ & $-$ & $-$ & $-$ & $-$ & $-$ & 3.45 & 3.97 & 4.65 \\ 
Hotelling $T^2$ & 14.66 & 14.83 & 15.45 & 14.32 & 14.52 & 15.28 & 13.04 & 13.32 & 14.35 & 8.42 & 8.75 & 9.94 & 5.28 & 5.52 & 6.27 \\ 

\cmidrule(l){1-16}

& \multicolumn{3}{c}{{$\boldsymbol{(\mu = 0.7, \ b^2 =  0.1)}$}} & \multicolumn{3}{c}{$\boldsymbol{(\mu = 0.7, \ b^2 =  0.3)}$} & \multicolumn{3}{c}{$\boldsymbol{(\mu = 0.7, \ b^2 =  1)}$} & \multicolumn{3}{c}{$\boldsymbol{(\mu = 0.7, \ b^2 =  4)}$} & \multicolumn{3}{c}{$\boldsymbol{(\mu = 0.7, \ b^2 =  9)}$}\\
ARL & 500 & 1000 & 2000 & 500 & 1000 & 2000 & 500 & 1000 & 2000 & 500 & 1000 & 2000 & 500 & 1000 & 2000 \\
\cmidrule(l){2-4} \cmidrule(l){5-7} \cmidrule(l){8-10} \cmidrule(l){11-13} \cmidrule(l){14-16}
Proposed & \textbf{3.81} & \textbf{3.86} & \textbf{3.99} & \textbf{3.9} & \textbf{3.93} & \textbf{4.03} & \textbf{4.58} & \textbf{4.63} & \textbf{4.96} & \textbf{3.61} & \textbf{3.64} & \textbf{3.89} & \textbf{2.25} & \textbf{2.25} & \textbf{2.38} \\ 
Scan $B$ & 8.94 & 9.23 & 10.45 & 9.5 & 9.87 & 11.02 & 10.97 & 11.36 & 12.71 & 8.6 & 8.94 & 10.11 & 6.53 & 6.72 & 7.35 \\ 
KCUSUM & $-$ & $-$ & $-$ & $-$ & $-$ & $-$ & $-$ & $-$ & $-$ & $-$ & $-$ & $-$ & 3.28 & 3.73 & 4.39 \\ 
Hotelling $T^2$ & 9.69 & 9.98 & 10.94 & 9.33 & 9.66 & 10.78 & 8.51 & 8.82 & 9.93 & 6.41 & 6.62 & 7.49 & 4.6 & 4.79 & 5.4 \\ 

\cmidrule(l){1-16}

& \multicolumn{3}{c}{{$\boldsymbol{(\mu = 1.5, \ b^2 =  0.1)}$}} & \multicolumn{3}{c}{$\boldsymbol{(\mu = 1.5, \ b^2 =  0.3)}$} & \multicolumn{3}{c}{$\boldsymbol{(\mu = 1.5, \ b^2 =  1)}$} & \multicolumn{3}{c}{$\boldsymbol{(\mu = 1.5, \ b^2 =  4)}$} & \multicolumn{3}{c}{$\boldsymbol{(\mu = 1.5, \ b^2 =  9)}$}\\
ARL & 500 & 1000 & 2000 & 500 & 1000 & 2000 & 500 & 1000 & 2000 & 500 & 1000 & 2000 & 500 & 1000 & 2000 \\
\cmidrule(l){2-4} \cmidrule(l){5-7} \cmidrule(l){8-10} \cmidrule(l){11-13} \cmidrule(l){14-16}
Proposed & \textbf{2} & \textbf{2} & \textbf{2} & \textbf{2} & \textbf{2} & \textbf{2} & \textbf{2.03} & \textbf{2.03} & \textbf{2.06} & \textbf{2.25} & \textbf{2.27} & \textbf{2.43} & \textbf{2.03} & \textbf{2.04} & \textbf{2.07} \\ 
Scan $B$ & 5 & 5 & 5.77 & 5 & 5.03 & 5.95 & 5.47 & 5.68 & 6.28 & 6.26 & 6.44 & 7.1 & 6.05 & 6.15 & 7 \\ 
KCUSUM & 2.01 & 2.03 & 2.05 & 2.07 & 2.11 & 2.18 & 2.54 & 2.72 & 2.94 & 3.52 & 3.99 & 4.81 & 2.48 & 2.75 & 3.06 \\ 
Hotelling $T^2$ & 4.01 & 4.05 & 4.93 & 4.06 & 4.12 & 4.76 & 4.06 & 4.14 & 4.51 & 3.66 & 3.74 & 4.11 & 3.12 & 3.21 & 3.57 \\ 

\bottomrule[1pt]
\end{tabular}
}

\end{small}


\end{table}

\begin{table}
\caption{\label{table:EDD_uniform}EDDs for given ARLs under more settings. The change is from standard Gaussian distribution  ${\cN}(\mathbf{0}_{20}, I_{20})$ to Uniform distribution. The best results are highlighted in bold fonts.
}
\centering
\begin{small}

\resizebox{.9\textwidth}{!}{%
\begin{tabular}{lccccccccccccccccc}
\multicolumn{16}{c}{\large{Setting (iii): the change from ${\cN}(\mathbf{0}_{20}, I_{20})$ to
$U[(a-b) \mathbf{1}_{20},  (a+b) \mathbf{1}_{20}]$ occurs at $\kappa = 50$.}} \\ 
\toprule[1pt]
& \multicolumn{3}{c}{{$\boldsymbol{(a =  0.1, \ b^2 =  0.1)}$}} & \multicolumn{3}{c}{$\boldsymbol{(a =  0.1, \ b^2 =  0.3)}$} & \multicolumn{3}{c}{$\boldsymbol{(a =  0.1, \ b^2 =  1)}$} & \multicolumn{3}{c}{$\boldsymbol{(a =  0.1, \ b^2 =  4)}$} & \multicolumn{3}{c}{$\boldsymbol{(a =  0.1, \ b^2 =  9)}$}\\
ARL & 500 & 1000 & 2000 & 500 & 1000 & 2000 & 500 & 1000 & 2000 & 500 & 1000 & 2000 & 500 & 1000 & 2000 \\
\cmidrule(l){2-4} \cmidrule(l){5-7} \cmidrule(l){8-10} \cmidrule(l){11-13} \cmidrule(l){14-16}
Proposed & \textbf{6} & \textbf{6} & \textbf{6.93} & \textbf{5.26} & \textbf{5.39} & \textbf{5.99} & \textbf{4} & \textbf{4} & \textbf{4} & \textbf{2.06} & \textbf{2.07} & \textbf{2.18} & \textbf{2} & \textbf{2} & \textbf{2} \\ 
Scan $B$ & 16 & 16.02 & 18 & 13.96 & 14.22 & 15.23 & 10.36 & 10.74 & 11.66 & 6.17 & 6.41 & 7.08 & 5.07 & 5.21 & 5.98 \\ 
KCUSUM & $-$ & $-$ & $-$ & $-$ & $-$ & $-$ & $-$ & $-$ & $-$ & 3.67 & 4.2 & 5.08 & 2.18 & 2.23 & 2.33 \\ 
Hotelling $T^2$ & 29.1 & 29.46 & 30.77 & 20.28 & 20.61 & 22.21 & 12.3 & 12.65 & 13.41 & 5.19 & 5.31 & 5.95 & 3.95 & 4 & 4.26 \\ 

\cmidrule(l){1-16}

& \multicolumn{3}{c}{{$\boldsymbol{(a =  0.3, \ b^2 =  0.1)}$}} & \multicolumn{3}{c}{$\boldsymbol{(a =  0.3, \ b^2 =  0.3)}$} & \multicolumn{3}{c}{$\boldsymbol{(a =  0.3, \ b^2 =  1)}$} & \multicolumn{3}{c}{$\boldsymbol{(a =  0.3, \ b^2 =  4)}$} & \multicolumn{3}{c}{$\boldsymbol{(a =  0.3, \ b^2 =  9)}$}\\
ARL & 500 & 1000 & 2000 & 500 & 1000 & 2000 & 500 & 1000 & 2000 & 500 & 1000 & 2000 & 500 & 1000 & 2000 \\
\cmidrule(l){2-4} \cmidrule(l){5-7} \cmidrule(l){8-10} \cmidrule(l){11-13} \cmidrule(l){14-16}
Proposed & \textbf{4} & \textbf{4} & \textbf{4.51} & \textbf{4} & \textbf{4} & \textbf{4} & \textbf{2.92} & \textbf{2.99} & \textbf{3.52} & \textbf{2} & \textbf{2} & \textbf{2} & \textbf{2} & \textbf{2} & \textbf{2} \\ 
Scan $B$ & 12.63 & 13 & 13.97 & 10.66 & 10.98 & 11.91 & 7.54 & 7.98 & 9.06 & 5.48 & 5.79 & 6.25 & 5 & 5.01 & 5.61 \\ 
KCUSUM & $-$ & $-$ & $-$ & $-$ & $-$ & $-$ & $-$ & $-$ & $-$ & 2.4 & 2.55 & 2.83 & 2.05 & 2.07 & 2.1 \\ 
Hotelling $T^2$ & 15.77 & 15.99 & 16.89 & 13.03 & 13.17 & 14.03 & 7.55 & 7.92 & 9.17 & 4.8 & 4.89 & 5.05 & 3.5 & 3.64 & 3.98 \\ 

\cmidrule(l){1-16}

& \multicolumn{3}{c}{{$\boldsymbol{(a =  0.7, \ b^2 =  0.1)}$}} & \multicolumn{3}{c}{$\boldsymbol{(a =  0.7, \ b^2 =  0.3)}$} & \multicolumn{3}{c}{$\boldsymbol{(a =  0.7, \ b^2 =  1)}$} & \multicolumn{3}{c}{$\boldsymbol{(a =  0.7, \ b^2 =  4)}$} & \multicolumn{3}{c}{$\boldsymbol{(a =  0.7, \ b^2 =  9)}$}\\
ARL & 500 & 1000 & 2000 & 500 & 1000 & 2000 & 500 & 1000 & 2000 & 500 & 1000 & 2000 & 500 & 1000 & 2000 \\
\cmidrule(l){2-4} \cmidrule(l){5-7} \cmidrule(l){8-10} \cmidrule(l){11-13} \cmidrule(l){14-16}
Proposed & \textbf{2.01} & \textbf{2.02} & \textbf{2.95} & \textbf{2} & \textbf{2} & \textbf{2} & \textbf{2} & \textbf{2} & \textbf{2} & \textbf{2} & \textbf{2} & \textbf{2} & \textbf{2} & \textbf{2} & \textbf{2} \\ 
Scan $B$ & 7 & 7 & 8.1 & 6.28 & 6.96 & 7.03 & 5.67 & 5.98 & 6.27 & 5 & 5 & 5.21 & 4.97 & 5 & 5 \\ 
KCUSUM & 6.71 & $-$ & $-$ & 3.88 & 4.77 & 6.2 & 2.37 & 2.55 & 2.84 & 2.01 & 2.02 & 2.03 & \textbf{2} & \textbf{2} & \textbf{2} \\ 
Hotelling $T^2$ & 7 & 7 & 8.07 & 6 & 6.06 & 7 & 5 & 5 & 5.4 & 3.92 & 3.97 & 4.04 & 3 & 3.01 & 3.16 \\ 

\cmidrule(l){1-16}

& \multicolumn{3}{c}{{$\boldsymbol{(a =  1.5, \ b^2 =  0.1)}$}} & \multicolumn{3}{c}{$\boldsymbol{(a =  1.5, \ b^2 =  0.3)}$} & \multicolumn{3}{c}{$\boldsymbol{(a =  1.5, \ b^2 =  1)}$} & \multicolumn{3}{c}{$\boldsymbol{(a =  1.5, \ b^2 =  4)}$} & \multicolumn{3}{c}{$\boldsymbol{(a =  1.5, \ b^2 =  9)}$}\\
ARL & 500 & 1000 & 2000 & 500 & 1000 & 2000 & 500 & 1000 & 2000 & 500 & 1000 & 2000 & 500 & 1000 & 2000 \\
\cmidrule(l){2-4} \cmidrule(l){5-7} \cmidrule(l){8-10} \cmidrule(l){11-13} \cmidrule(l){14-16}
Proposed & \textbf{2} & \textbf{2} & \textbf{2} & \textbf{2} & \textbf{2} & \textbf{2} & \textbf{2} & \textbf{2} & \textbf{2} & \textbf{2} & \textbf{2} & \textbf{2} & \textbf{2} & \textbf{2} & \textbf{2} \\ 
Scan $B$ & 5 & 5 & 5 & 5 & 5 & 5 & 4.12 & 4.99 & 5 & 4 & 4.09 & 5 & 4.05 & 4.69 & 5 \\ 
KCUSUM & \textbf{2} & \textbf{2} & \textbf{2} & \textbf{2} & \textbf{2} & \textbf{2} & \textbf{2} & \textbf{2} & \textbf{2} & \textbf{2} & \textbf{2} & \textbf{2} & \textbf{2} & \textbf{2} & \textbf{2} \\ 
Hotelling $T^2$ & 4 & 4 & 4 & 4 & 4 & 4 & 3 & 3.02 & 3.97 & 2.96 & 2.99 & 3 & 2.05 & 2.1 & 2.6 \\ 

\cmidrule(l){1-16}

& \multicolumn{3}{c}{{$\boldsymbol{(a =  2, \ b^2 =  0.1)}$}} & \multicolumn{3}{c}{$\boldsymbol{(a =  2, \ b^2 =  0.3)}$} & \multicolumn{3}{c}{$\boldsymbol{(a =  2, \ b^2 =  1)}$} & \multicolumn{3}{c}{$\boldsymbol{(a =  2, \ b^2 =  4)}$} & \multicolumn{3}{c}{$\boldsymbol{(a =  2, \ b^2 =  9)}$}\\
ARL & 500 & 1000 & 2000 & 500 & 1000 & 2000 & 500 & 1000 & 2000 & 500 & 1000 & 2000 & 500 & 1000 & 2000 \\
\cmidrule(l){2-4} \cmidrule(l){5-7} \cmidrule(l){8-10} \cmidrule(l){11-13} \cmidrule(l){14-16}
Proposed & \textbf{2} & \textbf{2} & \textbf{2} & \textbf{2} & \textbf{2} & \textbf{2} & \textbf{2} & \textbf{2} & \textbf{2} & \textbf{2} & \textbf{2} & \textbf{2} & \textbf{2} & \textbf{2} & \textbf{2} \\ 
Scan $B$ & 4 & 4 & 5 & 4 & 4 & 5 & 4 & 4 & 5 & 4 & 4 & 5 & 4.01 & 4.45 & 5 \\ 
KCUSUM & \textbf{2} & \textbf{2} & \textbf{2} & \textbf{2} & \textbf{2} & \textbf{2} & \textbf{2} & \textbf{2} & \textbf{2} & \textbf{2} & \textbf{2} & \textbf{2} & \textbf{2} & \textbf{2} & \textbf{2} \\ 
Hotelling $T^2$ & 3 & 3 & 3 & 3 & 3 & 3 & 3 & 3 & 3 & 2.01 & 2.04 & 2.74 & \textbf{2} & \textbf{2} & 2.01 \\ 
\bottomrule[1pt]
\end{tabular}
}

\end{small}

\end{table}

\newpage

\subsection{Real-world example: MNIST Dataset}\label{appendix:more_real_data_exp_MNIST} 
\begin{figure}[!htp]
\centerline{
\includegraphics[width = 0.99\textwidth]{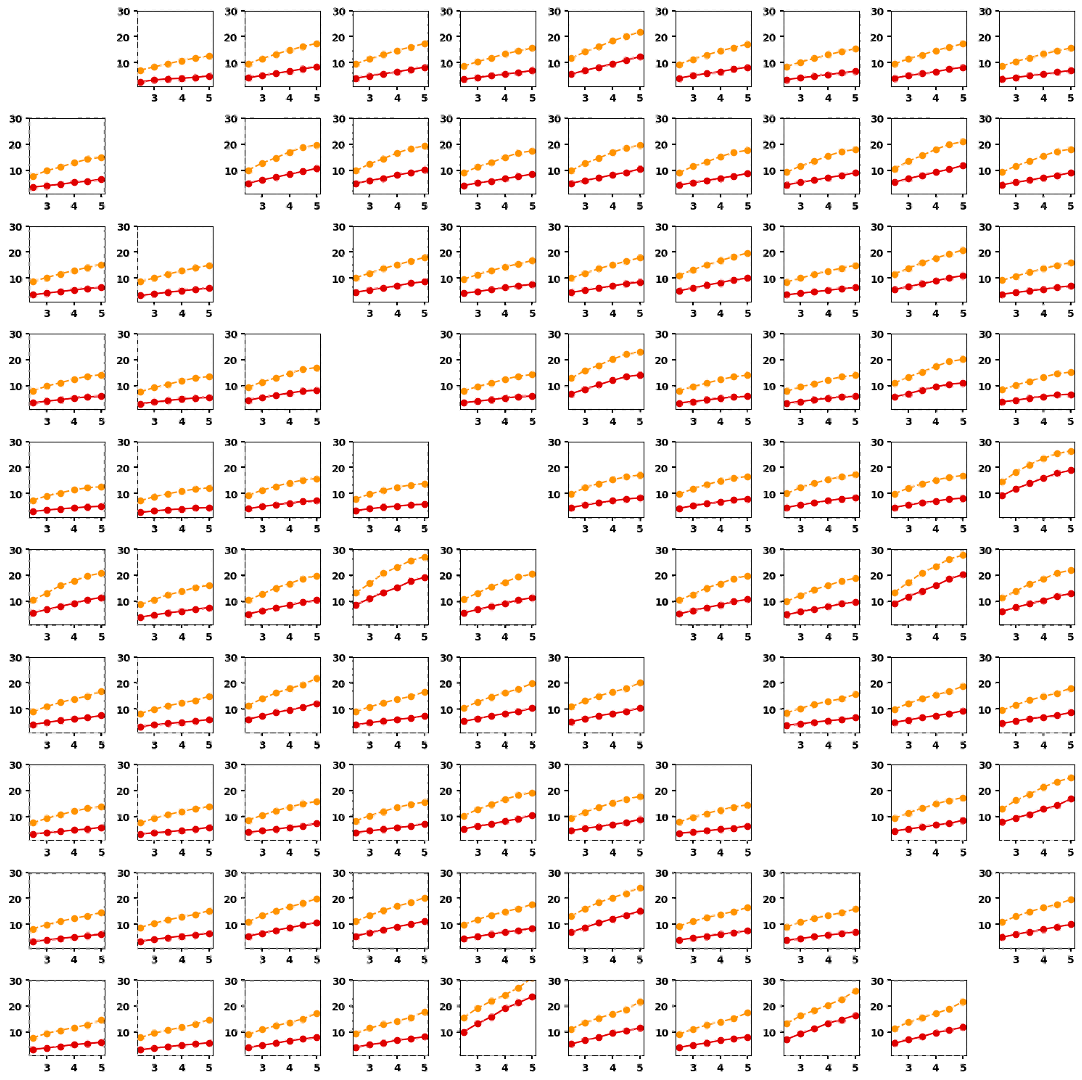} }
\caption{MNIST: Comparison of EDD for given ARL between our proposed (red) and the Scan $B$ (orange) detection procedures in detecting the transition of MNIST handwritten digits. In the 10-by-10 figure above, the $(i,j)$-th sub-figure corresponds to the setting where the reference data is uniform random samples from digit-$i$ and the sequential observations are digit-$j$ uniform random samples. In each sub-figure, the x- and y-axes correspond to the log-ARL and EDD, respectively.
}
\label{fig:MNIST}
\end{figure}

Here, we report the EDD comparison result between our proposed procedure and the Scan $B$-procedure on the raw MNIST dataset in Figure~\ref{fig:MNIST}.

\color{black}

\color{black}

\subsection{Real-World example: Stock market dataset}\label{appendix:more_real_data_exp_ABIDES}

In this experiment, the raw dataset is generated from ABIDES \citep{amrouni2021abides}, {which is available at \url{https://github.com/jpmorganchase/abides-jpmc-public}}.
However, the financial time series is known for its low signal-noise-ratio, and therefore, we use the causal effect feature extracted by \citet{wei2023intags} (see Section 3.2 and Appendix D.2 therein for empirical evidence that supports this feature). In short, this feature estimates the effect \citep{rosenbaum1983central} from placing a limit order to market return with the help of inverse probability weighting. We refer readers to \citet{wei2023intags} for complete details of this extracted feature and the necessary financial background.

\subsection{Real-world example: HASC Dataset}\label{appendix:more_real_data_exp}

The HASC dataset is publicly available at \url{http://hasc.jp/hc2011}. It contains measurements of human activities for six human subjects (indexed by 101, 102, 103, 105, 106, 107) and 6 types of activities (jog, walk, skip, stay, stair down, stair up). 
For those six human subjects, there are 10, 15, 10, 15, 15, 15 repeats taken, respectively; for each single repeat, the sequence length is around $2000$. Those repeats may correspond to activity under different conditions, e.g., floor types (asphalt or carpet), weather (for outdoor activities, including fine, cloudy, rain, and snow), and so on; however, we treat them as different repeats of the same for simplicity.
We need to mention that there are only 3 repeats taken for subject 107 starting up, and thus, this scenario is left out in our experiment.

In this experiment, the Gaussian kernel bandwidth is chosen by using the median heuristic on the whole reference data sequence, which gives us ${{r}} = 0.865$.

For completeness, we carry out the aforementioned experiment for the rest 5 human subjects and report the EDD over successfully detected changes and failures in Table~\ref{table:real_exp_therest}, where the detection threshold is again chosen as the $80 \%$ quantile of the maximum detection statistics under $H_0$ (i.e., the post-change activity is still walking). In this table, we highlight the best procedure for each human subject based on the failures (i.e., fewer failures means a better procedure).

\begin{table}
\caption{\label{table:real_exp_therest}EDD and failure of human activity change detection in HASC dataset. The pre-change activity is walking. The best results are highlighted in bold fonts.
}

\centering
\begin{small}

\resizebox{.8\textwidth}{!}{%
\begin{tabular}{lcccccccccc}
\multicolumn{11}{c}{\large{Subject 102 (in total 15 repeats).}} \\ 
\toprule[1pt]
activity & \multicolumn{2}{c}{{Jog}} & \multicolumn{2}{c}{{Skip}} & \multicolumn{2}{c}{{Stay}} & \multicolumn{2}{c}{{Stair down}} & \multicolumn{2}{c}{{Stair up}}\\
 & EDD & Failure & EDD & Failure & EDD & Failure & EDD & Failure & EDD & Failure \\
\cmidrule(l){2-3} \cmidrule(l){4-5} \cmidrule(l){6-7} \cmidrule(l){8-9} \cmidrule(l){10-11}
Proposed & \textbf{320} & \textbf{13} & $-$ & 15 & $-$ & 15 & $-$ & 15 & $-$ & 15 \\ 
Scan $B$ & \textbf{320} & \textbf{13} & $-$ & 15 & $-$ & 15 & $-$ & 15 & $-$ & 15 \\ 
\textbf{KCUSUM} & 886 & \textbf{13} & \textbf{1034} & \textbf{12} & $-$ & 15 & \textbf{1707} & \textbf{13} & $-$ & 15 \\ 
Hotelling $T^2$ & $-$ & 15 & $-$ & 15 & $-$ & 15 & $-$ & 15 & $-$ & 15 \\ 
\bottomrule[1pt] 

\rule{0pt}{.05in} \\

\multicolumn{11}{c}{\large{Subject 103 (in total 10 repeats).}} \\ 
\toprule[1pt]
activity & \multicolumn{2}{c}{{Jog}} & \multicolumn{2}{c}{{Skip}} & \multicolumn{2}{c}{{Stay}} & \multicolumn{2}{c}{{Stair down}} & \multicolumn{2}{c}{{Stair up}}\\
 & EDD & Failure & EDD & Failure & EDD & Failure & EDD & Failure & EDD & Failure \\
\cmidrule(l){2-3} \cmidrule(l){4-5} \cmidrule(l){6-7} \cmidrule(l){8-9} \cmidrule(l){10-11}
Proposed & 483.5 & 6 & 391.5 & 8 & \textbf{855.25} & \textbf{6} & $-$ & 10 & $-$ & 10 \\ 
Scan $B$ & 637 & 1 & \textbf{95.67} & 7 & 861 & \textbf{6} & $-$ & 10 & $-$ & 10 \\ 
\textbf{KCUSUM} & 348.6 & \textbf{0} & 333.2 & \textbf{0} & $-$ & 10 & \textbf{1266} & \textbf{9} & \textbf{80} & \textbf{9} \\ 
Hotelling $T^2$ & \textbf{320.67} & 7 & 104 & 9 & 875.25 & \textbf{6} & $-$ & 10 & $-$ & 10 \\ 
\bottomrule[1pt] 

\rule{0pt}{.05in} \\

\multicolumn{11}{c}{\large{Subject 105 (in total 15 repeats).}} \\ 
\toprule[1pt]
activity & \multicolumn{2}{c}{{Jog}} & \multicolumn{2}{c}{{Skip}} & \multicolumn{2}{c}{{Stay}} & \multicolumn{2}{c}{{Stair down}} & \multicolumn{2}{c}{{Stair up}}\\
 & EDD & Failure & EDD & Failure & EDD & Failure & EDD & Failure & EDD & Failure \\
\cmidrule(l){2-3} \cmidrule(l){4-5} \cmidrule(l){6-7} \cmidrule(l){8-9} \cmidrule(l){10-11}
\textbf{Proposed} & 260.93 & \textbf{0} & \textbf{49.6} & \textbf{0} & 3426.5 & \textbf{13} & \textbf{206.89} & \textbf{6} & 490.1 & \textbf{5} \\ 
Scan $B$ & \textbf{108} & 3 & 87.13 & \textbf{0} & \textbf{1591} & \textbf{13} & 775.43 & 8 & 884.29 & 8 \\ 
KCUSUM & 395.67 & 3 & 362.53 & \textbf{0} & $-$ & 15 & $-$ & 15 & \textbf{98} & 14 \\ 
Hotelling $T^2$ & 905 & 13 & 178.25 & 3 & 3328.5 & \textbf{13} & 521.2 & 10 & 931 & 8 \\ 
\bottomrule[1pt] 

\rule{0pt}{.05in} \\

\multicolumn{11}{c}{\large{Subject 106 (in total 15 repeats).}} \\ 
\toprule[1pt]
activity & \multicolumn{2}{c}{{Jog}} & \multicolumn{2}{c}{{Skip}} & \multicolumn{2}{c}{{Stay}} & \multicolumn{2}{c}{{Stair down}} & \multicolumn{2}{c}{{Stair up}}\\
 & EDD & Failure & EDD & Failure & EDD & Failure & EDD & Failure & EDD & Failure \\
\cmidrule(l){2-3} \cmidrule(l){4-5} \cmidrule(l){6-7} \cmidrule(l){8-9} \cmidrule(l){10-11}
\textbf{Proposed} & \textbf{121.1} & \textbf{5} & 124.4 & \textbf{0} & \textbf{19.5} & \textbf{13} & 584.5 & \textbf{11} & 430.25 & 7 \\ 
Scan $B$ & 368.14 & 8 & 162.4 & \textbf{0} & \textbf{19.5} & \textbf{13} & \textbf{334} & 12 & \textbf{137.4} & \textbf{5} \\ 
KCUSUM & 563.5 & 7 & 722.29 & 1 & $-$ & 15 & 884 & 13 & $-$ & 15 \\ 
Hotelling $T^2$ & 176.11 & 6 & \textbf{66.4} & \textbf{0} & 23 & \textbf{13} & 629 & 12 & 333.25 & 7 \\ 
\bottomrule[1pt] 

\rule{0pt}{.05in} \\

\multicolumn{11}{c}{\large{Subject 107 (in total 15 repeats).}} \\ 
\toprule[1pt]
activity & \multicolumn{2}{c}{{Jog}} & \multicolumn{2}{c}{{Skip}} & \multicolumn{2}{c}{{Stay}} & \multicolumn{2}{c}{{Stair down}} & \multicolumn{2}{c}{{Stair up}}\\
 & EDD & Failure & EDD & Failure & EDD & Failure & EDD & Failure & EDD & Failure \\
\cmidrule(l){2-3} \cmidrule(l){4-5} \cmidrule(l){6-7} \cmidrule(l){8-9} \cmidrule(l){10-11}
Proposed & $-$ & 15 & \textbf{641.4} & 10 & $-$ & 15 & $-$ & 15 & $-$ & $-$ \\ 
Scan $B$ & $-$ & 15 & $-$ & 15 & $-$ & 15 & $-$ & 15 & $-$ & $-$ \\ 
\textbf{KCUSUM} & \textbf{814.31} & \textbf{2} & 813.78 & \textbf{6} & $-$ & 15 & \textbf{1151.2} & \textbf{10} & $-$ & $-$ \\ 
Hotelling $T^2$ & 1917 & 13 & 1038 & 14 & $-$ & 15 & $-$ & 15 & $-$ & $-$ \\ 
\bottomrule[1pt] 
\end{tabular}
}

\end{small}

\end{table}

From Table~\ref{table:real_exp_therest}, we can make the following additional observations: (i) The KUSUM-type procedures based on non-parametric kernel MMD statistics do perform pretty well. (ii) The detection can be difficult for certain human subjects and certain repeats (since those repeats are carried out under different settings); in contrast, it is easier to detect the change for subjects 105 as well as 106, and we can see our procedure outperforms other benchmarks in the sense that it is the most robust one (i.e., it have the smallest number of failures), and oftentimes it has the smallest EDD. These findings further demonstrate the usefulness of our proposed online kernel CUSUM procedure in practice.

\end{document}